\documentclass[12pt, a4paper, notitlepage]{article}

\usepackage{amssymb, amsthm}
\usepackage{amsmath}
\usepackage[utf8]{inputenc} % allow utf-8 input
\usepackage[T1]{fontenc}    % use 8-bit T1 fonts
\usepackage{hyperref}       % hyperlinks
\usepackage{url}            % simple URL typesetting
\usepackage{booktabs}       % professional-quality tables
\usepackage{amsfonts}       % blackboard math symbols
\usepackage{nicefrac}       % compact symbols for 1/2, etc.
\usepackage{lipsum}		% Can be removed after putting your text content
\usepackage{graphicx}
\usepackage{titlesec}
\usepackage{cancel}
\usepackage{titling}
\usepackage [autostyle, english = american]{csquotes}
\usepackage{changepage}
\usepackage{esvect}
\usepackage{enumitem}

\usepackage{geometry}
\renewenvironment{abstract}
 {\small
  \begin{center}
  \bfseries \abstractname\vspace{-.5em}\vspace{0pt}
  \end{center}
  \list{}{
    \setlength{\leftmargin}{.3cm}%
    \setlength{\rightmargin}{\leftmargin}%
  }%
  \item\relax}
 {\endlist}

\MakeOuterQuote{"}

\title{Gravity Theories via Algebra Gauging}

\author{K. N. Lian \\
	Department of Mathematics\\
	National University of Singapore\\
	Singapore, 119077  \\
}

\date{\today}

\begin{document} 	
\begin{titlingpage}
    \maketitle
\begin{abstract}
	This work presents instructive, yet comprehensive derivation of quantized gravity theories in relativistic, classical, and semi-classical spacetime structure based on the Poincaré, Galilean, and Bargmann algebra, respectively.  The technique of algebra gauging to construct the spacetime dynamics – inspired by the approach of notable previous works – is introduced to complement the standard vielbein formulation.  The key characteristics and anomalies of Galilean gravity will then be analyzed: the degenerate metric structure, the additional degree of freedom in metric connection and the additional necessary conditions of Galilean invariance among others.  General metric connection solution in Galilean spacetime differs fundamentally from that of general relativity; this will be thoroughly investigated and an explicit formula for such solution – equivalent to the parameterization by Hartong and Obers (2015) – shall be derived. Multiple derivations of the Bargmann algebra will be provided, together with both physical and algebraic motivation for the extended Bargmann frame bundle.  Finally, the physical impact of constraining temporal torsion in classical spacetime will be discussed with emphasis on the geometrical interpretation of time foliations. 
\end{abstract}
\end{titlingpage}

\section{Introduction}\label{sec1}

 \paragraph{}
In the 1920s, a geometric formulation of the Newtonian gravity – a geometric theory based on non-relativistic gravity and Newtonian potential – similar in structure to general relativity (GR) was derived by Cartan\cite{Cartan} and Friedrichs \cite{Friedrichs}.  The prominence of these works might be undervalued as at a glance it sounds like a simple limting procedure that is $c\rightarrow\infty$ of GR.  However, the success Cartan and Friedrichs' formulation contradicts Einstein's original idea \cite{MTW} of general covariance being an intrinsic property of GR spacetime.  Despite rising certain symmetry issues (a consequence of dropping Lorentz invariance in place of Galilean), subtracting the relativity from Einstein's gravity does not take away our ability to express the theory in a general covariant sense.  Furthermore, while such non-relativistic theory can be seen as a step backwards from the elegance of GR, it holds a great potential to be incorporated with quantum mechanics.  In the recent pursuit of a quantum theory of gravity, it eludes the so-called "problem of time" by allowing different – and in fact, rather arbitrary – treatments and transformations on the separated dimension that is time.  After decades of further development (for example, \cite{Havas,Kunzle,Dixon}) and generalization beyond the Newtonian potential, this theory – which today goes by Newton-Cartan theory – indeed provides suitable background for some quantum gravity theory candidates, most notably the Hořava–Lifshitz gravity\cite{Horava}.

\paragraph{}
The fundamental difference between classical and Einstein's spacetime is their symmetry group: in a classical theory physics is invariant under spatial rotation and coordinate displacement/translation while in a relativistic one the role of spatial rotation is replaced by Lorentz rotation, which also includes frame boosting – a "rotation with time".  Despite being a basic concept, this fact is the basis of the formulation of geometric theory of Newtonian gravity: the breaking of Lorentz symmetry is the only necessary $c\rightarrow\infty$ limiting procedure from GR to the Newton Cartan theory.  This is true since the knowledge of the algebra/commutation of generators suffices to determine the symmetry transformations of frames as well as connections.  The connection here refers to a general $\Gamma^\lambda_{\mu \nu}$ – not necessarily coordinate basis – which describes how an (inertial) observer's frame transform as function of space and time following a curved spacetime.  Hence, a change of the underlying symmetry would significantly affect the global geometry.  In Newton Cartan for isntance, this is manifested in the notion of "time foliation" as a consequence of time being separated from the spacetime symmetry.

\paragraph{}
The symmetry group of GR is essentially the Poincaré group (with modification on translations, as we shall discussed) whose algebra involves few yet highly symmetrical generators.  In classical spacetime, it follows that each of these generators are split into two, since it is projected onto the separated space and time components.  This classical algebra is called Galilean as it contains Galilean rotations and boosts (these are the spatial and temporal projections of $c\rightarrow\infty$ Lorentz rotation).  Separated from the symmetry of rotation, the boosts now have significantly different algebras.  Translation generators in space and time are also treated differently, where the former becomes the momentum operator and the latter becomes time evolution operator, the Hamiltonian \footnote{In the Bargmann algebra, the time translator becomes Hamiltonian plus the central element representing the rest mass.}.  Just like the boosts, this modification introduces new algebra rules.  Thus, despite being a classical theory, Newton-Cartan would have a much more convoluted algebra.  Roughly speaking, deriving this theory requires twice the effort of deriving GR.  

\paragraph{}
Other than the splitting of generators, quantities such as the vielbeins i.e. the frame of references, connections and metrics will also be projected onto their spatial and temporal components.  For the metric, the result of the split is the introduction of a degenerate and separate spatial metric and temporal metric.  With these geometrical components being degenerate, one encounters computational complications in the theory as the old notion of covariant spacetime metric, inverse metric and inverse vielbeins becomes obsolete. To inverse a general vielbein indexed quantity for example, one requires the incorporation  of its separate spatial and temporal inverses (see the note on Galilean vielbeins on chapter 2). 

\paragraph{}
Furthermore, the formulation of affine connection in classical spacetime has significant deviations from its GR counterpart.  On top of requiring an updated form of $\Gamma^\lambda_{\mu \nu}$ from its "Levi-Civita term + contortion term" form in GR, new terms such as the Newton-Coriolis two-form arise.  Torsion falls under a different light in this theory and its manifestation on $\Gamma^\lambda_{\mu \nu}$ would involve not only the contortion, but also a part of the Levi-Civita term which is closely related to the time vielbeins i.e. the time foliations.  Hence in a classical spacetime, constraints of torsion directly opens up a whole new field of study of time foliations: in mainstream literatures on Newton-Cartan e.g. \cite{HartongObers, Geracie} various constraints are studied which include torsional Newton-Cartan geometry (TNC), twistless torsional Newton-Cartan geometry (TTNC) and torsionless Newton-Cartan geometry (TLNC).  We shall provide a derivation and discussion regarding these conditions in section \ref{sec3.7}. 

\paragraph{}
Finally, unlike GR – whose main brilliance is the fact that Lorentz invariance is automatic – a Galilean formulated spacetime is not inherently Galilean invariant.  For instance, in GR, quantities such as the metric and metric connection are Lorentz invariant by formulation.  This is not the case in the Newton-Cartan geometry, whose covariant metric cannot be Galilean invariant (in particular under boosts) and as such various conditions need to be imposed for the metric connection.  

\paragraph{}
Once the Galilean theory is established, we shall consider the so-called Bargmann algebra \cite{Bargmann}.  This algebra is a central extension of the Galilean algebra.  Mathematically, this merely involves an additional generator of $U(1)$ class – called the central charge – into the mix and postulating it to be a commutation of the Galilean boost and the space translation generator; no other changes are imposed on the Galilean theory as an $U(1)$ generator commutes with all generators.  Physically, it is a corrected version of Galilean algebra which inaccuracy lies in either the assumption of zero mass or the negligence of the quantum operators transformation law, in particular an incorrect momentum transformation under boosts.

\paragraph{}
The methods used to derive these theories include the algebra gauging method which is inspired mainly by the work of Hartong et al. (see for example \cite{HartongObers, HartongObers2}) to complement the usual direct vielbein bundle analysis.

%%%%%%%%%%%%%%%%%%%%%%%%%%%%%%%%%%%%%%%%%%%%%%%%%%%%%%%%%%%%

\section{Poincaré Algebra}
\label{chap1}

\paragraph{}
The Poincaré symmetry group ${\rm {\rm SO(1,d)}}$ contains only two generators\footnote{We shall follow the conventional index notation as follows: $\{a,b,...\}$ denotes spacetime vielbein index, $\{\mu,\nu,...\}$ denotes coordinate spacetime index.  Furthermore, our $\eta$ shall follow the $(+,-,-,-)$ convention.}:  the translation $P_a$ and the Lorentz transformation/rotation $M_{ab}$.

\begin{align*}
[P_a,P_b]&=0\\
[M_{ab},P_c]&=\eta_{bc}P_a- \eta_{ac}P_b\\
[M_{ab},M_{cd}]&=\eta_{ad}M_{bc} - \eta_{ac}M_{bd}+\eta_{bc}M_{ad}-\eta_{bd}M_{ac}
\end{align*}

\paragraph{}
To construct the vielbein formalism and ultimately the spacetime geometry, one can directly define the vielbein bundle and their connections. We shall however include a section where these vielbeins are constructed by gauging the Poincaré algebra, starting only from the above Lie algebra and the gauge transformation rule.   

\paragraph{} 
The above algebra (with corrected translations - see later sections) would recover GR once we gauge it or give it a compatible vielbein system.  This does sound like we have reconciled GR and the relativistic field theories (hence establishing a quantum gravity theory) since the Poincaré elements form the symmetry group of the latter.  The "problem of time" however comes when one tries to treat for example, $P_0$ differently than other $P_a$ where $a\neq0$ as in the case of the Schrodinger's algebra \cite{Schrodinger1, Schrodinger2}.  This motivates the separation of temporal generators from the spatial ones i.e. reverting back to the symmetry groups of non-relativistic physics.  

\paragraph{}
Standard equations and some results pertaining to the frame bundle, vielbein gauge fixing and gauge transformations follow basic textbook formulations, see for example \cite{Isham,Nakahara}.

\subsection{The (co-)frame bundle with Poincaré group structure}
\label{sec1.1}

\begin{figure}[h!]
	\centering
	\includegraphics[width=6.5cm]{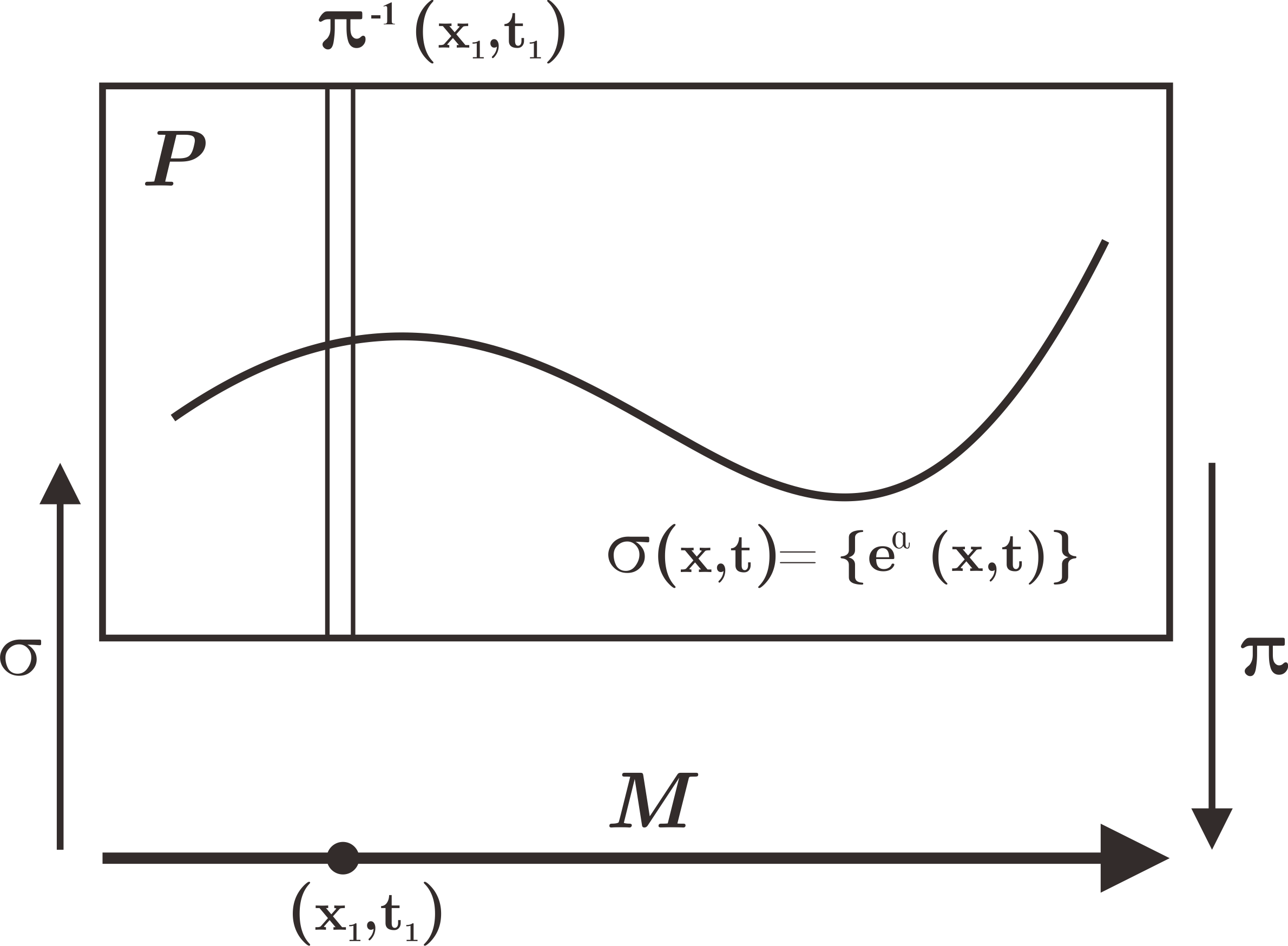}
	\caption{The co-frame bundle of vielbeins.  At every point in the base manifold $M$ that is our spacetime, one chooses a section from the bundle to describe the local frame}
	\label{fig1.1}
\end{figure}

\paragraph{}
Consider a co-frame bundle of spacetime vielbeins $e^a_\mu$, which is a principal fiber bundle with ${\rm SO(1,d)}$ as its group structure; this is illustrated in Fig\ref{fig1.1}.  In this illustration, the base manifold $M$ is the spacetime itself, the set of all vielbeins in the spacetime forms the bundle space and the set of all vielbeins for a particular point $(x,t)$ constitute a fiber $\pi^{-1}(x,t)$.  The section $\sigma$ in the bundle space thus represents our vielbein choices as a function of $(x,t)$.  Explicitly, choosing a set $e^a_\mu$ corresponds to choosing a co-frame of observation $dx^a=(e^a_0dx^0, e^a_1dx^1,\dots, e^a_ddx^d)$ where $1\leq a \leq d+1$.  Of course, one can analogously consider a bundle of inverse vielbeins $e^\mu_a$ i.e. a frame bundle where a choice of $e^\mu_a$ defines the frame of observation.  The fact that we can always choose a vielbein basis whose local metric is $\eta_{ab}$ (Minkowski) from the bundle is a statement of the equivalence principle: every point in the space or spacetime is locally flat.

\paragraph{}
The group structure comes into the picture when one considers the transformation from one vielbein to another i.e. from two different choices of sections $\sigma(x,t)$.  The group elements are referred to as the symmetries of the spacetime.

\paragraph{}
In particular, under Lorentz transformation $\Lambda^a_b$ (Poincaré without translations), vielbeins by definition transform as

\begin{align} \label{eq1}
e^a\rightarrow \Lambda^a_b e^b.
\end{align}

\paragraph{}
Transformations between the vielbein's spin connections on the other hand  follows the transformation rule of local connection representative\footnote{$\sigma^*\omega$ denotes the pullback of the form $\omega$ under the section function $\sigma : U \rightarrow \pi^{-1}(U)$, $U$ being a local neighbourhood in $M$.  Thus, $\sigma^*\omega$ is a form that lives in the spacetime.} $\sigma^*\omega(\partial_\mu)$.  If $\sigma_2(x,t)=\sigma_1(x,t)\Omega(x,t)$ where $\Omega(x,t)$ denotes a group element, then

\begin{equation} \label{eq2}
	\sigma_2^{ *}\omega(\partial_\mu)(x,t)=\Omega(x,t)^{-1}\sigma_1^{ *}\omega(\partial_\mu)(x,t)\Omega(x,t)+\Omega(x,t)^{-1}\big(\partial_\mu\Omega(x,t)\big).
\end{equation}

\paragraph{}
Eq.\ref{eq2} resembles a gauge transformation; indeed, we shall see their equivalence in section \ref{sec1.2} with a more general gauge transformation with general ${\rm SO(1,d)}$ valued connections.  

\paragraph{}
There is however a slight problem with $\Omega(x,t)$: if we consider the full ${\rm SO(1,d)}$, $\sigma_1(x,t)\Omega(x,t)$ is really $\sigma_2(x',t')$ and not $\sigma_2(x,t)$.  In other words, ${\rm SO(1,d)}$ transforms between sections corresponding  to different points in the base manifold.  This displacement is the result of the translations; after all, translations are not internal symmetries in general spacetime.  This demands a fundamental revision to our vilebeins: their internal symmetry should be based on Lorentz-like group elements instead of SO(1,d). 

\paragraph{}
Previous works \cite{HartongObers, Bergshoeff} suggested however, that it is possible to compromise this problem by replacing Poincaré's translation by local spacetime diffeomorphism and to identify the rest with Lorentz's $\Lambda(x,t)$ so that Eq.\ref{eq2} becomes valid in general; this is done in section \ref{sec1.2}.

\subsection{Vielbeins via Poincaré algebra gauging}
\label{sec1.2}

\paragraph{}
Leaving the whole discussion of vielbein bundles, let us consider the most general ${\rm SO(1,d)}$ valued connection

\begin{equation} \label{eq3}
A_\mu=P_a f^a_\mu+\frac{1}{2}M_{ab}\omega^{\text{ } ab}_\mu ,
\end{equation}

\paragraph{}
where $f^a_\mu$ and $\omega^{\text{ } ab}_\mu$ are the gauge fields of the translation and Lorentz rotation generators, respectively.  The $1/2$ factor is a mere convention due to the fact that $M_{ab}$ and $\omega^{ab}$ are anti-symmetric in $\{a,b\}$.  We are interested in the change of $A_\mu$ under local infinitesimal gauge transformation as this will allow us to determine the transformation rules of the gauge fields.  First, we define our infinitesimal transform parameter $\Pi$ as

\begin{align} \label{eq4}
\Pi &=P_a\zeta^a+\frac{1}{2}M_{ab}\sigma^{ab} \nonumber \\
&= \xi^\mu (P_a f^a_\mu ) + \frac{1}{2}  M_{ab} \sigma^{ab} \nonumber \\
&= \xi^\mu A_\mu  +\frac{1}{2}  M_{ab} \lambda^{ab}
\end{align}

where $\zeta^a$ and the anti-symmetric $\sigma^{ab}$ are the infinitesimal translation and Lorentz rotation parameters, respectively; $\xi^\mu$ is a spacetime coordinate parameter s.t. $\zeta^a= \xi^\mu  f^a_\mu$.  Furthermore, $\lambda^{ab}=\sigma^{ab}-\xi^\mu \omega^{\text{ } ab}_\mu$ is the new generalized rotation parameter whose second term cancels the translation.

\paragraph{}
Using the gauge transformation $A'_\mu=\Omega^{-1}A_\mu\Omega+\Omega^{-1}(\partial_\mu\Omega)$ with $\Omega = (I+\Pi)$, one can derive the expression for the infinitesimal transform

\begin{align}
A'_\mu&=(I+\Pi)^{-1}A_\mu(I+\Pi)+(I+\Pi)^{-1}\big(\partial_\mu(I+\Pi)\big) \nonumber \\
&\approx(I-\Pi)A_\mu(I+\Pi)+(I-\Pi)\big(\partial_\mu(I+\Pi)\big) \nonumber\\ 
&=A_\mu-\Pi A_\mu+A_\mu \Pi+\partial_\mu \Pi +O(\Pi^2) \nonumber\\ 
&\approx A_\mu+[A_\mu,\Pi]+\partial_\mu \Pi \nonumber\\ 
\text{hence,   }\hspace{0.2cm}\delta A_\mu&=\partial_\mu \Pi+[A_\mu,\Pi].   \label{eq5}
\end{align}

\paragraph{}
While it is tempting to use Eq.\ref{eq5} directly as the transformation rule, one must realize that it is not a proper internal transformation yet as discussed in the last section.  To obtain a proper internal gauge transformation rule, one must isolate the troublesome translation part from the rotation, an operation that is made possible by the fact that $\delta_P$ differs to a diffeomorphism transform by a curvature term and a $\delta_M$ term\footnote{Here, $\delta_P$ and $\delta_M$ refers to $\delta$ when $\Pi$ is purely translational and rotational, respectively.} \cite{Bergshoeff}.  More precisely, with the help of the curvature 2-form defined as

\begin{equation} \label{eq6}
F_{\mu \nu}=\partial_\mu A_\nu - \partial_\nu A_\mu +[A_\mu,A_\nu],
\end{equation}

one can define $\bar{\delta}$ based on $\delta$ as

\begin{align} 
\bar{\delta}A_\mu&= \delta A_\mu - \xi^\nu F_{\mu \nu} \nonumber \\
&=\partial_\mu \Pi + [A_\mu,\Pi]-\xi^\nu \partial_\mu A_\nu + \xi^\nu \partial_\nu A_\mu - \xi^\nu [A_\mu, A_\nu] \nonumber \\
&=\partial_\mu ( \xi^\nu A_\nu ) +\partial_\mu ( \frac{1}{2} M_{ab} \lambda^{ab} ) +  [A_\mu, \xi^\nu A_\nu+\frac{1}{2} M_{ab} \lambda^{ab}]-\xi^\nu \partial_\mu A_\nu + \pounds_\xi A_\mu - \xi^\nu [A_\mu, A_\nu] \nonumber \\
&=\partial_\mu \Sigma +  [A_\mu, \xi^\nu A_\nu+\Sigma]+ \pounds_\xi A_\mu - \xi^\nu [A_\mu, A_\nu] \nonumber \\
&=\pounds_\xi A_\mu + \partial_\mu \Sigma+[A_\mu,\Sigma]. \label{eq7}
\end{align}

\paragraph{}
Note that $\Sigma=\frac{1}{2}M_{ab}\lambda^{ab}$ now corresponds to pure rotation which is an internal gauge transformation; it is a transformation within sections of a single point $(x,t)$ in the context of frame bundles.

\paragraph{}
At this point, one might object the above seemingly unnatural definition of $\bar{\delta}$: it is not $A_\mu$'s infinitesimal change under the Poincaré group elements.  This objection is indeed justified: $\bar{\delta} A_\mu$ describes $A_\mu$'s infinitesimal change under under pure rotation and a self diffeomorphism $\pounds_\xi A_\mu$.  The latter term is the corrected translation, it replaces $\delta_P A_\mu$ precisely because translation $P$ is not a symmetry in a curved spacetime; in other words, Poincaré group itself is not a symmetry group for GR.  What we have done is simply adding the curvature correction term $\xi^\nu F_{\mu \nu}$ so that the translation does not simply shift coordinate: it translates following the manifold's curvature such that the coordinate essentially never shifts.  In short, $\bar{\delta}$ refers to the infnitesimal change under the corrected Poincaré group, the correct symmetry group for GR.  For the rest of this chapter, we shall denote $\bar{\delta}$ simply as $\delta$. 

\begin{figure}[h!]
	\centering
	\includegraphics[width=6.5cm]{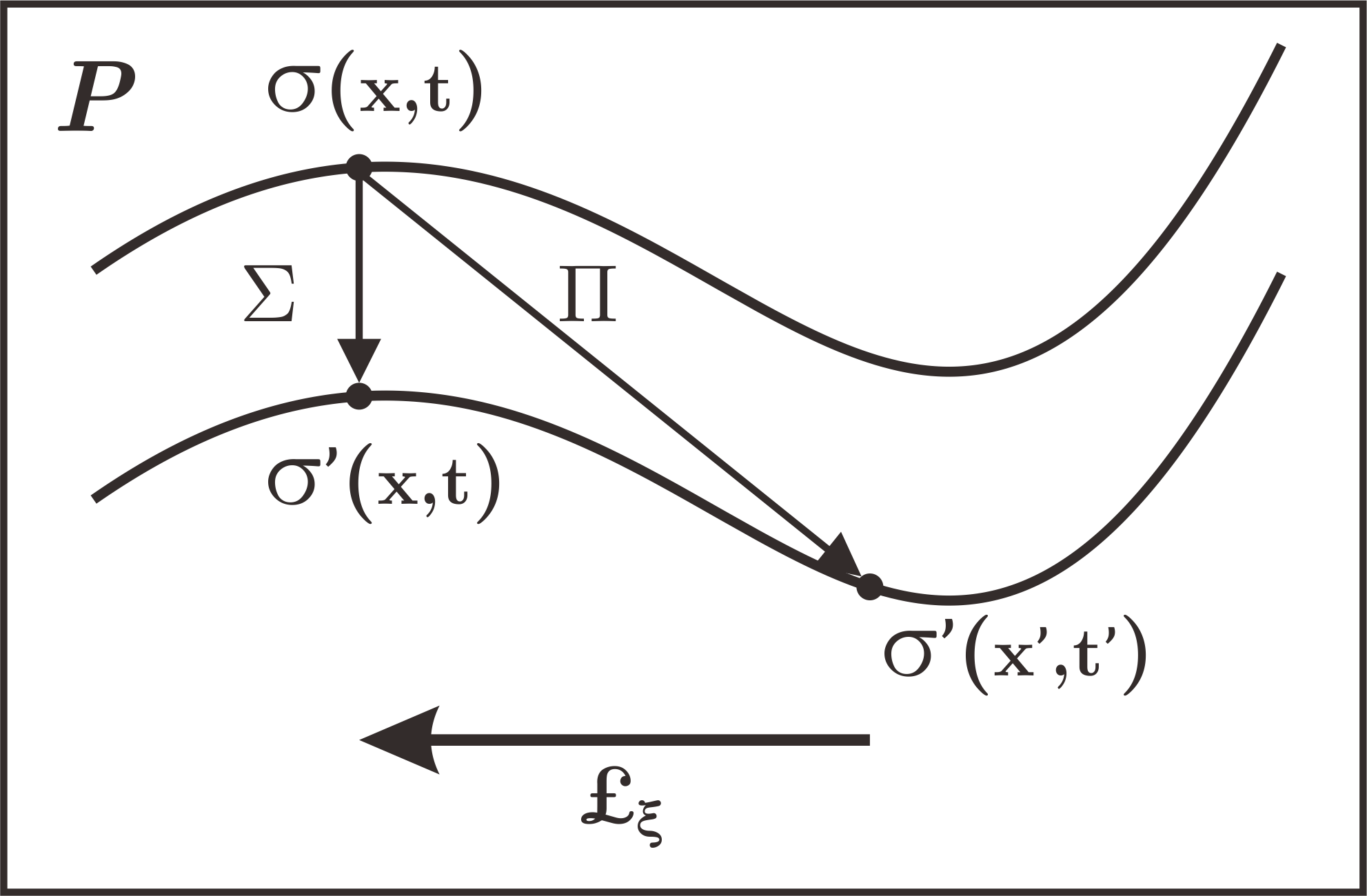}
	\caption{An illustration of a complete Poincaré transformation $\Pi$ in contrast with a proper local gauge transformation $\Sigma$ where translation is replaced by spacetime diffeomorphism}
	\label{fig2}
\end{figure}

\paragraph{}
Lastly, using Eq.\ref{eq7} it is straightforward to derive the expressions for $\delta f^a_\mu$ and $\delta \omega^{\text{ } ab}_\mu$; one simply needs to expand $A_\mu$ on both sides, separate the $P_a$ terms from the $M_{ab}$ terms, and equate accordingly.  We shall derive the expression for $\delta f^a_\mu$ and leave $\delta \omega^{\text{ } ab}_\mu$ for the reader: \footnote{The $P_a$ subscripts on the brackets below denotes "the $P_a$ component of".}

\begin{align} \label{eq8}
{\delta} f^a_\mu= {\delta}_P A_\mu  &=\delta_P ( \pounds_\xi A_\mu )+\delta_P ( \partial_\mu \Sigma) + \delta_P [A_\mu,\Sigma] \nonumber \\
&=(\partial_\nu f^a_\mu) \xi^\nu +0+[P_b f^b_\mu, \frac{1}{2} M_{cd} \lambda^{cd}]_{P_a} \nonumber \\
&=(\partial_\nu f^a_\mu) \xi^\nu + \Big( \frac{1}{2} (\eta_{cb}P_d-\eta_{db}P_c) f^b_\mu \lambda^{cd} \Big)_{P_a} \nonumber \\
&=(\partial_\nu f^a_\mu) \xi^\nu + \frac{1}{2} (\lambda^{ba}-\lambda^{ab})\eta_{bc} f^c_\mu 
 \nonumber \\
&=\pounds_\xi f^a_\mu  + f^c_\mu \lambda^a_{\text{ } \text{ }c},
\end{align}

\begin{equation} \label{eq9}
\delta \omega^{\text{ } ab}_\mu = {\delta}_M A_\mu  =\pounds_\xi \omega^{\text{ } ab}_\mu +\partial_\mu \lambda^{ab} + 2 \lambda^{[a}_{\text{ } \text{ } \text{ }c}  \omega_\mu^{|c|b]}.
\end{equation}
  
\paragraph{}
Eq.\ref{eq8} tells us that $f^a_\mu$ transform exactly in the manner of a vielbein (infinitesimal) transformation i.e. equivalent to Eq.\ref{eq1} under rotation.  In fact, they refer to the same entity as far as the mathematics is concerned.  Hence, using our freedom in defining the connection $A_\mu$, we shall choose $f^a_\mu=e^a_\mu$ from now on. Furthermore, the transformation of $\omega^{\text{ } ab}_\mu$ is the infinitesimal version of Eq.\ref{eq2} for spin connections; thus, we shall likewise choose it to be the spin connection of $e^a_\mu$.  The vielbein postulate in section\ref{sec1.4} is however still required to establish the relation between  $\omega^{\text{ } ab}_\mu$ and  $e^a_\mu$.

\paragraph{}
We would like to stress the importance of this identification: in this section we have defined $A_\mu$, an ${\rm SO(1,d)}$ valued one form without alluding to fiber bundles and eventually interpret the gauge fields as the vielbeins and spin connections.  With these, we are set to construct the full vielbein formalism i.e. applying vielbein postulate and ultimate recovers GR.  The fiber bundle structure then simply becomes a consequence of the postulation of $A_\mu$.  On the other hand, we can of course directly proceed from section \ref{sec1.1} to the vielbein postulate on section \ref{sec1.4} and ultimately recover GR with the resulting Lorentz or corrected-Poincaré vielbein formalism.  Thus strictly speaking, the gauging of Eq.\ref{eq3} is not necessary and the latter can in fact be seen as a consequence of Eq.\ref{eq2}.  Afterall, we can freely define $\omega$ so that $\sigma^* \omega$ can be shaped into any ${\rm SO(1,d)}$ valued one form.  The gauge transformation rules thus becomes a consequence of the rules in our vielbein bundle.

\paragraph{}
To conclude, both approaches are but one equivalent and inseparable concept. Here we have used a mixed approach: we view the vielbein bundle and the $A_\mu$ gauge fields as different entities and ultimately show that they are completely equivalent.  Even though the algebra gauging process can be skipped, we deem its introduction necessary as it is a powerful method when one encounters more convoluted Lie algebras e.g. the Bargmann algebra in chapter 3 where the vielbein transformation law is not easily postulated.

\subsection{The curvature 2-form}
\label{sec1.3}

\paragraph{}
It is imperative to study the physical interpretation of the curvature 2-form we involved in our local gauge transformation.  It turns out in the language of Riemannian geometry this 2-form describes both the spacetime's torsion and Riemann curvature tensor despite what its designation.  The first step to show this is to project $F_{\mu \nu}$ into its $P$ and $M$ components

\begin{equation}\label{eq10}
F_{\mu \nu}=P_a R^a_{\mu \nu}(P) + \frac{1}{2} M_{ab} R^{ab}_{\mu \nu}(M).
\end{equation}

\paragraph{}
Once again, we show a derivation for the expression of $R^a_{\mu \nu}(P)$ in terms of $e^a_\mu$ and $\omega^{\text{ } ab}_\mu$ and leave the one for $R^{ab}_{\mu \nu}(M)$ to the reader:

\begin{align} \label{eq11}
R^a_{\mu \nu}(P)&=(\partial_\mu A_\nu - \partial_\nu A_\mu)_{P_a} +[A_\mu,A_\nu]_{P_a} \nonumber \\
&=(\partial_\mu e^a_\nu - \partial_\nu e^a_\mu) +[P_b e^b_\mu+\frac{1}{2}M_{bc}\omega^{\text{ } bc}_\mu, P_b f^b_\nu+\frac{1}{2}M_{bc}\omega^{\text{ } bc}_\nu]_{P_a}  \nonumber \\
&=2\partial_{[\mu}e^{a}_{\nu]} +[P_b e^b_\mu, \frac{1}{2}M_{bc}\omega^{ \text{ } bc}_\nu]_{P_a}+[\frac{1}{2}M_{bc}\omega^{\text{ } bc}_\mu, P_b e^b_\nu]_{P_a}  \nonumber \\
&=2\partial_{[\mu}e^{a}_{\nu]} + e^d_{[\mu}\omega_{\nu]}^{\text{ } bc} [P_d,\frac{1}{2}M_{bc}]_{P_a}  \nonumber \\
&=2\partial_{[\mu}e^{a}_{\nu]} + e^d_{[\mu}\omega_{\nu]}^{\text{ } bc} \frac{1}{2} (\eta_{bd}P_c-\eta_{cd}P_b)_{P_a}  \nonumber \\
&=2\partial_{[\mu}e^{a}_{\nu]}-2\omega_{[\mu}^{\text{ } \text{ } ab}e_{\nu]b},
\end{align}

\begin{equation} \label{eq12}
R^{ab}_{\mu \nu}(M) = 2\partial_{[\mu}\omega^{\text{ } ab}_{\nu]}-2\omega_{[\mu}^{\text{ } \text{ } ca}\omega_{\nu]c} ^{\text{ } b}.
\end{equation}

\paragraph{}
Without the notion of linear connection $\Gamma^\lambda_{\mu \nu}$, it is difficult to see how Eq.\ref{eq11} and Eq.\ref{eq12} describes torsion and spacetime curvature.  However, it is only possible to express them in terms of $\Gamma^\lambda_{\mu \nu}$  after the application of the vielbein postulate.

\subsection{Vielbein postulate}
\label{sec1.4}

\paragraph{}
It is common in the literatures, e.g. \cite{HartongObers} to define a general covariant derivative $D_\mu$ of the vielbeins containing a linear connection $\Gamma^\lambda_{\mu \nu}$ and the spin connection $\omega^{\text{ } ab}_\mu$ and applying the vielbein postulate by setting this to zero i.e.

\begin{equation} \label{eq13}
D_\mu e^a_\nu = \partial_\mu e^a_\nu -  \Gamma^\lambda_{\mu \nu} e^a_\lambda - \omega_{\mu \text{ } b}^{\text{ } \text{ } a} e^b_\nu = 0. 
\end{equation}

Here, we would like to stress that this vielbein postulate is nothing but a necessary condition for $\omega^{\text{ } ab}_\mu$ to be the spin connection of $e^a_\mu$; it is simply an equation relating $\Gamma^\lambda_{\mu \nu}$ and $\omega^{\text{ } ab}_\mu$ in the vielbein/non-coordinate basis formulation.  More precisely, given a non-coordinate orthonormal basis $e^a_\mu$, the connection in this basis satisfies

\begin{equation} \label{eq14}
\omega_{\mu \text{ } b}^{\text{ }a} e^b_\nu=\nabla_\mu e^a_\nu =  \partial_\mu e^a_\nu -  \Gamma^\lambda_{\mu \nu} e^a_\lambda.
\end{equation}

\paragraph{}
We shall also define the inverse vielbeins, which are the corresponding choice of section from the frame bundle s.t. $e^a_\mu e^\mu_b=\eta^a_b$.  For $e^\mu_b$, the transformation rule can be derived from Eq.\ref{eq8} by demanding $\delta(e^a_\mu e^\mu_b)=\delta(eta^a_{\text{ } b})=0$:

\begin{equation} \label{eq15}
\delta e^\mu_a=\pounds_\xi e^\mu_a  - \lambda^c_{\text{ }a} e^\mu_c.
\end{equation}

Its vielbein postulate on the other hand is

\begin{equation} \label{eq16}
D_\mu e^\nu_a = \partial_\mu e^\nu_a +  \Gamma^\nu_{\mu \lambda} e^\lambda_a + \omega_{\mu \text{ } a}^{\text{ } \text{ } b} e^\nu_b = 0.
\end{equation}

\paragraph{}
Some other works (e.g. \cite{DennisHansen}) also interpret Eq.\ref{eq13} and Eq.\ref{eq14} as consequence of the isomorphism between our vielbein co-frame bundle and the co-tangent bundle $T^*M$ (and between the corresponding frame bundle and the tangent bundle $TM$).  This isomorphism implies that one can always switch from the vielbein and coordinate representation; moreover, contractions such as $A^\mu B_\mu$ is representation invariant i.e. $A^\mu B_\mu=A^aB_a$.  

\paragraph{}
For later calculations, it is useful to express  $\Gamma^\lambda_{\mu \nu}$ in terms of $e^a_\mu$ and $\omega^{\text{ } ab}_\mu$:

\begin{equation} \label{eq17}
\Gamma^\lambda_{\mu \nu}=e^\lambda_a (\partial_\mu e^a_\nu -\omega_{\mu \text{ } b}^{\text{ } \text{ } a} e^b_\nu).
\end{equation}

\subsection{$R^a_{\mu \nu}(P)$ and $R^{ab}_{\mu \nu}(M)$ as torsion $2\Gamma_{[{\mu}{\nu}]}^\lambda$ and curvature $R_{{\mu}{\nu}{\sigma}}^\lambda$ }
\label{sec1.5}

\paragraph{}
Presently, we can use Eq.\ref{eq17} to express Eq.\ref{eq11} and Eq.\ref{eq12} in terms of $\Gamma^\lambda_{\mu \nu}$ or more precisely in terms of the torsion tensor $2\Gamma^\lambda_{[\mu \nu]}$ and the Riemann curvature tensor $R_{\mu \nu \sigma}^\lambda$ defined as

\begin{equation} \label{eq18}
R_{\mu \nu \sigma}^\lambda=-\partial_\mu \Gamma^\lambda_{\nu \sigma}+\partial_\nu \Gamma^\lambda_{\mu \sigma} - \Gamma^\lambda_{\mu \rho} \Gamma^\rho_{\nu \sigma} +\Gamma^\lambda_{\nu \rho} \Gamma^\rho_{\mu \sigma}.
\end{equation}

\paragraph{}
First, note that 

\begin{align}\label{eq19}
\Gamma^\lambda_{[\mu \nu]} &=   e^\lambda_a (\partial_{[\mu} e^a_{\nu]} -\omega^{\text{ } \text{ } \text{ } a}_{[\mu \text{ } |b|}e^b_{\nu]}) \nonumber \\
&= e^\lambda_a (\partial_{[\mu} e^a_{\nu]} -\omega^{ \text{ } \text{ } a b}_{[\mu}e_{\nu] b}),
\end{align}

from which we deduce

\begin{equation} \label{eq20}
R^a_{\mu \nu}(P)=2 e^a_\lambda \Gamma^\lambda_{[\mu \nu]} 
\end{equation}

\paragraph{}
The calculation for $R^{ab}_{\mu \nu}(M)$ is also straightforward albeit much more tedious. We shall provide the details of the proof in the Appendix.  Essentially, substituting Eq.\ref{eq17} into Eq.\ref{eq18} one would find

\begin{equation} \label{eq21}
R_{\mu \nu \sigma}^\lambda=-e^\lambda_b e_{\sigma a}  (2\partial_{[\mu}\omega^{\text{ } ab}_{\nu]}-2\omega_{[\mu}^{\text{ } \text{ } ca}\omega_{\nu]c} ^{\text{ } b}),
\end{equation}

and hence

\begin{equation} \label{eq22}
R^{ab}_{\mu \nu}(M)=-e^b_\lambda e^{\sigma a} R_{\mu \nu \sigma}^\lambda.
\end{equation}

\paragraph{}
The curvature 2-form is a fundamental entity that holds a significant role in the analysis of the constraints of the base manifold.  For instance, a torsionless spacetime – which one demands in GR – in the present case corresponds to constraining $R^a_{\mu \nu}(P)=0$.  This concept will be carried on when one deals with Galilean and Bargmann algebra, where more components would arise and the above conventional notion of torsion and curvature will be generalized. 

\subsection{Metric compatibility and the general $\Gamma_{{\mu}{\nu}}^\lambda$ solution}
\label{sec1.6}

\paragraph{}
We are now in position to express an important result: that the vielbein postulate (where $\omega^{\text{ } ab}_\mu$ is anti-symmetric) implies metric compatibility i.e. $\Gamma^\lambda_{\mu \nu}$ that satisfies Eq.\ref{eq17} is guaranteed to be metric compatible.  In fact, metric compatibility is mathematically equivalent to the condition that $\Gamma^\lambda_{\mu \nu}$ is related to some anti-symmetric $\omega^{\text{ } ab}_\mu$ by Eq.\ref{eq17}.  

\begin{align}\label{eq23}
\Gamma^\lambda_{\mu \nu}=e^\lambda_a (\partial_\mu e^a_\nu -\omega^{ \text{ }a}_{\mu \text{ } b}e^b_\nu)  \text{  for some vielbeins with anti-symmetric $\omega^{\text{ } ab}_\mu$} \nonumber \\
 \Longleftrightarrow \nabla_\rho (g_{\mu \nu})=\nabla_\rho (\eta_{ab} e^a_\mu e^b_\nu)=0
\end{align}

\paragraph{}
\emph{Proof}: Note that 
\begin{align*}
\nabla_\rho (g_{\mu \nu})&=\partial_\rho g_{\mu \nu} - \Gamma^\lambda_{\rho \mu} g_{\lambda \nu}- \Gamma^\lambda_{\rho \nu} g_{\lambda \mu}\\
&=0 - \Gamma^\lambda_{\rho \mu} \eta_{ab} e^a_\lambda e^b_\nu - \Gamma^\lambda_{\rho \nu} \eta_{ab} e^a_\lambda e^b_\mu \\
&=-e_{\lambda a} ( e^a_\nu \Gamma^\lambda_{\rho \mu} -  e^a_\mu \Gamma^\lambda_{\rho \nu}).
\end{align*}

From here, the "$\Rightarrow$" direction is straightforward:

\begin{align*} 
\nabla_\rho (g_{\mu \nu})&=-e_{\lambda a} ( e^a_\nu \Gamma^\lambda_{\rho \mu} -  e^a_\mu \Gamma^\lambda_{\rho \nu}) \\
&=-e_{ \lambda a} \big( e^a_\nu e^\lambda_c (\partial_\rho e^c_\mu -\omega^{\text{ }\text{ } c}_{\rho \text{ } b}e^b_\mu) -  e^a_\mu e^\lambda_c (\partial_\rho e^c_\nu -\omega^{\text{ }\text{ } c}_{\rho \text{ } b}e^b_\nu)\big) \\
&=-\delta_{ac} ( e^a_\nu \partial_\rho e^c_\mu + e^a_\mu \partial_\rho e^c_\nu) + \delta_{ac} ( e^a_{\nu} e_{\mu b} \omega^{ \text{ } cb}_{\rho} + e^a_{\mu} e_{\nu b} \omega^{ \text{ } cb}_{\rho}) \\
&=-\partial_\rho(e_{\nu c} e^c_\mu) + ( e_{\nu c} e_{\mu b} \omega^{\text{ } cb}_{\rho} + e_{\mu c} e_{\nu b} \omega^{\text{ } cb}_{\rho}) \\
&=-\partial_\rho(\delta_{\mu \nu}) + ( e_{\nu c} e_{\mu b} \omega^{ \text{ } cb}_{\rho} -  e_{\mu b}e_{\nu c} \omega^{ \text{ } cb}_{\rho})\\
&=0.
\end{align*}

From the above, the "$\Leftarrow$" direction is also clear; given a vielbein frame, $\Gamma^\lambda_{\mu \nu}$ is always related to its spin connection $\omega^{\text{ } ab}_\mu$ by Eq.\ref{eq17}.  The only way to make $\nabla_\rho (g_{\mu \nu})=0$ is to have $\omega^{\text{ } ab}_\mu$ anti-symmetric.  \qed

\paragraph{}
Finally, one can derive the general solutions for metric compatible connections.  Going from $\nabla_\rho (g_{\mu \nu})=0$ to Eq.\ref{eq24} is a standard procedure and we shall express the latter without derivation.  

\begin{equation} \label{eq24}
\Gamma^\lambda_{\mu \nu}=\frac{1}{2}g^{\lambda \sigma} (\partial_\mu h_{\nu \sigma} + \partial_\nu h_{\mu \sigma} - \partial_\sigma h_{\mu \nu}) - \frac{1}{2}g^{\lambda \sigma} ( 2 \Gamma^\rho_{[\mu \sigma]} g_{\nu \rho} +2 \Gamma^\rho_{[\nu \sigma]} g_{\mu \rho} - 2 \Gamma^\rho_{[\mu \nu]} g_{\rho \sigma})
\end{equation}

\paragraph{}
For later references, we shall refer the first term of Eq.\ref{eq24} as the Levi-Civita term and the second as the contortion term.  Since the latter is still an arbitrary – albeit constrained – quantity, this $\Gamma^\lambda_{\mu \nu}$ is not uniquely determined given a set of vielbeins (that is, given a metric).  If the contortion term is also given however\footnote{Which means one fixes a value for $\omega^{\text{ } \text{ } ab}_{[\mu}e_{\nu] b}$ in Eq.\ref{eq19}, the torsion of the spin connection.}, one will have definite pairwise relations between $\Gamma^\lambda_{\mu \nu}$, $e^a_\mu$ (and  $e^\mu_a$), and $\omega^{\text{ } ab}_\mu$.  

\subsection{Invariance of general relativity}
\label{sec1.7}

\paragraph{}
To close our discussion on Poincaré algebra, we shall explicitly show what we meant by GR being an automatic Lorentz invariant theory.  For starters, note that the metric constructed using the vielbeins, $g_{\mu \nu}=\eta_{ab} e^a_\mu e^b_\nu$ is invariant under the Lorentz transformation

\paragraph{}
\emph{Proof}:
\begin{align*}
\delta_M g_{\mu \nu}&= \delta_M(\eta_{ab}  e^a_\mu e^b_\nu) \nonumber \\
&=\eta_{ab}  e^b_\nu ( \delta_M e^a_\mu) +\eta_{ab} e^a_\mu (\delta_M e^b_\nu ) \nonumber \\
&=\eta_{ab}  e^b_\nu \lambda^a_{\text{ } c} e^c_\mu +\eta_{ab} e^a_\mu \lambda^b_{\text{ } c} e^c_\nu  \nonumber \\
&= e^b_\nu \lambda_{b c} e^c_\mu +e^a_\mu \lambda_{ac} e^c_\nu  \nonumber \\
&= e^c_\nu e^a_\mu (\lambda_{ca} + \lambda_{ac} )  \nonumber \\
&=0. \qed
\end{align*}

\paragraph{}
This may look trivial when we talk about Einstein's spacetime; after all, the Lorentz transformation is designed to preserve the notion of an invariant measure – defined by the metric itself – across different frame of references.  The significance of this discussion shall however be apparent in the next chapter.  Another invariant is given by the inverse metric $g^{\mu \nu}$ which is also easily proven using the transformation rule of $e^\mu_a$.

\paragraph{}
Perhaps rather less obvious is the fact that a general metric connection – that is, all the $\Gamma^\lambda_{\mu \nu}$ satisfying Eq.\ref{eq17} and hence Eg\ref{eq24} – is also automatically Lorentz invariant.  

\begin{align} \label{eq25}
\delta_M \Gamma^\lambda_{\mu \nu} =\delta_M \big[ e^\lambda_a (\partial_\mu e^a_\nu -\omega_{\mu \text{ } b}^{\text{ } \text{ } a} e^b_\nu) \big]=0
\end{align}

\paragraph{}
\emph{Proof}: we simply apply the transformations of $e^a_\mu$ and $\omega^{\text{ } ab}_\mu$, 

\begin{align*}
\delta_M \Gamma^\lambda_{\mu \nu} &=(\delta_M e^\lambda_a) (\partial_\mu e^a_\nu -\omega_{\mu \text{ } b}^{\text{ } \text{ } a} e^b_\nu)+e^\lambda_a \Big(\partial_\mu (\delta_M e^a_\nu) -(\delta_M \omega^{\text{ } ab}_\mu ) e_{\nu b} -\omega^{\text{ } ab}_\mu \big( \delta_M ( \eta_{bc}  e^c_\nu)\big) \Big)   \\
&= -\lambda^b_{\text{ } a} e^\lambda_b  (\partial_\mu e^a_\nu -\omega_{\mu \text{ } c}^{\text{ } \text{ } a} e^c_\nu) + e^\lambda_a \big(\partial_\mu ( \lambda^a_{\text{ } b} e^b_\nu)-( \partial_\mu \lambda^{ab} + 2 \lambda^{[a}_{\text{ } \text{ } \text{ }c}  \omega_\mu^{|c|b]} ) e_{\nu b} -\omega^{\text{ } ab}_\mu (\eta_{bc} \lambda^c_{\text{ } d} e^d_{\nu} )\big).
\end{align*}

The terms not containing $\omega^{\text{ } ab}_\mu$ are

\begin{align*}
&e^\lambda_b  \big( -\lambda^b_{\text{ } a}(\partial_\mu e^a_\nu) + \partial_\mu ( \lambda^b_{\text{ } a} e^a_\nu ) -  (\partial_\mu \lambda^{ba}) e_{\nu a}\big)\\
= \text{ } & e^\lambda_b \big(  e^a_\nu ( \partial_\mu \lambda^b_{\text{ } a})- (\partial_\mu  \lambda^b_{\text{ } a}) e^a_\nu \big)\\
= \text{ } & 0.
\end{align*}

While the terms with $\omega^{\text{ } ab}_\mu$

\begin{align*}
&\lambda^b_{\text{ } a} e^\lambda_b \omega^{\text{ } ac}_\mu e_{\nu c} - e^\lambda_a  2 \lambda^{[a}_{\text{ } \text{ } \text{ }c}  \omega_\mu^{|c|b]} e_{\nu b} - e^\lambda_a \omega^{\text{ } ab}_\mu \lambda_b^{\text{ } d} e_{\nu d}  \\ 
= \text{ }&\lambda^a_{\text{ } c} e^\lambda_a \omega^{\text{ } cb}_\mu e_{\nu b} - e^\lambda_a \lambda^a_{\text{ } c} \omega^{\text{ }cb}_\mu e_{\nu b}+ e^\lambda_a \lambda^b_{\text{ } c} \omega^{\text{ }ca}_\mu e_{\nu b} - e^\lambda_a \omega^{\text{ } ab}_\mu  \lambda_b^{\text{ } d} e_{\nu d} \\
= \text{ } &0 + e^\lambda_a e_{\nu b} (  \lambda^b_{\text{ } c} \omega^{\text{ }ca}_\mu -  \omega^{\text{ } ac}_\mu  \lambda_c^{\text{ } b} ) \\
=\text{ } &e^\lambda_a e_{\nu b} \lambda^b_{\text{ } c} ( \omega^{\text{ }ca}_\mu +  \omega^{\text{ } ac}_\mu ) \\
= \text{ } &0.
\end{align*}

Thus, $\delta_M \Gamma^\lambda_{\mu \nu} =0$. \qed

\paragraph{}
To conclude, the construction of gravity theory in relativistic background is foolproof; physics is invariant to all boosted and rotated observers as it should be, independent of the spacetime torsion and curvature.\footnote{Notice however that under the corrected translation – that is, under diffeomorphism in our formulation – $\Gamma^\lambda_{\mu \nu}$ varies in general i.e. $\pounds_\xi \Gamma^\lambda_{\mu \nu}\neq0$.  This doesn't imply non-symmetry however, since the transformation is merely that of general coordinate transformation.}

%%%%%%%%%%%%%%%%%%%%%%%%%%%%%%%%%%%%%%%%%%%%%%%%%%%%%%%%%%%%

\section{Galilean Algebra}
\label{chap2}

\paragraph{} 
In this chapter we essentailly repeat the whole derivation of gravity in chapter 1 based on the Galilean algebra, which is the classical algebra analogous to Poincaré algebra.  It consists of separate space and time components of the rotations and translations\footnote{With the separate notion of space and time, our vielbein index notation shall be updated as follows: $\{A,B,...\}$ denotes the whole vielbeinspacetime index, $\{a,b,...\}$ denotes the vielbein spatial index and $0$ denotes the vielbein temporal index.}: $M_{AB}$ is projected into spatial rotations $J_{ab}$ and boosts $G_a={\rm lim}_{c \to \infty} \frac{1}{c} M_{0a}=-{\rm lim}_{c \to \infty} \frac{1}{c} M_{a0}$ whereas $P_A$ is projected into the spatial translations $P_a$ and temporal translation $P_0={\rm lim}_{c \to \infty} \frac{1}{c} H$.  

\paragraph{}
Indeed, the Galilean algebra can be easily derived from Poincaré's by employing the (naive) Inonu-Wigner contraction \cite{InonuWigner}.  Note that our $\eta$ has the $(+,-,-,-)$ signature and the limit $c\rightarrow \infty$ is assumed below

\begin{align*}
&[P_A,P_B]=0 \\
&\Rightarrow  \hspace{8.47cm} \mathbf{ [P_a,P_b]=0} \\
&\Rightarrow [P_a, P_0]=[P_a, \frac{1}{c} H]=0  \hspace{4.42cm} \mathbf{  [P_a, H]=0 }\\
\\
&[M_{AB},P_C]=\eta_{BC}P_A-\eta_{AC}P_B \\
&\Rightarrow [M_{a0}, P_0]=[-cG_a, \frac{1}{c} H]=\eta_{00}P_a \hspace{2.82cm} \mathbf{  [H, G_a]=P_a}\\
&\Rightarrow [M_{a0}, P_c]=[-cG_a, P_c]=-\eta_{ac}P_0 \hspace{2.8cm} \mathbf{  [G_a, P_c]=-\delta_{ac} \frac{1}{c^2} H \approx0}\\
&\Rightarrow [M_{ab}, P_0]= [J_{ab},  \frac{1}{c} H]   \hspace{4.88cm}\mathbf{  [J_{ab}, H]=0}\\
&\Rightarrow [M_{ab}, P_c]= [J_{ab}, P_c]   \hspace{5.18cm}\mathbf{  [J_{ab}, P_c]=\delta_{ac}P_b-\delta_{bc}P_a}\\
\\
&[M_{AB},M_{CD}]=\eta_{AD}M_{BC}-\eta_{AC}M_{BD}+\eta_{BC}M_{AD}-\eta_{BD}M_{AC} \\
&\Rightarrow [M_{a0}, M_{c0}]=[-cG_a, -cG_c]=\eta_{00}J_{ac} \hspace{2.33cm} \mathbf{  [G_a, G_c]=\frac{1}{c^2}J_{ac}\approx0}\\
&\Rightarrow [M_{a0}, M_{cd}]=[-cG_a, J_{cd}]=\eta_{ad} M_{0c} -\eta_{ac} M_{0d}  \hspace{0.88cm} \mathbf{  [G_a, J_{cd}]= \delta_{ad} G_c-\delta_{ac} G_d}\\
&\Rightarrow   [M_{ab}, M_{cd}]=  [J_{ab}, J_{cd}] \hspace{4.93cm} \mathbf{  [J_{ab}, J_{cd}]= \delta_{ac}J_{bd}-\delta_{ad}J_{bc}+\delta_{bd}J_{ac} - \delta_{bc}J_{ad}}
\end{align*}

\subsection{The (co-)frame bundle with Galilean structure group}
\label{sec2.1}

\paragraph{}
Now that the base manifold is a classical spacetime, the corresponding vielbein bundle must provide both spatial and temporal frames.  In other words, our vielbein section $\sigma(x)$ now corresponds to the choice of spatial vielbein $e^a_\mu$ whose local metric is a flat $\delta_{ab}$ and a temporal one $\tau_\mu$.  As in the previous chapter, these vielbeins determines our co-frame of observation which now consists of a form corresponding to a spacelike vector $dx^a=(e^a_0 dx^0, e^a_1dx^1, \dots , e^a_ddx^d)$ and one corresponding to a timelike vector $d\tau=(\tau_0 dx^0, \tau_1dx^1, \dots, \tau_ddx^d)$; the latter can be thought as an observer's personal clock as it measures time corresponding to a certain rule, some literatures in fact refers to it as the clock form. 

Under translationless Galilean transformation and denoting rotation and boost as $\Lambda^a_b$ and $\Lambda^a$, respectively, these vielbeins by definition transform as:

\begin{align}  
\tau&\rightarrow\tau,  \label{eq26} \\    
e^a&\rightarrow \Lambda^a_b e^b +\Lambda^a \tau.  \label{eq27}
\end{align}

\subsection{Galilean algebra valued connection}
 \label{sec2.2}

\paragraph{}
As in the previous chapter, the expression for a general  Galilean algebra valued connection is given by the generators and their corresponding gauge fields. The new gauge fields corresponding to temporal translation and Galilean boosts are $t_\mu$ and $\omega^a_\mu$, respectively.

\begin{equation} \label{eq28}
A_\mu=H t_\mu + P_a f^a_\mu +G_a \omega^a_\mu + \frac{1}{2} J_{ab} \omega^{\text{ }ab}_\mu 
\end{equation}

Eq.\ref{eq28} is basically Eq.\ref{eq3} with split $P$ and $M$ generators. 

\paragraph{}
We then extend the infinitesimal transformation parameter $\Pi$ containing the infinitesimal set $(\delta, \zeta^a, \sigma^a, \sigma^{ab})$ as

\begin{align} \label{eq29}
\Pi&= H \Delta + P_a \zeta^a  + G_a \sigma^a +\frac{1}{2}  J_{ab} \sigma^{ab} \nonumber \\
&= \xi^\mu ( H t_\mu + P_a f^a_\mu ) + G_a \sigma^a +\frac{1}{2}  J_{ab} \sigma^{ab} \nonumber \\
&= \xi^\mu A_\mu + G_a \lambda^a +\frac{1}{2}  J_{ab} \lambda^{ab}.
\end{align}

In the above, the spacetime parameter $\xi^\mu$ is chosen such that $\xi^\mu t_\mu=\Delta$ and $\xi^\mu f^a_\mu=\zeta^a$ whereas $\lambda^a=\sigma^a-\xi^\mu \omega^a_\mu$ and $\lambda^{ab}=\sigma^{ab}-\xi^\mu \omega^{ab}_\mu$ are the general boost and rotation parameters that has zero translational shift.   

\paragraph{}
Next, as we have argued in the last chapter we define the proper infinitesimal gauge transformation $\bar{\delta}A_\mu=\delta A_\mu -\xi^\nu F_{\mu \nu}$ with $\delta$ given by Eq.\ref{eq5} and $F_{\mu \nu}$ by Eq.\ref{eq6}.  The resulting expression is essentially the same as Eq.\ref{eq7}:

\begin{align*}
 \bar{\delta}A_\mu =\pounds_\xi A_\mu + \partial_\mu \Sigma+[A_\mu,\Sigma],
\end{align*}

with $\Sigma$ updated to $\frac{1}{2}  J_{ab}  \lambda^{ab} + G_a \lambda^a$.  The gauge transformation of $\bar{\delta}$ – which we shall refer simply as $\delta$ from this point on – is generated only by internal symmetries of spatial rotations and boosts (on top of a spacetime diffeomorphism).  

\paragraph{}
With the same method as the one to derive Eq.\ref{eq8} and Eq.\ref{eq9}, we can easily derive the gauge fields transformations:\footnote{We leave the derivations for the rotational and boost transformations for the reader.}

\begin{align} 
{\delta}t_\mu = {\delta}_H A_\mu &=\delta_H ( \pounds_\xi A_\mu )+\delta_H ( \partial_\mu \Sigma) + \delta_H [A_\mu,\Sigma] \nonumber \\
&=(\partial_\nu f^a_\mu) \xi^\nu  \nonumber \\
&=  \pounds_\xi t_\mu, \label{eq30}
\end{align}

\begin{align} 
{\delta} f^a_\mu= {\delta}_P A_\mu  &=\delta_P ( \pounds_\xi A_\mu )+\delta_P ( \partial_\mu \Sigma) + \delta_P [A_\mu,\Sigma] \nonumber \\
&=(\partial_\nu f^a_\mu) \xi^\nu +0+[H t_\mu, G_b \lambda^b]_{P_a} + [P_c f^c_\mu, \frac{1}{2} J_{bd} \lambda^{bd}]_{P_a} \nonumber \\
&=(\partial_\nu f^a_\mu) \xi^\nu +(\lambda^b P_b t_\mu)_{P_a} + \Big( \frac{1}{2} (\delta_{dc}P_b-\delta_{bc}P_d) f^c_\mu \lambda^{bd} \Big)_{P_a} \nonumber \\
&=(\partial_\nu f^a_\mu) \xi^\nu +\lambda^a t_\mu+ \frac{1}{2} (\lambda^{ab}-\lambda^{ba})\delta_{bc} f^c_\mu 
 \nonumber \\
&=\pounds_\xi f^a_\mu   +\lambda^a t_\mu + f^c_\mu \lambda^a_{\text{ } \text{ }c},  \label{eq31}
\end{align}

\begin{align} 
\delta \omega^a_\mu &= {\delta}_G A_\mu  =\pounds_\xi \omega^a_\mu +\partial_\mu \lambda^a +  \lambda^a_{\text{ }b} \omega^b_\mu + \lambda^b  \omega^{\text{ }a}_{\mu b}, \label{eq32}\\
\delta \omega^{\text{ } ab}_\mu &= {\delta}_J A_\mu  =\pounds_\xi \omega^{\text{ } ab}_\mu +\partial_\mu \lambda^{ab} + 2 \lambda^{[a}_{\text{ } \text{ } \text{ }c}  \omega_\mu^{|c|b]}. \label{eq33}
\end{align}

\paragraph{}
Finally, we shall use our freedom in defining $A_\mu$ to identify $t_\mu=\tau_\mu$ and $f^a_\mu=e^a_\mu$ as they transform exactly in the manner of the vielbeins (see Eq.\ref{eq26} and Eq.\ref{eq27}); the $\omega^a_\mu$ and $\omega^{\text{ }ab}_\mu$'s transformation similarly suggest that they can be identified as the spin connections (see Eq.\ref{eq2}).

\subsection{The curvature 2-form}
\label{sec2.3}

\paragraph{}
The projection of the curvature 2-form in this case can be written as

\begin{equation}\label{eq34}
F_{\mu \nu}=H R_{\mu \nu}(H) + P_a R^a_{\mu \nu}(P) + G_a R^a_{\mu \nu}(G) + \frac{1}{2} J_{ab} R^{ab}_{\mu \nu}(J).
\end{equation}

\paragraph{}
The expressions for $R_{\mu \nu}(H)$, $R^a_{\mu \nu}(P)$, $R^a_{\mu \nu}(G)$ and $R^{ab}_{\mu \nu}(J)$ are given by\footnote{We leave the derivations for the rotational boost curvatures for the reader.}

\begin{align} \label{eq35}
R_{\mu \nu}(H)&=(\partial_\mu A_\nu - \partial_\nu A_\mu)_{H} +[A_\mu,A_\nu]_{H} \nonumber \\
&=(\partial_\mu \tau_\nu - \partial_\nu \tau_\mu) +0  \nonumber \\
&=\partial_{[\mu} \tau_{\nu]},
\end{align}

\begin{align} \label{eq36}
R^a_{\mu \nu}(P)&=(\partial_\mu A_\nu - \partial_\nu A_\mu)_{P_a} +[A_\mu,A_\nu]_{P_a} \nonumber \\
&=(\partial_\mu e^a_\nu - \partial_\nu e^a_\mu) +[H \tau_\mu + G_b \omega^b_\mu, H \tau_\nu + G_b \omega^b_\nu]_{P_a}  + [P_d e^d_\mu+\frac{1}{2}J_{bc}\omega^{\text{ } bc}_\mu,P_d e^d_\nu+\frac{1}{2}J_{bc}\omega^{\text{ } bc}_\nu]_{P_a} \nonumber \\
&=2\partial_{[\mu}e^{a}_{\nu]} +[H \tau_\mu, G_b \omega^b_\nu]_{P_a}+ [G_b \omega^b_\nu,H \tau_\mu]_{P_a}+[P_d e^d_\mu,\frac{1}{2}J_{bc}\omega^{\text{ } bc}_\nu]_{P_a}+[\frac{1}{2}J_{bc}\omega^{\text{ } bc}_\mu,P_d e^d_\nu]_{P_a}  \nonumber \\
&=2\partial_{[\mu}e^{a}_{\nu]} +2 \tau_{[\mu}\omega_{\nu]}^b [H,G_b]_{P_a} + 2e^d_{[\mu}\omega_{\nu]}^{\text{ } bc} [P_d,\frac{1}{2}J_{bc}]_{P_a} \nonumber \\
&=2\partial_{[\mu}e^{a}_{\nu]} + 2\tau_{[\mu}\omega_{\nu]}^b \delta^a_b +e^d_{[\mu}\omega_{\nu]}^{\text{ } bc} (\delta_{dc}P_b-\delta_{db}P_c)_{P_a}  \nonumber \\
&=2\partial_{[\mu}e^{a}_{\nu]} - 2\omega_{[\mu}^a \tau_{\nu]} -2\omega_{[\mu}^{\text{ } \text{ } ab}e_{\nu]b},
\end{align}

\begin{equation} \label{eq37}
R^a_{\mu \nu}(G) = 2\partial_{[\mu}\omega^a_{\nu]} -2\omega_{[\mu}^{\text{ } \text{ } ab}\omega_{\nu]b},
\end{equation}

\begin{equation} \label{eq38}
R^{ab}_{\mu \nu}(J) = 2\partial_{[\mu}\omega^{\text{ } ab}_{\nu]}-2\omega_{[\mu}^{\text{ } \text{ } ca}\omega_{\nu]c} ^{\text{ } b}.
\end{equation}

\paragraph{}
Once again, one will see in section \ref{sec2.5} that $R_{\mu \nu}(H)$ and $R^a_{\mu \nu}(P)$ describes temporal and spatial torsion whereas $R^a_{\mu \nu}(G)$ and $R^{ab}_{\mu \nu}(J)$ describes temporal and spatial curvatures.

\subsection{Vielbein postulate}
\label{sec2.4}

\paragraph{}
Our next step is once again to give a complete vielbein structure to the problem by demanding the vielbein postulate relating $\omega^a_\mu$ and $\omega^{\text{ }ab}_\mu$ with the affine connection $\Gamma^\lambda_{\mu \nu}$: 

\begin{equation} \label{eq39}
D_\mu \tau_\nu = \partial_\mu \tau_\nu -  \Gamma^\lambda_{\mu \nu} \tau_\lambda = 0,
\end{equation}

\begin{equation} \label{eq40}
D_\mu e^a_\nu = \partial_\mu e^a_\nu -  \Gamma^\lambda_{\mu \nu} e^a_\lambda - \omega^a_\mu \tau_\nu- \omega_{\mu \text{ } b}^{\text{ } \text{ } a} e^b_\nu = 0.
\end{equation}

\paragraph{}
To express $\Gamma^\lambda_{\mu \nu}$ in terms of the vielbeins and connections, we write 

\begin{align*} 
\Gamma^\lambda_{\mu \nu} e^a_\lambda e^\sigma_a &= e^\sigma_a( \partial_\mu e^a_\nu - \omega^a_\mu \tau_\nu-  \omega_{\mu \text{ } b}^{\text{ } \text{ } a} e^b_\nu) \hspace{0.3cm} \text{and} \\
\Gamma^\lambda_{\mu \nu} \tau_\lambda  v^\sigma &= v^\sigma \partial_\mu \tau_\nu,
\end{align*}

where $v^\mu$ and $e^\mu_a$ are the temporal and spatial inverse vielbeins, respectively (see the note on Galilean vielbein formalism at the end of this chapter).  Substracting the equation and using the vielbein completeness relation, we obtain

\begin{equation} \label{eq41}
\Gamma^\lambda_{\mu \nu}= -v^\lambda \partial_\mu \tau_\nu+ e^\lambda_a( \partial_\mu e^a_\nu - \omega^a_\mu \tau_\nu-  \omega_{\mu \text{ } b}^{\text{ } \text{ } a} e^b_\nu).
\end{equation}

\subsection{$R(H)$, $R(P)$, $R(G)$ and $R(J)$ in terms of $\Gamma_{[{\mu}{\nu}]}^\lambda$ and $R_{{\mu}{\nu}{\sigma}}^\lambda$ }
\label{sec2.5}

With Eq.\ref{eq41}, we can now re-express Eq.\ref{eq35} to \ref{eq38} in terms of the torsion tensor $\Gamma_{[{\mu}{\nu}]}^\lambda$ and the Riemann curvature tensor $R_{{\mu}{\nu}{\sigma}}^\lambda$.  The process is similar to the one in the previous chapter; we first write 

\begin{align*}
\Gamma^\lambda_{[\mu \nu]} &=  -v^\lambda \partial_{[\mu} \tau_{\nu]} + e^\lambda_a (\partial_{[\mu} e^a_{\nu]} -\omega^a_{[\mu}\tau_{\nu]} - \omega^{\text{ } \text{ } \text{ } a}_{[\mu \text{ } |b|}e^b_{\nu]}) \nonumber \\
&= -v^\lambda \partial_{[\mu} \tau_{\nu]} + e^\lambda_a (\partial_{[\mu}e^{a}_{\nu]} - \omega_{[\mu}^a \tau_{\nu]} -\omega_{[\mu}^{\text{ } \text{ } ab}e_{\nu]b})\\
&=-\frac{1}{2}v^\lambda R_{\mu\nu} (H) +\frac{1}{2} e^\lambda_a R^a_{\mu \nu}(P),
\end{align*}

that is, 

\begin{align}
R_{\mu\nu} (H)&= 2\tau_\lambda \Gamma^\lambda_{[\mu \nu]},  \label{eq42}\\
 R^a_{\mu \nu}(P)&= 2 e^a_\lambda \Gamma^\lambda_{[\mu \nu]}. \label{eq43}
\end{align}

\paragraph{}
Hence as we mentioned, Galilean's $R_{\mu\nu} (H)$ and $ R^a_{\mu \nu}(P)$ are nothing but the temporal and spatial components of the spacetime torsion, which in a relativistic spacetime is compactly contained in $ R^a_{\mu \nu}(P)$ where $P$ is spacetime translations. 

\paragraph{}
On the other hand, we can show (see Appendix for derivation) that the Riemann curvature tensor satisfies

\begin{align*}
R_{{\mu}{\nu}{\sigma}}^\lambda&=e^\lambda_a \tau_\sigma (2\partial_{[\mu}\omega^a_{\nu]} -2\omega_{[\mu}^{\text{ } \text{ } ab}\omega_{\nu]b}) - e_{\sigma a} e^\lambda_b (2\partial_{[\mu}\omega^{\text{ } ab}_{\nu]}-2\omega_{[\mu}^{\text{ } \text{ } ca}\omega_{\nu]c} ^{\text{ } b}) \\
&=e^\lambda_a \tau_\sigma R^a_{\mu \nu}(G) - e_{\sigma a} e^\lambda_b R^{ab}_{\mu \nu}(J).
\end{align*}

That is, 

\begin{align}
 R^a_{\mu \nu}(G)&= e^a_\lambda v^\sigma R_{{\mu}{\nu}{\sigma}}^\lambda,   \label{eq44} \\
 R^{ab}_{\mu \nu}(J)&=  -  e^b_\lambda e^{\sigma a} R_{{\mu}{\nu}{\sigma}}^\lambda . \label{eq45}
\end{align}

\paragraph{}
Once again, this shows that Galilean's $R^a_{\mu \nu}(G)$ and $  R^{ab}_{\mu \nu}(J)$ are simply projections of the Riemann curvature tensor along the temporal and spatial directions, respectively; in the relativistic case, they are a unity in $  R^{ab}_{\mu \nu}(J)$ where $J$ is spacetime rotations.

\paragraph{}
As in last chapter, we stress that these components of the curvature 2-form are vital when one aims to impose certain constraints in the background spacetime.  As an example, for a completely tosionless Galilean spacetime it is necessary to constrain both $R_{\mu\nu} (H)$ and $ R^a_{\mu \nu}(P)$ to zero whereas the condition $R_{\mu\nu} (H)=0$ corresponds only to a torsionless time that physically implies the existence of an absolute time measure.  We shall discuss more about torsion in the next section.

\subsection{Galilean metric compatibility and general $\Gamma_{{\mu}{\nu}}^\lambda$ solution}
\label{sec2.6}

In this section we find the first telling characteristic deviation Galilean spacetime from GR: its general solution of its metric connection.  While in chapter 1GR's vielbein postulate is equivalent to metric compatibility, in the present case the former implies only the following compatibilities:

\begin{align}
 \nabla_\rho(h^{\mu \nu})&=\nabla_\rho( \delta^{ab}  e^\mu_a e^\nu_b)=0 \hspace{0.3cm} \text{and}  \label{eq46} \\
\nabla_\rho(\tau_\mu)&=0. \label{eq47}
\end{align} 

In fact, one can show that the conventional metric compatibility is violated:

\begin{equation}\label{eq48}
\nabla_\rho(h_{\mu \nu})=\nabla_\rho( \delta_{ab}  e^a_\mu e^b_\nu)=2\omega^a_\rho \tau_{(\mu} e_{\nu ) a}.
\end{equation} 

\emph{Proof}: From the vielbein postulate of $e^a_\mu$,

\begin{align*}
\nabla_\rho(h_{\mu \nu})&=\delta_{ab} (\omega^a_\rho \tau_\mu +\omega^{\text{ }\text{ }a}_{\rho \text{ } c} e^c_\mu) e^b_\nu + \delta_{ab} (\omega^b_\rho \tau_\nu +\omega^{\text{ }\text{ }b}_{\rho \text{ } c} e^c_\nu) e^a_\mu \\
&=\omega^a_\rho \tau_\mu e_{\nu a} +\omega^{\text{ }\text{ }a}_{\rho \text{ } c} e^c_\mu e_{\nu a}  + \omega^a_\rho \tau_\nu e_{\mu a} +\omega^{\text{ }\text{ }b}_{\rho \text{ } c} e^c_\nu e_{\mu b}    \\
&=(\omega^a_\rho \tau_\mu e_{\nu a} +\omega^a_\rho \tau_\nu e_{\mu a} ) + (\omega^{\text{ }\text{ }a}_{\rho \text{ } c} e^c_\mu e_{\nu a} + \omega^{\text{ }\text{ }c}_{\rho \text{ } a} e^a_\nu e_{\mu c}  )\\
&=2\omega^a_\rho \tau_{(\mu} e_{\nu ) a} + e^c_\mu e_{\nu a} (\omega^{\text{ }\text{ }a}_{\rho \text{ } c} +\omega^{\text{ }\text{ }\text{ }a}_{\rho c} ) \\
&=2\omega^a_\rho \tau_{(\mu} e_{\nu ) a}. \qed
\end{align*}

Furthermore, it is also clear that $\nabla_\rho(v^\mu)= \omega^a_\rho e^\mu_a \neq 0$ and hence $v^\mu$ is also not compatible with $\nabla$. 

\paragraph{}
 The non-compatibility of metric however does not stop us from deriving $\Gamma^\lambda_{\mu \nu}$ from $\nabla_\rho(h_{\mu \nu})$ and $\nabla_\rho(\tau_\mu)$ as usual.  Summing up the following terms,

\begin{align*}
\nabla_\mu(h_{\nu \rho})&= \partial_\mu h_{\nu \rho} - \Gamma^\lambda_{\mu \nu} h_{\lambda \rho} -  \Gamma^\lambda_{\mu \rho} h_{\nu \lambda} = 2\omega_{\mu a} \tau_{(\nu} e^a_{\rho )} \\
\nabla_\nu(h_{\mu \rho})&= \partial_\nu h_{\mu \rho} - \Gamma^\lambda_{\nu \mu} h_{\lambda \rho} - \Gamma^\lambda_{\nu \rho} h_{\mu \lambda} = 2\omega_{\nu a} \tau_{(\mu} e^a_{\rho )}\\
-\nabla_\rho(h_{\mu \nu})&= -\partial_\rho h_{\mu \nu} + \Gamma^\lambda_{\rho \mu} h_{\lambda \nu} + \Gamma^\lambda_{\rho \nu} h_{\mu \lambda} = -2\omega_{\rho a} \tau_{(\mu} e^a_{\nu )},
\end{align*}

we obtain

\begin{align*}
(\partial_\mu h_{\nu \rho} + \partial_\nu h_{\mu \rho}-\partial_\rho h_{\mu \nu}) - ( 2\Gamma^\lambda_{[\mu \rho]} h_{\nu \lambda} +2 \Gamma^\lambda_{[\nu \rho]} h_{\mu \lambda} +2\Gamma^\lambda_{(\mu \nu)} h_{\lambda \rho} )=  2\omega_{\mu a} \tau_{(\nu} e^a_{\rho )} + 2\omega_{\nu a} \tau_{(\mu} e^a_{\rho )} -2\omega_{\rho a} \tau_{(\mu} e^a_{\nu )}. \\
\end{align*}

\newpage
In other words,

\begin{align*}
 2\Gamma^\lambda_{\mu \nu} h_{\lambda \rho}=(\partial_\mu h_{\nu \rho} + \partial_\nu h_{\mu \rho}-\partial_\rho h_{\mu \nu}) &-(2\Gamma^\lambda_{[\mu \rho]} h_{\nu \lambda} +2 \Gamma^\lambda_{[\nu \rho]} h_{\mu \lambda} -2\Gamma^\lambda_{[\mu \nu]} h_{\lambda \rho} ) \\
&- (2\omega_{\mu a} \tau_{(\nu} e^a_{\rho )} + 2\omega_{\nu a} \tau_{(\mu} e^a_{\rho )} -2\omega_{\rho a} \tau_{(\mu} e^a_{\nu )} ).
\end{align*}

From Eq.\ref{eq39} or Eq.\ref{eq47} we also obtain

\begin{align*}
\Gamma^\lambda_{\mu \nu} \tau_\lambda = -  \partial_\mu \tau_\nu.
\end{align*}

The last two equations describe the spatial and temporal components of $\Gamma_{{\mu}{\nu}}^\lambda$; that is, $\Gamma_{{\mu}{\nu}}^\lambda =-\Gamma^\rho_{\mu \nu} \tau_\rho  v^\lambda+ \Gamma^\rho_{\mu \nu} h_{\rho \sigma} h^{\sigma \lambda}$ which is clear from the identity $h_{\mu \sigma}h^{\sigma \nu}=e^\mu_a e^a_\nu=\delta^\mu_\nu+v^\mu \tau_\nu$.  Thus, we have

\begin{align*}
\Gamma_{{\mu}{\nu}}^\lambda= \text{"Levi-Civita"} - \frac{1}{2} h^{\lambda \sigma} (2\Gamma^\rho_{[\mu \sigma]} h_{\nu \rho} +2 \Gamma^\rho_{[\nu \sigma]} h_{\mu \rho} -2 \Gamma^\rho_{[\mu \nu]} h_{\rho \sigma} ) - \frac{1}{2} h^{\lambda \sigma} X_{\mu \nu \sigma},
\end{align*}

where\footnote{Note that while this "Levi-Civita" resembles the original Riemannian Levi-Civita connection, they are dissimilar on a fundamental level; for instance the present "Levi-Civita" is not torsionless as the original is.}

\begin{align*} 
\text{"Levi-Civita"} = -  v^\lambda 	\partial_\mu \tau_\nu  +  \frac{1}{2}  h^{\lambda \sigma} (\partial_\mu h_{\nu \sigma} + \partial_\nu h_{\mu \sigma}-\partial_\sigma h_{\mu \nu}),
\end{align*}

and 

\begin{align*}
X_{\mu \nu \sigma}&=2\omega_{\mu a} \tau_{(\nu} e^a_{\sigma )} + 2\omega_{\nu a} \tau_{(\mu} e^a_{\sigma )} -2\omega_{\sigma a} \tau_{(\mu} e^a_{\nu )} \\
&=\omega_{\mu a} \tau_{\nu} e^a_{\sigma } +\omega_{\mu a} \tau_{\sigma} e^a_{\nu } + \omega_{\nu a} \tau_{\mu} e^a_{\sigma } + \omega_{\nu a} \tau_{\sigma} e^a_{\mu } - \omega_{\sigma a} \tau_{\mu} e^a_{\nu } - \omega_{\sigma a} \tau_{\nu} e^a_{\mu } \\
&=2 \tau_\nu e^a_{[\sigma} \omega_{\mu ] a} + 2 \tau_\sigma e^a_{(\mu} \omega_{\nu ) a} + 2 \tau_\mu e^a_{[\sigma} \omega_{\nu ] a}.
\end{align*}

Finally, 

\begin{align} 
\Gamma_{{\mu}{\nu}}^\rho&= \text{"Levi-Civita"} - \frac{1}{2} h^{\lambda \sigma} (2\Gamma^\rho_{[\mu \sigma]} h_{\nu \rho} +2 \Gamma^\rho_{[\nu \sigma]} h_{\mu \rho} -2 \Gamma^\rho_{[\mu \nu]} h_{\rho \sigma} )\nonumber  \\
&\hspace{3cm}- \frac{1}{2} h^{\lambda \sigma} ( 2 \tau_\mu e^a_{[\sigma} \omega_{\nu ] a}  + 2 \tau_\nu e^a_{[\sigma} \omega_{\mu ] a} + 2 \tau_\sigma e^a_{(\mu} \omega_{\nu ) a})  \nonumber \\
&=\text{"Levi-Civita"} - \frac{1}{2} h^{\lambda \sigma} (2\Gamma^\rho_{[\mu \sigma]} h_{\nu \rho} +2 \Gamma^\rho_{[\nu \sigma]} h_{\mu \rho} -2 \Gamma^\rho_{[\mu \nu]} h_{\rho \sigma}   + 2 \tau_\mu e^a_{[\sigma} \omega_{\nu ] a} +  2 \tau_\nu e^a_{[\sigma} \omega_{\mu ] a}). \label{eq49}
\end{align}

Note that the non-Levi-Civita terms include the spatial contortion term and $2 \tau_\mu e^a_{[\sigma} \omega_{\nu ] a}$ – the latter known as the Newton-Coriolis two-form – is anti-symmetric in $\{ \sigma, \nu \}$ while the last term is anti-symmetric in $\{ \sigma, \mu \}$.  Both of these terms are semi-arbitrary; hence, the above expression for $\Gamma_{{\mu}{\nu}}^\rho$ can be re-parameterized as 

\begin{equation} \label{eq50}
\Gamma_{{\mu}{\nu}}^\lambda= \text{"Levi-Civita"} + \frac{1}{2} h^{\lambda \sigma} Y_{\mu \nu \sigma},
\end{equation}

where

\begin{equation*}
Y_{\mu \nu \sigma}= C_{\sigma \mu \nu} +\tau_\nu B_{\sigma \mu} \hspace{0.1cm} \text{where} \hspace{0.1cm}  C_{\sigma \mu \nu}= - C_{\nu \mu \sigma} \hspace{0.1cm} \text{and} \hspace{0.1cm}  B_{\mu \nu}= -B_{\nu \mu} \hspace{0.1cm} \text{and otherwise arbitrary.}
\end{equation*}

However, it turns out that any quantity $Y_{\mu \nu \sigma}$ such that 

\begin{equation}\label{eq51}
(h^{\lambda \sigma} h^{\xi \nu} + h^{\xi \sigma} h^{\lambda \nu} ) Y_{\mu \nu \sigma}=0
\end{equation}

is also qualified for Eq.\ref{eq50}.  This is true since such $Y_{\mu \nu \sigma}$ would have a general solution of the form 

\begin{equation*}
Y_{\mu \nu \sigma}=C'_{\sigma \mu \nu} + \tau_\nu B'_{\sigma \mu}+  \tau_\sigma A'_{\mu \nu} \hspace{0.1cm} \text{where} \hspace{0.1cm}  C'_{\sigma \mu \nu}= - C'_{\nu \mu \sigma} \hspace{0.1cm} \text{but} \hspace{0.1cm}  B'_{\sigma \mu} \hspace{0.1cm} \text{and} \hspace{0.1cm} A'_{\mu \nu} \hspace{0.1cm} \text{are completely arbitrary.}
\end{equation*}

The arbitrariness of $B'_{\sigma \mu}$ and $ A'_{\mu \nu}$ poses no problems since we can always write

\begin{equation*}
Y_{\mu \nu \sigma}= (C'_{\sigma \mu \nu} +  \tau_\nu B'_{(\sigma \mu)} - \tau_\sigma B'_{\nu \mu}) + \tau_\nu B'_{[\sigma \mu]}+  \tau_\sigma A'_{\mu \nu}+ \tau_\sigma B'_{(\mu \nu)},
\end{equation*}

where the first term can be chosen as $C_{\sigma \mu \nu}$, the second as $ B_{\sigma \mu}$, whereas the terms with $\tau_\sigma$ are simply irrelevant as $Y_{\mu \nu \sigma}$ will be contracted with $ h^{\rho \sigma}$ anyway.  Hence, if one asks for a Galilean metric connection, the straightforward answer would be Eq.\ref{eq49} while the simplest answer would be "any $\Gamma_{{\mu}{\nu}}^\lambda$ in Eq.\ref{eq50} whose $Y_{\mu \nu \sigma}$ satisfies Eq.\ref{eq51}".

\vspace{1cm}

\paragraph{}
Yet a different parameterization for $Y_{\mu \nu \sigma}$ is given by \cite{HartongObers}; this shall be referred to as Hartong-Obers' parameterization. 

\begin{equation*}
Y_{\mu \nu \sigma}=\tau_\mu K_{\sigma \nu} +\tau_\nu K_{\sigma \mu} + L_{\sigma \mu \nu} \hspace{0.1cm} \text{where} \hspace{0.1cm}  L_{\sigma \mu \nu}= - L_{\nu \mu \sigma} \hspace{0.1cm} \text{and} \hspace{0.1cm}  K_{\mu \nu}= -K_{\nu \mu } \hspace{0.1cm} \text{and otherwise arbitrary.}
\end{equation*}

This is made possible by taking $B_{\sigma \mu}=K_{\sigma \mu}$ and $C_{\sigma \mu \nu}  = L_{\sigma \mu \nu} + \tau_\mu  K_{\sigma \nu}$.  

\paragraph{}
The form of Hartong-Obers' parameterization has advantages over Eq.\ref{eq49} and Eq.\ref{eq50} when one imposes the Galilean boost invariant condition in section\ref{sec2.7}; in particular, it is easier to impose the required constraints on $K_{\mu \nu}$ and $L_{\sigma \mu \nu}$ than on $\Gamma^\lambda_{[\mu \nu]}$, $2 \tau_\mu e^a_{[\sigma} \omega_{\nu ] a}$, or $Y_{\mu \nu \sigma}$.

\paragraph{}
As we mentioned, the degrees of freedom in choosing $\Gamma^\lambda_{\mu \nu}$ come from the arbitrariness of two quantities.  The first one is the spatial contortion – the spatial component of the old contortion tensor to be exact – which we denote $\kappa_{\mu\nu\sigma} = -(2\Gamma^\lambda_{[\mu \sigma]} h_{\nu \lambda} +2 \Gamma^\lambda_{[\nu \sigma]} h_{\mu \lambda} -2 \Gamma^\lambda_{[\mu \nu]} h_{\lambda \sigma} )$ which boils down to the arbitrariness of the individual torsion tensors $\Gamma^\lambda_{[\mu \nu]}$.  The second is the Newton-Coriolis $2  e^a_{[\sigma} \omega_{\nu ] a}$ which is a brand new term not found in GR.  Since these are ultimately represented by $K_{\mu \nu}$ and $L_{\sigma \mu \nu}$ and vice versa, it is useful to write one representation in terms of the other.  In particular, one can show the following: (see Appendix for derivation)

\begin{align} 
2e^a_{[\nu} \omega_{\mu ] a} &= -L_{\sigma[\mu \nu ]} v^\sigma +K_{\mu \nu },  \label{eq52}\\
\kappa_{\mu\nu\sigma}&= L_{\sigma \mu \nu } - \tau_\mu v^\lambda L_{\lambda [\nu \sigma]} -  \tau_\nu v^\lambda L_{\lambda [\mu \sigma]}. \label{eq53}
\end{align}

Notice that while one can legally choose $K_{\mu \nu}$ and $L_{\sigma \mu \nu}$ to simply be $e^a_{[\nu} \omega_{\mu ] a} $ and $\kappa_{\mu\nu\sigma}$ to obtain the most straightforward interpretation of the parameters, this is not the case in general.  $K_{\mu \nu}$ is not equivalent to the Newton-Coriolis and $L_{\sigma \mu \nu}$ is not equivalent to the contortion.  It is true however that if $K_{\mu \nu}=0$ one has zero Newton-Coriolis and if $L_{\sigma \mu \nu}=0$ one has zero contortion (the converse is not true in general). 
  
\paragraph{}
Lastly, note that the torsion of Galilean's $\Gamma^\lambda_{\mu \nu}$ obtained by taking the $\{\mu, \nu\}$ anti-symmetric component of Eq.\ref{eq49} is 

\begin{equation}\label{eq54}
\Gamma^\lambda_{[\mu \nu]}= - v^\lambda \partial_{[\mu} \tau_{\nu]} + h^{\lambda \sigma} h_{\rho \sigma} \Gamma^\rho_{[\mu \nu]},
\end{equation}

which is just a projection equation of $\Gamma^\lambda_{[\mu \nu]}$ in time and space.  This is significant however, since the temporal component comes purely from the "Levi-Civita" part of Eq.\ref{eq49}. This fact makes Newton-Cartan theory seems like an inherently torsional theory since without the contortion term its spacetime is still torsional.  This is not entirely accurate though, since "contortion" in this case corresponds only to spatial contortion and clearly without demanding the temporal contortion $\partial_{[\mu} \tau_{\nu]}$ to also be zero, one would not obtain a torsionless spacetime.  The vital difference here is simply the fact that Galilean's temporal torsion is contained not in the contortion, but in the Levi-Civita term instead.  

\paragraph{}
It is also worth noting that if we demand zero spatial contortion but not zero $\partial_{[\mu} \tau_{\nu]}$, there are always some boosted observers that will nevertheless have spatial torsion\footnote{To see this, consider $\Gamma^\lambda_{[\mu \nu]}= -v^\lambda \partial_{[\mu} \tau_{\nu]} + e^\lambda_a (\partial_{[\mu}e^{a}_{\nu]} - \omega_{[\mu}^a \tau_{\nu]} -\omega_{[\mu}^{\text{ } \text{ } ab}e_{\nu]b})$.  Even its spatial component is zero in one frame, in a boosted frame where $e^a_\mu \rightarrow e^a_\mu + \Lambda^a \tau_\mu$, this component will be non zero again since it will definitely contain a non-vanishing $\Lambda^a \partial_{[\mu}\tau_{\nu]}$ term among others.}.  Hence, $\partial_{[\mu} \tau_{\nu]}=0$ is necessary (but not sufficient) if one is to have a completely torsionless spacetime for all observers like in the torsionless GR. 

\paragraph{}
Once the values of  $K_{\mu \nu}$ and $L_{\sigma \mu \nu}$ are chosen, one can in principle express all spacetime parameters from $\Gamma^\lambda_{\mu \nu}$ to the various curvature form components in terms of only the vielbeins.

\subsection{Imposing invariance on Galilean theory}
\label{sec2.7}

\paragraph{}
Our quest to derive a general Galilean metric connection does not end with Eq.\ref{eq49} and its re-parametrizations.  Another constraint must be imposed on it to ensure invariance under Galilean transformations, which is vital since we are unable to use a frame dependent connection to construct our physical system.  Going back to section \ref{sec1.7}, such constraint is not required on a relativistic background since $\delta_M \Gamma^\lambda_{\mu \nu}$ and $\delta_M g_{\mu \nu}$ always automatically vanish. It is the mere fact that Galilean boost separate itself from the combined spacetime rotation i.e. the Lorentz rotation that rises this difficulty.  

\paragraph{}
To show that under $\delta_J$ all these quantities do vanish is as easy as showing the same for $\delta_M$ in section \ref{sec1.7}; this is true since the anti-symmetricity of rotation is still preserved by $J$.  Thus, one simply needs to ensure the condition $\delta_G \Gamma^\rho_{\mu \nu}=0$ to obtain the legitimate solutions of Galilean metric connection.  In this section we shall employ Hartong-Obers parameterization to simplify calculations

\begin{equation}\label{eq55}
 \Gamma^\lambda_{\mu \nu}=-  v^\lambda  \partial_\mu \tau_\nu  +  \frac{1}{2}  h^{\lambda \sigma}(\partial_\mu h_{\nu \sigma} + \partial_\nu h_{\mu \sigma}-\partial_\sigma h_{\mu \nu}) +\frac{1}{2} h^{\lambda \sigma} (\tau_\mu K_{\sigma \nu} +\tau_\nu K_{\sigma \mu} + L_{\sigma \mu \nu}).
\end{equation}

\paragraph{}
First we shall note that our metric is not $G$ invariant itself; more precisely, only two out of the four metrics and inverse metrics are invariant: $h^{\mu \nu}$ and $\tau_\mu$.  It is no coincidence that these quantities are the ones that become legitimate "metrics" out of the vielbein postulate i.e. the ones that end up satisfying the metric connection.   Afterall, only quantities invariant under a change of reference frame can be qualified as a metric.

\begin{align}
\delta_G h_{\mu \nu}&=\delta_G( \delta_{ab} e^a_\mu e^b_\nu) \nonumber \\
&= \delta_{ab} \big( ( e^c_\mu \lambda^a_c + \lambda^a \tau_\mu) e^b_\nu   + e^a_\mu  ( e^c_\nu \lambda^b_c + \lambda^b \tau_\nu) \big)  \nonumber \\
&= \lambda_b e^b_\nu \tau_\mu +\lambda_a e^a_\mu \tau_\nu \nonumber  \\
&= \lambda_\nu \tau_\mu + \lambda_\mu \tau_\nu \label{eq56} \\
\delta_G v^\mu&=e^\mu_a \lambda^a = \delta^{ab} \lambda_b e_{\nu a} h^{\nu \mu} =\lambda_\nu  h^{\nu \mu} \label{eq57}
\end{align}

\paragraph{}
In the above we have written $e^a_\mu \lambda_a$ as $\lambda_\mu$.  This is however an abuse of notation since $e^a_\mu$ – being an incomplete vielbein – cannot transform a Galilean vielbein indexed vector into a coordinate indexed vector (see the note on Galilean vielbein formalism).  Thus, one has to keep in mind that $\lambda_\mu$ is simply a shorthand for $e^a_\mu \lambda_a$.

\begin{align}
\delta_G h^{\mu \nu}&=\delta_G( \delta^{ab} e^\mu_a e^\nu_b) \nonumber \\
&= \delta^{ab} \big( (- \lambda^c_a e^\mu_c) e^\nu_b   + (- \lambda^c_b e^\nu_c) e^\mu_a \big)  \nonumber \\
&= 0 \label{eq58} \\
\delta_G \tau_\mu &= 0 \hspace{0.3cm} \text{from Eq.\ref{eq30}} \label{eq59}
\end{align}

Based on these, the change of $\Gamma^\lambda_{\mu \nu}$ under Galilean boost is simply the sum of: (see derivation in the Appendix)

\begin{align*}
 \delta_G \big[ -  v^\lambda  \partial_\mu \tau_\nu  +  \frac{1}{2}  h^{\lambda \sigma} (\partial_\mu h_{\nu \sigma} + \partial_\nu h_{\mu \sigma}-\partial_\sigma h_{\mu \nu})\big]=&\frac{1}{2}  h^{\lambda \sigma} \big[ 2 \tau_\nu \partial_{[\mu} \lambda_{\sigma]} + 2 \tau_\mu \partial_{[\nu} \lambda_{\sigma]}  -2\lambda_\sigma \partial_{[\mu} \tau_{\nu]}  \\ 
&+ 2\lambda_\mu \partial_{[\nu} \tau_{\sigma]} + 2 \lambda_\nu \partial_{[\mu} \tau_{\sigma]} \big]
\end{align*}
and
\begin{equation*}
\delta_G \big[ \frac{1}{2}  h^{\lambda \sigma} (\tau_\mu K_{\sigma \nu} +\tau_\nu K_{\sigma \mu} + L_{\sigma \mu \nu})\big]= \frac{1}{2}  h^{\lambda \sigma} (\tau_\mu \delta_G K_{\sigma \nu}  + \tau_\nu \delta_G K_{\sigma \mu} + \delta_G L_{\sigma \mu \nu})
\end{equation*}

Hence, to $\delta_G \Gamma^\lambda_{\mu \nu}=0$ is achieved if and only if 

\begin{align}
\delta_G K_{\mu \nu} &= 2 \partial_{[\mu} \lambda_{\nu]}, \label{eq60} \\
\delta_G L_{\sigma \mu \nu}&= 2\lambda_\sigma \partial_{[\mu} \tau_{\nu]} - 2\lambda_\mu \partial_{[\nu} \tau_{\sigma]} -2 \lambda_\nu \partial_{[\mu} \tau_{\sigma]}.  \label{eq61}
\end{align}

\paragraph{}
The above shows us that $\delta_G$ is always non-zero for Galilean's "Levi-Civita" term due to: 1) the non-invariance of $h_{\mu \nu}$ corresponding to the $\partial_{[\mu} \tau_{\nu]}$ terms and 2) the non-invariance of $v^\lambda$  corresponding to the $\partial_{[\mu} \lambda_{\sigma]}$ terms.  Consequently, to have a Galilean invariant theory additional terms in the form of $K_{\mu \nu}$ and $L_{\sigma \mu \nu}$ are required to cancel off these  $\delta_G$'s; this implies that the contortion and Newton-Coriolis are essential in such theory.  Of course, when one demands $\partial_{[\mu} \tau_{\nu]}$ in a torsionless spacetime, $L_{\sigma \mu \nu}$ is allowed to be zero.  $K_{\mu \nu}$ on the other hand is always required, implying (from Eq.\ref{eq52}) that the Newton Coriolis must never be set to zero in this theory.  

\newpage

%%%%%%%%%%%%%%%%%%%%%%%%%%%%%%%%%%%%%%%%%%%%%%%%%%%%%%%%%%%%%%%%%%%%%%%%%%%%%%%%%%%%%%%%%%%%%%%%%%%%%%%%%%%%%%%%%%%%%%%%%%%%
\noindent\fbox{%
    \parbox{\textwidth}{\section*{\Large{Galilean Vielbein Formalism}}

Let $v^\mu$ and $e^\mu_a$ be denote the inverses of $\tau_\mu$ and $e^a_\mu$, respectively.  Note that these inverses are by no means uniquely determined; they are simply any spacetime forms that satisfy the following: 
   
 \begin{align*} 
 &v^\mu \tau_\mu=-1      &e^\mu_a e^b_\mu=\delta^a_b \\
 &e^\mu_a \tau_\mu=0         &v^\mu e^a_\mu=0,
\end{align*}

and the completeness relation

\begin{equation*}
e^\mu_a e^a_\nu=\delta^\mu_\nu+v^\mu \tau_\nu.
\end{equation*}

Using Eq.\ref{eq31} and $\delta(e^a_\mu e^\mu_b)=\delta(\delta^a_{\text{ } b})=0$ one deduces

\begin{equation*}
\delta e^\mu_a=\pounds_\xi e^\mu_a  - \lambda^c_{\text{ }a} e^\mu_c.
\end{equation*}
\\
Similarly, usingEq.\ref{eq30} and $\delta(v^\mu \tau_\nu )=\delta(e^\mu_a e^a_\nu-\delta^\mu_\nu)=\delta(e^\mu_a e^a_\nu)=e^\mu_a \lambda^a \tau_\nu$,

\begin{equation*} 
\delta v^\mu=\pounds_\xi  v^\mu + \lambda^a e^\mu_a.
\end{equation*}

The vielbein postulates are also satisfied by these vielbein inverses:

\begin{align*}
D_\mu v^\nu =\partial_\mu v^\nu +\Gamma^\nu_{\mu \sigma} v^\sigma -\omega^a_\mu e^\nu_a =0, \text{and} \\
D_\mu e^\nu_a =\partial_\mu e^\nu_a +\Gamma^\nu_{\mu \sigma} e^\sigma_a -\omega^{\text{ }\text{ }b}_{\mu \text{ } a}e^\nu_b=0.
\end{align*}

One can introduce the spatial metric $h_{\mu \nu}=\delta_{ab} e^a_\mu e^b_\nu$ and its inverse $h^{\mu \nu}=\delta^{ab} e^\mu_a e^\nu_b$.  Note that these are spacetime entities despite being an exclusively spatial metric.  The temporal metric is simply given by $\tau_\mu$.  
\\ \\
Being a degenerate (incomplete) metric, $h_{\mu \nu}$ and $h^{\mu \nu}$ cannot be used for index raising and lowering of general spacetime vectors $f_\mu$ and forms $f^\mu$, unlike $g_{\mu \nu}$ and $g^{\mu \nu}$ in GR:

\begin{equation*}
h^{\mu \nu} f_\mu \neq f^\nu \hspace{0.5cm} \text{and} \hspace{0.5cm}  h_{\mu \nu} f^\mu \neq f_\nu.
\end{equation*}  

They do however, work on purely spatial entities such as the spatial vielbeins $e^a_\mu$ and $e^\mu_a$:

\begin{equation*}
h^{\mu \nu} e^a_\mu=e^{\nu a} \hspace{0.5cm} \text{and} \hspace{0.5cm} h_{\mu \nu} e^\mu_a = e_{\nu a}  .
\end{equation*}  

Due to the same reason, spatial vielbeins are not adequate to transform a coordinate indexed quantity to a vielbein one:

 \begin{align*} 
 &e^a_\mu f_a \neq f_\mu  \hspace{0.5cm} \text{and} \hspace{0.5cm}  e^\mu_a f^a \neq f^\mu .
\end{align*}

Although they are able to do the opposite direction since the coordinate index contraction is whole in spacetime

 \begin{align*} 
 &e^a_\mu f^\mu = f^a      \hspace{0.5cm} \text{and} \hspace{0.5cm}  e^\mu_a f_\mu = f_a .
\end{align*}

    }
}

%%%%%%%%%%%%%%%%%%%%%%%%%%%%%%%%%%%%%%%%%%%%%%%%%%%%%%%%%%%%%%%%%%%%%%%%%%%%%%%%%%%%%%%%%%%%%%%%%%%%%%%%%%%%%%%%%%%%%%%%%%%%

%%%%%%%%%%%%%%%%%%%%%%%%%%%%%%%%%%%%%%%%%%%%%%%%%%%%%%%%%%%%%%%%%%%%%%%%%%%%%%%%%%%%%%%%%%%%%%%%%%%%%%%%%%%%%%%%%%%%%%%%%%%%

\noindent\fbox{%
    \parbox{\textwidth}{\section*{\Large{Galilean Vielbein Formalism (cont.)}}

The Galilean spatial metric $\delta_{ab}$ on the other hand is legitimate, non-degenerate spacetime metric and as such are able to bring Galilean indexes up and down

\begin{equation*}
\delta_{ab} f^a=f_b\hspace{0.5cm} \text{and} \hspace{0.5cm} \delta^{ab} f_a=f^b   .
\end{equation*}  

Some identities of $h_{\mu \nu}$ and $h^{\mu \nu}$ that might be useful on common calculations include

\begin{equation*}
 f^af_a=h^{\mu \nu} f_\mu f_\nu,
\end{equation*}

\begin{equation*}
h_{\mu \sigma}h^{\sigma \nu}=e^\mu_a e^a_\nu=\delta^\mu_\nu+v^\mu \tau_\nu.
\end{equation*}

		} 
}

%%%%%%%%%%%%%%%%%%%%%%%%%%%%%%%%%%%%%%%%%%%%%%%%%%%%%%%%%%%%%%%%%%%%%%%%%%%%%%%%%%%%%%%%%%%%%%%%%%%%%%%%%%%%%%%%%%%%%%%%

%%%%%%%%%%%%%%%%%%%%%%%%%%%%%%%%%%%%%%%%%%%%%%%%%%%%%%%%%%%%

\section{Bargmann Algebra}
\label{chap3}

\paragraph{} 
In this chapter we shall finally analyze the Bargmann algebra, which is the so-called central extension of the Galilean algebra of chapter 2.  Just like Galilean, Bargmann algebra aims to quantize the geometric theory of Newtonian gravity and we shall indeed arrive at the exact same set of spatial and temporal metrics, as well as the general solution for metric compatible connections $\Gamma^\lambda_{\mu \nu}$.  Why then, would we go through all the hassle to perform this central extension and construct a new algebra?  There are two reasons: first, loosely speaking, our Galilean algebra of the previous chapter is an incorrect algebra, at least for massive particles.  Second, the gauge field corresponding to the new central element provides us with a simple class of solution for Galilean invariant metric connections (provides us with simple $K_{\mu \nu}$ and $L_{\sigma \mu \nu}$ satisfying Eq.\ref{eq60} and Eq.\ref{eq61}).  

\paragraph{} 
In the previous chapter, we incorrectly – and rather purposely – assert $\lim_{c \to \infty} c P_0= H$.  This is incorrect since $P_0=+\sqrt{(mc)^2+\mathbf{P}^2}\approx \big( mc+\frac{\mathbf{P}^2}{2m}+\dots \big)$ i.e. the total energy whereas $H$ represents only the kinetic energy.  This means $\lim_{c \to \infty} c P_0= H + mc^2$ and $H=\big(\frac{\mathbf{P}^2}{2m} +\dots\big)$.  Note that this differs fundamentally from the relativistic field theories; in the latter, the time evolution generator is given by $P_0$ since $H=cP_0$.  This is possibly the most vital step in eluding the "problem of time": we have decomposed the troublesome $P_0$ into $H$ and the mass terms and hence it no longer represents time evolution.  

\paragraph{}
If we now redo the Inonu-Wigner contraction of the Poincaré algebra while keeping the rest mass correction term in mind, we shall obtain an algebra in which this mass is represented by a central element $M$:

\begin{align*}
&[P_A,P_B]=0 \\
&\Rightarrow  \hspace{9.8cm} \mathbf{ [P_a,P_b]=0} \\
&\Rightarrow [P_a, P_0]=[P_a, \frac{1}{c^2} (Mc^2+cH)]=0  \hspace{3.75cm} \mathbf{  [P_a, H]=0 }\\
& \hspace{10.5cm} \mathbf{  [P_a, M]=0 }
\\
&[M_{AB},P_C]=\eta_{BC}P_A-\eta_{AC}P_B \\
&\Rightarrow [M_{a0}, P_0]=[-cG_a, \frac{1}{c^2} (Mc^2+cH)]=\eta_{00}P_a \hspace{2.18cm} \mathbf{  [H, G_a]=P_a}\\
& \hspace{10.55cm} \mathbf{  [G_a, M]=0 } \\
&\Rightarrow [M_{a0}, P_c]=[-cG_a, P_c]=-\eta_{ac}P_0 = \delta_{ac} \frac{1}{c^2} (Mc^2+cH) \hspace{0.5cm} \mathbf{  [G_a, P_c] \approx -\delta_{ac}M}\\
&\Rightarrow [M_{ab}, P_0]= [J_{ab},  \frac{1}{c^2} (Mc^2+cH)]   \hspace{4.33cm}\mathbf{  [J_{ab}, H]=0}\\
& \hspace{10.68cm} \mathbf{  [J_{ab}, M]=0 } \\
&\Rightarrow [M_{ab}, P_c]= [J_{ab}, P_c]   \hspace{6.58cm}\mathbf{  [J_{ab}, P_c]=\delta_{ac}P_b-\delta_{bc}P_a}\\
\\
&[M_{AB},M_{CD}]=\eta_{AD}M_{BC}-\eta_{AC}M_{BD}+\eta_{BC}M_{AD}-\eta_{BD}M_{AC} \\
&\Rightarrow [M_{a0}, M_{c0}]=[-cG_a, -cG_c]=\eta_{00}J_{ac} \hspace{3.35cm} \mathbf{  [G_a, G_c]=\frac{1}{c^2}J_{ac}\approx0}\\
&\Rightarrow [M_{a0}, M_{cd}]=[-cG_a, J_{cd}]=\eta_{ad} M_{0c} -\eta_{ac} M_{0d}  \hspace{1.9cm} \mathbf{  [G_a, J_{cd}]= \delta_{ad} G_c-\delta_{ac} G_d}\\
&\Rightarrow   [M_{ab}, M_{cd}]=  [J_{ab}, J_{cd}] \hspace{5.95cm} \mathbf{  [J_{ab}, J_{cd}]= \delta_{ac}J_{bd}-\delta_{ad}J_{bc}+\delta_{bd}J_{ac} - \delta_{bc}J_{ad}}
 \end{align*}

\paragraph{}
Perhaps a more intuitive view on the significance of the central extension is via analyzing\footnote{Note that all $i$ and $\hbar$ factors are suppressed in our algebra notation.}  the transformation of the momentum generator under boost, $e^{ \frac{i}{\hbar}  \Lambda^b G_b} \mathbf{P} e^{- \frac{i}{\hbar}  \Lambda^b G_b}$, which action should certainly be $\mathbf{P}\rightarrow \mathbf{P}+M \mathbf{\Lambda}$ where $ \mathbf{\Lambda}$ is the vector of $\Lambda^b $. Using the identity

\begin{align*}
	e^{ \frac{i}{\hbar}  \Lambda^b G_b} P_a  e^{- \frac{i}{\hbar}  \Lambda^b G_b}&= P_a +  \frac{i}{\hbar}  v_b [G_b,P_a]- \frac{ v_b v_c}{2! \hbar^2}[G_b [G_c,P_a]]+\dots,
\end{align*}

we immediately see that $[G_b, P_a]= - i\hbar \delta_{ab}  M$ does result in the correct transformation.  Finally, it is useful to note as well the commutation $[G_b, x_a]= - i\hbar \delta_{ab}  t$, which can be deduced by comparing the correct transformation $\mathbf{x}\rightarrow \mathbf{x} +\mathbf{\Lambda} t$ with 

\begin{align*}
	e^{ \frac{i}{\hbar}  \Lambda^b G_b} x_a  e^{- \frac{i}{\hbar}  \Lambda^b G_b}&= x_a +  \frac{i}{\hbar}  v_b [G_b,x_a]- \frac{ v_b v_c}{2! \hbar^2}[G_b [G_c,x_a]]+\dots . 
\end{align*}

\subsection{The extension of Galilean vielbein bundle}
\label{sec3.1}

\paragraph{}
With the addition of the central element $M$, the Galilean vielbein bundle can be extended to include one other vielbein field $-m=-m_\mu \partial_\mu$ i.e. on the whole becomes $[\mathbf{e}, \tau, -m]$.  This means the old Galilean vielbein system consisting $(d+1)$-dimensional co-vectors is extended with another frame vector $m$, satisfying its own transformation law under the translationless Galilean elements \cite{GeracieBargmann}:

\begin{equation}\label{eq62}
-m\rightarrow -m- \frac{1}{2} \Lambda^a \Lambda_a \tau - \Lambda_a \Lambda^a_b e^b.
\end{equation}

\paragraph{}
The above transformation resembles the four-momentum transformation of a massive particle of mass $m$.  Choosing a section $\sigma(x,t)$ on the extended vielbein now corresponds – on top of a coframe $e^a$ and $\tau$ – to choosing a measurement frame that is the rest mass for this energy. 

\paragraph{}
As we mentioned in chapter 1, it is not straightforward to see why Eq.\ref{eq62} is true whereas for the original vielbeins, Galilean boost and rotations have obvious effects.  Nevertheless, one can justify this equation \footnote {For a detailed discussion on Galilean unitary representation and its extensions, see for example \cite{Dawson}.} by considering the unitary representation of a general Galilean group $B=(\Delta_\theta, \Lambda^a, \Delta^a_x, \Delta_t)$ corresponding to rotation $\Delta_\theta$, boost vector $\Lambda^a$, spatial translation vector $\Delta^a_x$, and temporal translation $\Delta_t$.  Group element corresponding to internal $M$ transformation can also be added on top of these, but we are interested only on Galilean group action for now.  We shall apply $B$ as follows: first, the system initially at $(x,t)$ is translated to the origin $(0,0)$, followed by $\Delta_\theta$ rotation, followed by forward spatial translation to $\mathbf{x}+\Delta_x$, followed by $\Lambda^a$ boosting and finally translated forward in time to $t+\Delta_t$.

\begin{align}
U_{B}(x,t)= &{\rm exp} \Big(- \frac{i}{\hbar} g(x,t)\Big) \text{ } U_H(t+\Delta_t) \text{ } U_G (\Lambda^a) \text{ } \text{ } U_P(\mathbf{x}+\Delta_x)  \text{ } U_J(\Delta_\theta) \text{ } U_P(-x)  \text{ } U_H(-t)  \nonumber \\
= &{\rm exp} \Big(- \frac{i}{\hbar}M( \frac{1}{2} \Lambda^a \Lambda_a t + \Lambda_a \Delta^a_x) \Big) {\rm exp} \big(-\frac{i}{\hbar} H \Delta_t\big)  {\rm exp} \big(-\frac{i}{\hbar} \Delta^a_x P_a\big)  {\rm exp} \big(- \frac{i}{\hbar} \Lambda^a G_a \big) {\rm exp} \big(- \frac{i}{\hbar} \Delta_\theta \mathbf{\hat{n} \cdot \hat {J}}\big)\nonumber \\
=&{\rm exp} \Big( -\frac{i}{\hbar}M( \frac{1}{2} \Lambda^a \Lambda_a t  + \Lambda_a \Delta^a_x) \Big) {\rm exp} \big(-\frac{i}{\hbar} H \Delta_t\big)  {\rm exp} \big(-\frac{i}{\hbar} \Delta^a_x P_a\big){\rm exp} \Big( -\frac{i}{\hbar}( P_a t-M \Lambda_a x'^a ) \Big) \nonumber \\
&\text{ }{\rm exp} \big(- \frac{i}{\hbar} \Delta_\theta \mathbf{\hat{n} \cdot \hat {J}} \big) \nonumber\\
=&{\rm exp} \Big( \frac{i}{\hbar}M( \frac{1}{2} \Lambda^a \Lambda_a t + \Lambda_a \Delta^a_x + \Lambda_a \Lambda^a_b x^b )\Big) {\rm exp} \big(-\frac{i}{\hbar} H \Delta_t\big) {\rm exp} \Big(-\frac{i}{\hbar} (\Delta^a_x + \Lambda^a t) P_a\Big)  {\rm exp} \big(- \frac{i}{\hbar} \Delta_\theta \mathbf{\hat{n} \cdot \hat {J}} \big) \label{eq63}
\end{align}

The phase factor ${\rm exp} \big(- \frac{i}{\hbar} g(x,t)\big)$ where $g(x,t)=M( \frac{1}{2} \Lambda^a \Lambda_a t + \Lambda_a \Delta^a_x) $ is necessary since the representation of this extended Galilean group is not faithful\footnote{The Poincaré group on the other hand is always faithful, whereas in the Galilean classical system there are no such thing as two states differing by a phase factor; faithfulness is therefore assumed a priori.  This is the algebraic point of view of why $M$ is assumed to be zero in Galilean yet arises in the Bargmann algebra.} i.e. $U_{BB'}\neq U_BU_{B'}$ for two general elements $B$ and $B'$.  The corrected multiplication rule includes this phase factor whose value was derived by Bargmann \cite{Bargmann}.  From the second to the third line, we have used $G_a=P_a t -M x_a$ which is deduced from its commutation $[G_b, P_a]= - i \hbar \delta_{ab}  M$ and $[G_b, x_a]= - i \hbar \delta_{ab} t$.  When $U_P$ acts however, the system has been rotated, boosted and translated forward so we use $x'^a= \Lambda^a_b x^b+ \Lambda^a t +\Delta_x^a$. 

\paragraph{}
If we  now set $\Delta_x$ and $\Delta_t$ to zero, Eq.\ref{eq63} implies that a general translationless Bargmann unitary representation is

\begin{equation}
U_{B}(x,t)={\rm exp} \Big(\frac{i}{\hbar}M( \frac{1}{2} \Lambda^a \Lambda_a t + \Lambda_a \Lambda^a_b x^b)\Big) {\rm exp} \big(-\frac{i}{\hbar}  \Lambda^a t P_a\big)  {\rm exp} \big(- \frac{i}{\hbar} \Delta_\theta \mathbf{\hat{n} \cdot \hat {J}} \big). \label{eq64}
\end{equation}

\paragraph{}
The last two operators are just rotation by $\Delta_\theta$ and translation by $\Lambda^a t$ whose impact is only on the position vector or in our case, the spatial vielbeins: $e^a \rightarrow \Lambda^a_b e^b+ \Lambda^a \tau$.  The temporal vielbein on the other hand is unaffected by both transformations.  There is however an additional term ${\rm exp} \big(\frac{i}{\hbar}M( \frac{1}{2} \Lambda^a \Lambda_a t + \Lambda_a \Lambda^a_b x^b)\big)$ corresponding to the new generator $M$.  This implies that there is a certain quantity describing the system's state – other than space and time – that is translated by $-(\frac{1}{2} \Lambda^a \Lambda_a t +\Lambda_a \Lambda^a_b x^b)$.  This mysterious quantity motivates our invention of the new vielbein, and we name it $-m$.  To conclude, the matrix representation Eq.\ref{eq64} can be written as:

\begin{equation} \label{eq65}
\begin{bmatrix}
    x'    \\
   t'      \\
-m'\\
\end{bmatrix}
=
\begin{bmatrix}
    \Lambda^a_b     & \Lambda^a    &0\\
   1       & 0     &0   \\
-\frac{1}{2} \Lambda^a_b \Lambda_a & -\frac{1}{2} \Lambda^a \Lambda_a &1\\
\end{bmatrix}
\begin{bmatrix}
    x    \\
   t      \\
-m\\
\end{bmatrix}.\\
\end{equation}

In section \ref{sec3.2}, Eq.\ref{eq62} simply becomes the consequence of gauging the Bargmann algebra by identifying $m$ (not the vielbein $-m$) as the gauge field of $M$, showing the superior efficiency of the technique.

\subsection{The Bargmann gauge field}
\label{sec3.2}
 
\paragraph{}
The most general Bargmann algebra valued form is the extension of Eq.\ref{eq28} with the term $M n_\mu$.  Being a central element means $M$ commutes all other generators and we shall see that $M n_\mu$ does not affect the original Galilean gauge transformation rules.

\begin{equation} \label{eq66}
A_\mu=H t_\mu + P_a f^a_\mu +G_a \omega^a_\mu + \frac{1}{2} J_{ab} \omega^{\text{ }ab}_\mu + M n_\mu.
\end{equation}

\paragraph{}
The $\Pi$ in Eq. \ref{eq28} now becomes 

\begin{align}  \label{eq67}
\Pi&= H \Delta + P_a \zeta^a  + G_a \sigma^a +\frac{1}{2}  J_{ab} \sigma^{ab} + M \sigma \nonumber  \\
&= \xi^\mu A_\mu + G_a \lambda^a +\frac{1}{2}  J_{ab} \lambda^{ab} + M \lambda
\end{align}
 
for an infinitesimal $\sigma$, and $\lambda=\sigma-\xi^\mu n_\mu$.  As such, in 

\begin{align*}
 \bar{\delta}A_\mu =\pounds_\xi A_\mu + \partial_\mu \Sigma+[A_\mu,\Sigma],
\end{align*}

the internal infinitesimal transform parameter $\Sigma$ is given by  $\frac{1}{2}  J_{ab}  \lambda^{ab} + G_a \lambda^a+M \lambda$.  It is then straightforward to see that Eq.\ref{eq30} to Eq.\ref{eq33} are still valid since the $M$ term in $\Sigma$ commutes with every other generators.  The $M$ component of $\delta A_\mu$ which is $\delta n_\mu$ – once again we refer $\bar{\delta}$ simply as $\delta$ – is given by\footnote{The $\delta_M$ here has nothing to do with $\delta_M$ in chapter one, which was referring to Lorentz rotation.  This is simply an abuse of notation.}

\begin{align}   \label{eq68}
\delta n_\mu = {\delta}_M A_\mu &=\delta_M ( \pounds_\xi A_\mu )+\delta_M ( \partial_\mu \Sigma) + \delta_M [A_\mu,\Sigma] \nonumber \\
&=(\partial_\nu m_\mu) \xi^\nu + \partial_\mu \lambda + [P_b f^b_\mu ,  G_c \lambda^c]_{M} \nonumber \\
&=(\partial_\nu m_\mu)  \xi^\nu + \partial_\mu \lambda + (\delta_{bc} \lambda^c f^b_\mu M )_{M} \nonumber \\
&=\pounds_\xi n_\mu + \partial_\mu \lambda+f^a_\mu \lambda_a.
\end{align}

When only rotations and boosts are concerned, this is exactly the infinitesimal version of Eq.\ref{eq62}.  Hence, we can identify this gauge field as the negative components of our vielbein $-m$: $n_\mu=m_\mu$. 

\paragraph{} 
Notice that Eq.\ref{eq68} implies that under $M$'s own internal transform parameterized by $\lambda$ – this is outside the normal Galilean group action on Eq.\ref{eq64} – there is also a translation of $m_\mu$ by $\partial_\mu \lambda$.  This means our gravity theory might not be invariant under $M$ translation, just as it isn't under normal translations..  We can however also elude this problem by using our gauge freedom to replace $m_\mu$ everywhere with $M_\mu=m_\mu -\partial_\mu \chi$ where the background field $\chi$ – called the Stueckelberg scalar – transforms as $\bar{\delta}\chi=\pounds_\xi \chi + \lambda$ as suggested by \cite{HartongObers}.  This guarantees $\bar{\delta}_M (M_\mu)=0$, leaving boosts and rotations the only transformation we need to look out for once again.  In this work, we shall keep using $m_\mu$ while keeping in mind that the $\chi$ term can always be added any time.  

\subsection{The curvature 2-form}

Once again, we write the curvature $F_{\mu \nu}$ in terms of its projection on the generators:

\begin{equation}\label{eq69}
F_{\mu \nu}=H R_{\mu \nu}(H) + P_a R^a_{\mu \nu}(P) + G_a R^a_{\mu \nu}(G) + \frac{1}{2} J_{ab} R^{ab}_{\mu \nu}(J) + M R_{\mu \nu}(M).
\end{equation}

\paragraph{}
Once again, the fact that $M$ is a central elements means that Eq.\ref{eq35} to Eq.\ref{eq38} are still valid.  $R_{\mu \nu}(M)$ on the other hand is given by

\begin{align} \label{eq70}
R_{\mu \nu}(M)&=(\partial_\mu A_\nu - \partial_\nu A_\mu)_{M} +[A_\mu,A_\nu]_{M} \nonumber \\
&=(\partial_\mu m_\nu - \partial_\nu m_\mu) +[P_c e^c_\mu + G_d \omega^d_\mu, P_c e^c_\nu + G_d \omega^d_\nu]_{M}  \nonumber \\
&=2\partial_{[\mu}m_{\nu]} +[P_c e^c_\mu, G_d \omega^d_\nu]_{M}+ [G_d \omega^d_\mu,P_c e^c_\nu]_{M}  \nonumber \\
&=2\partial_{[\mu}m_{\nu]} +2 e^c_{[\mu}\omega_{\nu]}^d [P_c,G_d]_{M} \nonumber \\
&= 2\partial_{[\mu}m_{\nu]} +2 e^c_{[\mu}\omega_{\nu]}^d [P_c,G_d]_{M} \nonumber \\
&=2\partial_{[\mu}m_{\nu]} + 2 e^c_{[\mu}\omega_{\nu]}^d (\delta_{dc} M)_{M} \nonumber \\
&=2\partial_{[\mu}m_{\nu]}+ 2 e^a_{[\mu} \omega_{\nu ] a}.
\end{align}

\subsection{The new and combination vielbeins}

Although the choices of spatial and temporal vielbeins in this algebra are nothing different than the Galilean's, the fact that another vielbein field $m_\mu$ transforming as Eq.\ref{eq68} exists allows us to construct some Galilean boost invariant vielbein combinations such as

\begin{align}
 \hat{e}^a_\nu &=e^a_\nu -m^a\tau_\nu,  \label{eq71}\\
\hat{v}^\nu &= v^\mu -h^{\nu \sigma} m_\sigma,  \label{eq72}
\end{align}

where 

\begin{equation}
m^a=e^{\sigma a} m_\sigma  \label{eq73}.  
\end{equation}

Its transformation, ignoring the diffeomorphism term is

\begin{align}
\delta m^a&= \delta^{ab} \big( m_\sigma  (\delta e^\sigma_b )+ e^\sigma_b  (\delta m_\sigma)\big) \nonumber \\
                &= \delta^{ab} ( -\lambda^c_{\text{ } b} e^\sigma_c   m_\sigma + e^\sigma_b  e^c_\sigma \lambda_c) \nonumber\\
                &=  -\lambda_c^{\text{ }a}m^c+\lambda^a \nonumber\\
                &=  \lambda^a_{\text{ }c} m^c+\lambda^a.  \label{eq74}
\end{align}

One can easily confirm that $(\hat{e}^a_\mu, e^\mu_a, \tau_\mu, \hat{v}^\mu)$ forms a vielbein set that is just as good as the original $(e^a_\mu, e^\mu_a, \tau_\mu, v^\mu)$.  See the note on Bargmann vielbein formalism at the end of this chapter for details.  The expressions of Eq.\ref{eq71} and Eq.\ref{eq72} are indeed invariants since $\delta_G (e^a_\nu)=\lambda^a \tau_\nu$ cancels with $\tau_\nu \delta_G (m^a)$ and $\delta_G v^\nu=\lambda^a e^\nu_a$ cancels with $h^{\nu \sigma} \delta_G (m_\sigma)=h^{\nu \sigma} e^a_\sigma \lambda_a$.  The reason these vielbeins are preferred will be clear as we construct new metric like quantities in section \ref{sec3.5}.

\paragraph{}
The inverse relation of Eq.\ref{eq73} i.e. $m_\sigma$ in terms of $m^a$ is given by

\begin{align}
m_\lambda&= \hat{e}^a_\lambda m_b - \hat{v}^\mu \tau_\lambda m_\mu \nonumber \\
&= ( e^a_\lambda -m^a \tau_\lambda) m_b -(v^\mu - h^{\mu \nu} m_\nu)  \tau_\lambda m_\mu \nonumber \\
&=  e^a_\lambda m_a -m^a m_a \tau_\lambda - v^\mu m_\mu \tau_\lambda + h^{\mu \nu} m_\mu  m_\nu \tau_\lambda \nonumber \\
&=  e^a_\lambda m_a - \frac{1}{2} m^a m_a \tau_\lambda + (-v^\mu m_\mu +  \frac{1}{2} h^{\mu \nu} m_\mu  m_\nu) \tau_\lambda \nonumber \\
&=e^a_\lambda m_a - \frac{1}{2} m_a m^a \tau_\lambda + \Phi \tau_\lambda  \label{eq75}
\end{align}

where we have used $m^a m_a = h^{\mu \nu} m_\mu  m_\nu $ on the fourth line above. We leave the reader to confirm that the quantity

\begin{equation}
\Phi= -v^\mu m_\mu +  \frac{1}{2} h^{\mu \nu} m_\mu  m_\nu \label{eq76}
\end{equation}

is a complete Galilean (rotation and boost) invariant in this algebra.  

\subsection{Galilean $\Gamma_{{\mu}{\nu}}^\lambda$ solutions via combination vielbeins}
\label{sec3.5}

\paragraph{}
We have shown that the Galilean metric $h_{\mu \nu}$ is not a Galilean boost invariant, which causes the non invariance of even the Levi-Civita part of the Galilean connection.  This problem requires the contortion and Newton-Coriolis terms (parameterized by $L_{\sigma \mu \nu }$ and $K_{\mu \nu}$) that satisfies Eq.\ref{eq60} and Eq.\ref{eq61} to cancel the deviation under boosts.  There are infinitely many solutions for these $L_{\sigma \mu \nu }$ and $K_{\mu \nu}$, yet the simplest ones can be easily obtained with the help of the new set of vielbeins defined by Eq.\ref{eq71} and Eq.\ref{eq72}.  

\paragraph{}
 First, we define two new invariant metric-like quantities.  The first one is obtained by attempting to cancel off the deviation of $h_{\mu \nu}$ in Eq.\ref{eq56} with the help of the new vielbeins $-m_\mu$.  This can be obtained by writing

\begin{equation} \label{eq77}
\bar{h}_{\mu \nu}= h_{\mu \nu} -\tau_\mu m_\nu - \tau_\nu m_\mu
\end{equation}

since $\delta_G m_\mu =  e^a_\mu \lambda_a$.  Invariance under rotations are also granted since none of the terms on the right hand side changes under $\delta_J$.  The second invariant is obtained by using $\hat{e}^a_\mu$ to construct another metric:

\begin{equation} \label{eq78}
\hat{h}_{\mu \nu}= \delta_{ab} \hat{e}^a_\mu \hat{e}^b_\nu.
\end{equation}

The boost invariance of $\hat{e}^a_\mu$ guarantees the same for $\hat{h}_{\mu \nu}$ whereas the fact that $\delta_J$ of $\hat{e}^a_\mu$ and $\hat{e}^b_\nu$ cancels each other (much like $\delta_J$ of $e^a_\mu$ and $e^b_\nu$ cancels each other) guarantees its rotational invariance.  Being tensors of the same rank and mutually invariant, $\bar{h}_{\mu \nu}$ and $\hat{h}_{\mu \nu}$ must only differ by another invariant.  Indeed, 

\begin{align}
\hat{h}_{\mu \nu} &= \delta_{ab} \hat{e}^a_\mu \hat{e}^b_\nu \nonumber\\
&=\delta_{ab} (e^a_\mu -m^a\tau_\mu) (e^b_\nu -m^b\tau_\nu)\nonumber \\
&=\delta_{ab} (e^a_\mu e^b_\nu  -m^a\tau_\mu e^b_\nu  +e^a_\mu m^b\tau_\nu + m^a\tau_\mu m^b\tau_\nu) \nonumber\\
&=h_{\mu \nu}  - m_a  e^a_\nu \tau_\mu + m_a e^a_\mu \tau_\nu + m_a m^a  \tau_\mu \tau_\nu \nonumber\\
&=h_{\mu \nu}   -\tau_\mu m_\nu - \tau_\nu m_\mu  + 2 \tau_\mu \tau_\nu  \Phi \nonumber\\
&=\bar{h}_{\mu \nu} + 2 \tau_\mu \tau_\nu  \Phi  \label{eq79}.
\end{align}

where we have used Eq.\ref{eq75} from the fourth to the fifth line.  Thus, this other invariant rank two covariant tensor is given by $ 2 \tau_\mu \tau_\nu  \Phi$. 

\paragraph{}
We are finally set to construct Galilean metric connections that are guaranteed to be invariant.  Clearly, Levi-Civita connections where the $h_{\mu \nu}$ metric is replaced by $\bar{h}_{\mu \nu}$ and $v^\lambda$ by $\hat{v}^\lambda$ has to qualify for these.

\begin{align*}
\bar {\Gamma}_{{\mu}{\nu}}^\lambda &= -  \hat{v}^\lambda  \partial_\mu \tau_\nu  +  \frac{1}{2}  h^{\lambda \sigma} (\partial_\mu \bar{h}_{\nu \sigma} + \partial_\nu \bar{h}_{\mu \sigma}-\partial_\sigma \bar{h}_{\mu \nu}) \\
&= -  v^\lambda  \partial_\mu \tau_\nu  +  \frac{1}{2}  h^{\lambda \sigma} (\partial_\mu h_{\nu \sigma} + \partial_\nu h_{\mu \sigma}- \partial_\sigma h_{\mu \nu}) +\frac{1}{2} h^{\lambda \sigma} (\tau_\mu \bar{K}_{\sigma \nu} +\tau_\nu \bar{K}_{\sigma \mu} + \bar{L}_{\sigma \mu \nu})
\end{align*}

where $\bar{K}_{\mu \nu}$ and $\bar{L}_{\sigma \mu \nu}$ are the $K_{\mu \nu}$ and $L_{\sigma \mu \nu}$ corresponding to the additional terms in $\hat{v}^\lambda$ and $\bar{h}_{\mu \nu}$.  For $\bar {\Gamma}_{{\mu}{\nu}}^\lambda$, it is not difficult to see that 

\begin{align}
\bar{K}_{\mu \nu} &= 2\partial_{[\mu} m_{\nu]},  \label{eq80} \\ 
\bar{L}_{\sigma \mu \nu} &= 2m_\sigma \partial_{[\mu} \tau_{\nu]} - 2m_\mu \partial_{[\nu} \tau_{\sigma]} -2 m_\nu \partial_{[\mu} \tau_{\sigma]} , \label{eq81}
\end{align}

and that they flawlessly satisfy Eq.\ref{eq60} and Eq.\ref{eq61}.  Furthermore, in terms of $m_\mu$, the Newton-Coriolis and the spatial contortion is given by: (using Eq.\ref{eq52} and Eq.\ref{eq53})

\begin{align}
\bar{e}^{\text{ }a}_{[\nu} \omega_{\mu ] a} &= - v^\sigma m_\sigma \partial_{[\mu} \tau_{\nu]} +  \partial_{[\mu} m_{\nu]},  \label{eq82}\\
\bar{\kappa}_{\mu\nu\sigma}&= 2 m_\sigma \partial_{[\mu} \tau_{\nu]} -2 e^a_\mu e^\rho_a  m_\rho \partial_{[\nu} \tau_{\sigma]} - 2 e^a_\nu e^\rho_a   m_\rho \partial_{[\mu} \tau_{\sigma]}.  \label{eq83}
\end{align}

\paragraph{}
The same argument holds if we replace $h_{\mu \nu}$ with $\hat{h}_{\mu \nu}$ instead:

\begin{align*}
\hat {\Gamma}_{{\mu}{\nu}}^\lambda &= -  \hat{v}^\lambda  \partial_\mu \tau_\nu  +  \frac{1}{2}  h^{\lambda \sigma} (\partial_\mu \hat{h}_{\nu \sigma} + \partial_\nu \hat{h}_{\mu \sigma}-\partial_\sigma \hat{h}_{\mu \nu})  \\
&= -  v^\lambda  \partial_\mu \tau_\nu  +  \frac{1}{2}  h^{\lambda \sigma} (\partial_\mu h_{\nu \sigma} + \partial_\nu h_{\mu \sigma}-\partial_\sigma h_{\mu \nu}) +\frac{1}{2} h^{\lambda \sigma} (\tau_\mu \hat{K}_{\sigma \nu} +\tau_\nu \hat{K}_{\sigma \mu} + \hat{L}_{\sigma \mu \nu}) .
\end{align*}

In this case, the choice of $\hat{K}_{\mu \nu}$ and $\hat{L}_{\sigma \mu \nu}$ corresponds to:

\begin{align}
\hat{K}_{\mu \nu} &= \bar{K}_{\mu \nu} + 2 \Phi \text{ } \partial_{[\mu} \tau_{\nu]} + \tau_{[\nu} \partial_{\mu]} \Phi,   \label{eq84} \\
\hat{L}_{\sigma \mu \nu} &=\bar{L}_{\sigma \mu \nu}, \label{eq85}
\end{align}

and hence

\begin{align}
\hat{e}^{\text{ }a}_{[\nu} \omega_{\mu ] a} &= \bar{e}^{\text{ }a}_{[\nu} \omega_{\mu ] a} + \Phi \text{ } \partial_{[\mu} \tau_{\nu]} + \frac{1}{2} \tau_{[\nu} \partial_{\mu]} \Phi, \label{eq86}\\
\hat{\kappa}_{\mu\nu\sigma}&= \bar{\kappa}_{\mu\nu\sigma}. \label{eq87}
\end{align}

\paragraph{}
While $\bar {\Gamma}_{{\mu}{\nu}}^\lambda$ and $\hat {\Gamma}_{{\mu}{\nu}}^\lambda$ are the most popular choice solutions of Galilean metric connection, notice that a general metric like quantity $ H_{\mu \nu}=\bar{h}_{\mu \nu} + k\tau_\mu \tau_\nu  \Phi$ where $k$ is any real number can also make a legitimate solution.  The corresponding parameters (the equations \ref{eq84} to \ref{eq87}) are obtained by simply changing $\Phi \rightarrow \frac{k}{2}\Phi$.  The set of all values of  $H_{\mu \nu}$ generates a solution class for Galilean $\Gamma_{{\mu}{\nu}}^\lambda$; in fact, it has been shown that Newton-Cartan limit for standard metrics such as the Schwarzschild metric are covered within this class \cite{Lian2,Bleeken}.

\subsection{$R(H)$, $R(P)$ and $R(M)$ under the metric class $ H_{\mu \nu}=\bar{h}_{\mu \nu} + k\tau_\mu \tau_\nu  \Phi$ }
\label{sec3.6}

Under the $ H_{\mu \nu}$ metric class, it is easy to express the simpler of the curvature components: $R(H)$, $R(P)$ and $R(M)$ in terms of the vielbeins only.  For instance, from Eq. \ref{eq43},

\begin{align}
R^a_{\mu \nu}(P)&= 2 e^a_\lambda \Gamma^\lambda_{[\mu \nu]} \nonumber\\
&=-2\hat{v}^\lambda \partial_{[\mu} \tau_{\nu ]} e^a_\lambda  = -2 (v^\lambda -h^{\lambda \sigma} m_\sigma ) \partial_{[\mu} \tau_{\nu ]} e^a_\lambda  \nonumber\\
&=-2e^{\sigma a}m_\sigma \partial_{[\mu} \tau_{\nu ]}  = 2 m^a \partial_{[\mu} \tau_{\nu ]}. \label{eq88}
\end{align}

On the other hand, Eq. \ref{eq42} must consistently imply $R^a_{\mu \nu}(H)=2 \partial_{[\mu} \tau_{\nu ]}$:

\begin{align}
R^a_{\mu \nu}(H)&= 2 \tau_\lambda \Gamma^\lambda_{[\mu \nu]} \nonumber\\
&=-2\hat{v}^\lambda \partial_{[\mu} \tau_{\nu ]} \tau_\lambda  = -2 (v^\lambda -h^{\lambda \sigma} m_\sigma ) \partial_{[\mu} \tau_{\nu ]} \tau_\lambda  \nonumber\\
&=2 \partial_{[\mu} \tau_{\nu ]}. \label{eq89}
\end{align}

Finally, the Newton-Coriolis can be substituted into Eq.\ref{eq70} from Eq.\ref{eq82} and Eq.\ref{eq86} (with $\Phi \rightarrow \frac{k}{2}\Phi$) to express $R(M)$ in terms of the vielbeins.  

\begin{align} 
R_{\mu \nu}(M)&=2\partial_{[\mu}m_{\nu]}+ 2 e^a_{[\mu} \omega_{\nu ] a} \nonumber \\
&=2\partial_{[\mu}m_{\nu]} + L_{\sigma[\mu \nu ]} v^\sigma - K_{\mu \nu } \\
&=2\partial_{[\mu}m_{\nu]} + 2 v^\sigma m_\sigma \partial_{[\mu} \tau_{\nu]} - 2 \partial_{[\mu} m_{\nu]}  - k \Phi \text{ } \partial_{[\mu} \tau_{\nu]} - \frac{k}{2}\tau_{[\nu} \partial_{\mu]}  \Phi \label{eq90}
\end{align}

\subsection{Time foliations in Newton-Cartan theory}
\label{sec3.7}

\paragraph{}
At last, we are in position to discuss some rather tangible physics in this semi-classical geometric theory of gravity.  We shall mainly consider the implications of constraining the temporal torsion $\partial_{[\mu} \tau_{\nu]}$.  This new term is particularly interesting since it is left non-zero when we demand zero contortion, yet we can demand some constraints on it without violating the theory's Galilean invariance.  The Newton-Coriolis on the other hand is always non-zero and no constraints with seemingly obvious physical interpretation can be imposed on it; investigations on its physical significance might be a discussion topic for future works.  

\paragraph{}
Note that the objects discussed in this section such as "absolute time", "torsionless space/time" and "time slices" are intrinsic to observers with the corresponding frame $[e,\tau,-m]$.  Observers boosted relative to that frame may draw different conclusions since their idea of space and time vielbeins $e$ and $\tau$ are different.   

\paragraph{}
We shall discuss three level of constraints of $\partial_{[\mu} \tau_{\nu]}$.  The loosest constraint is of course no constraint at all i.e. $\partial_{[\mu} \tau_{\nu]}$ is completely arbitrary; literatures name this situation the torsional Newton-Cartan (TNC).  At the other extreme end, one of the strictest condition one might impose is $\partial_{[\mu} \tau_{\nu]}=0$ i.e. ${\rm d}\tau=0$.  This is called the torsionless Newton-Cartan geometry (TLNC); in this case and when the $H_{\mu \nu}$ solution class is concerned, spatial torsion vanishes as well.

\paragraph{}
We will however begin our discussion with the intermediate constraint called the twistless condition since it is the one richest with physical interpretations; these would be vital to fully understand TNC and TLNC.  This case is named the twistless torsional Newton-Cartan (TTNC) and the mathematical condition imposed on $\partial_{[\mu} \tau_{\nu]}$ is the Frobenius condition\cite{Frobenius}:

\begin{align}
\tau_{[\sigma} \partial_\mu \tau_{\nu]}&=0, \nonumber \\
\text{or equivalently,}\hspace{0.3cm} \tau \wedge {\rm d}\tau&=0. \label{eq92}
\end{align}

\paragraph{}
An entire literature could be written on the Frobenius condition and theory, see for example \cite{Frobenius1,Frobenius2}.  Here, we only stress its implications pertaining to our spacetime.  The Frobenius theorem states that Eq.\ref{eq92} implies (and in fact, is equivalent to) the existence of a differentiable functions $f,g: M \rightarrow \mathbb{R}$ such that the surfaces $f={\rm const}$ under smooth deformation by $g$ are the integral surfaces for the time slices form $\tau$ i.e.,

\begin{equation}
\tau \wedge {\rm d}\tau=0 \Longleftrightarrow \text{there exists $f,g$ such that }\hspace{0.2cm} \tau=g\hspace{0.1cm} {\rm d}f. \label{eq93}
\end{equation}

\paragraph{}
Note that the definition of an integral surface of $\tau$ is a submanifold in which at every point, $\tau$ has vanishing contractions with all the vectors on the tangent spaces.  In this case, they are the hypersurfaces of equal time slice i.e. constant $\tau$.  These hypersurface are however not generally guaranteed to form a nice foliation on the base manifold\footnote{At least not when the base manifold dimension is higher than 2.}.  The Frobenius condition according to Eq.\ref{eq93} implies that these integral surfaces are – up to a smooth deformation – described by $f={\rm const}$ surfaces for a continuous differentiable $f$.  Since the $f={\rm const}$ surfaces certainly form foliations, this condition forces the time slices to form a foliation as well.  In other words, the foliation of $\tau$ is obtained by smoothly deforming the foliation of $f$ by a functional factor – called the integrating factor – $g$; under such deformation, there is no way a foliated manifold becomes non-foliated.

\begin{figure}[h!]
	\centering
	\includegraphics[width=6.5cm]{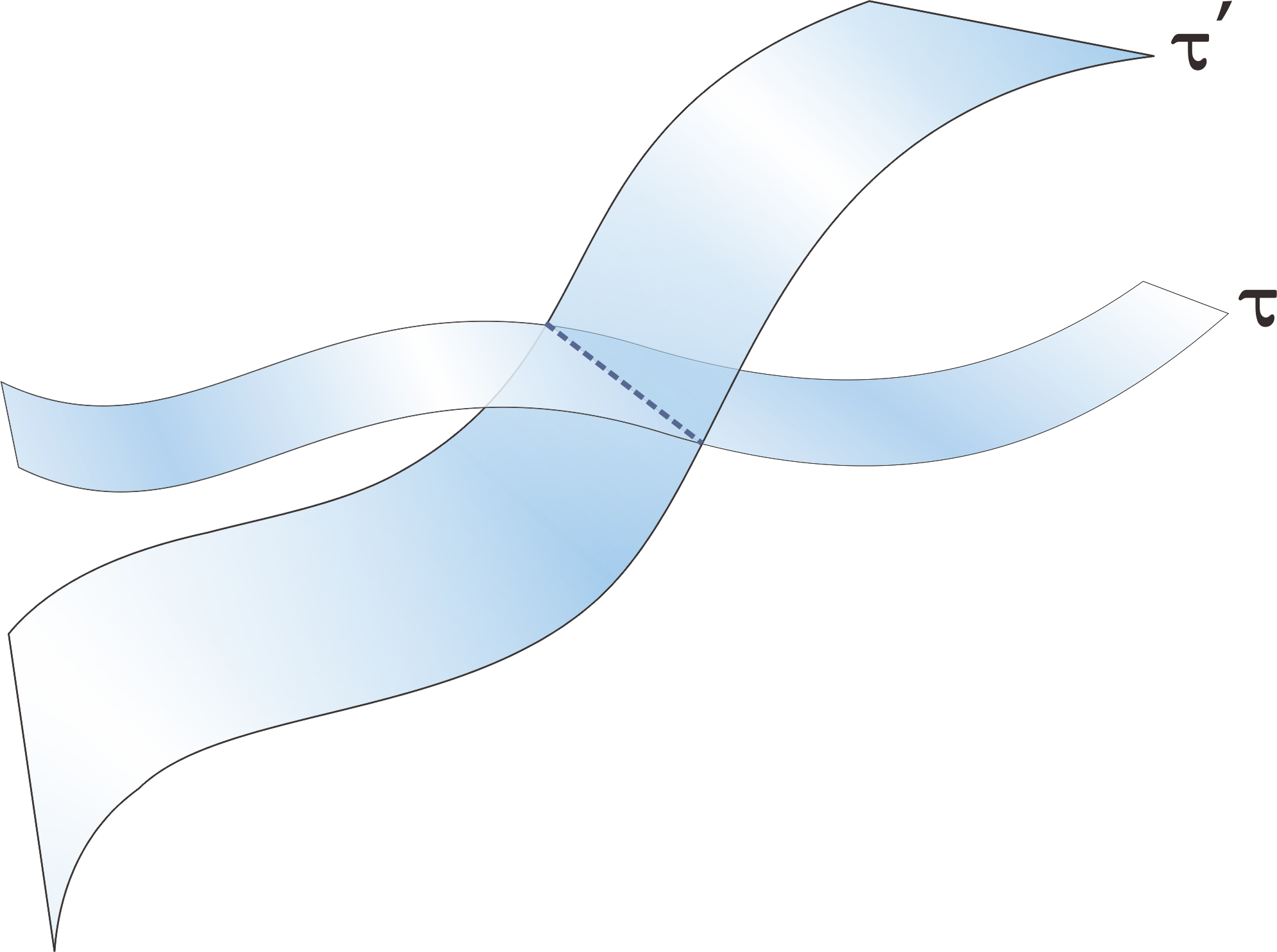}
	\caption{Illustration of $\tau$-hypersurfaces without the Frobenius condition e.g. in TNC.  There is a lack of causality as different observers may have different conclusion on whether $\tau$ or $\tau'$ is earlier}
	\label{fig3}
\end{figure}

\paragraph{}
Physically, the function $f$ is a clock that measures time smoothly and differs by a magnification factor $g$ from the clock of $\tau$.  It follows that situations such as in Fig.\ref{fig3} are not possible under the TTNC; an "earlier" hypersurface cannot intersect with a "later" hypersurface, preserving the notion of causality. TTNC thus might just describe real physical spacetimes.

\paragraph{}
It is not difficult to see that the right hand side of Eq.\ref{eq93} implies that one can write ${\rm d}\tau= a \wedge \tau$ for some form $a$.  Indeed, 

\begin{align}
&\tau=g \hspace{0.1cm}df \nonumber\\
\Rightarrow& \hspace{0.1cm} {\rm d}\tau=dg \wedge df \nonumber\\
\Rightarrow& \hspace{0.1cm} {\rm d}\tau=\frac{dg}{g} \wedge \tau \nonumber\\
\Rightarrow& \text{ there exists $a$, for example $a=\frac{dg}{g}$ such that ${\rm d}\tau= a \wedge \tau$} .\label{eq94}
\end{align}

\begin{figure}[h!]
	\centering
	\includegraphics[width=8.5cm]{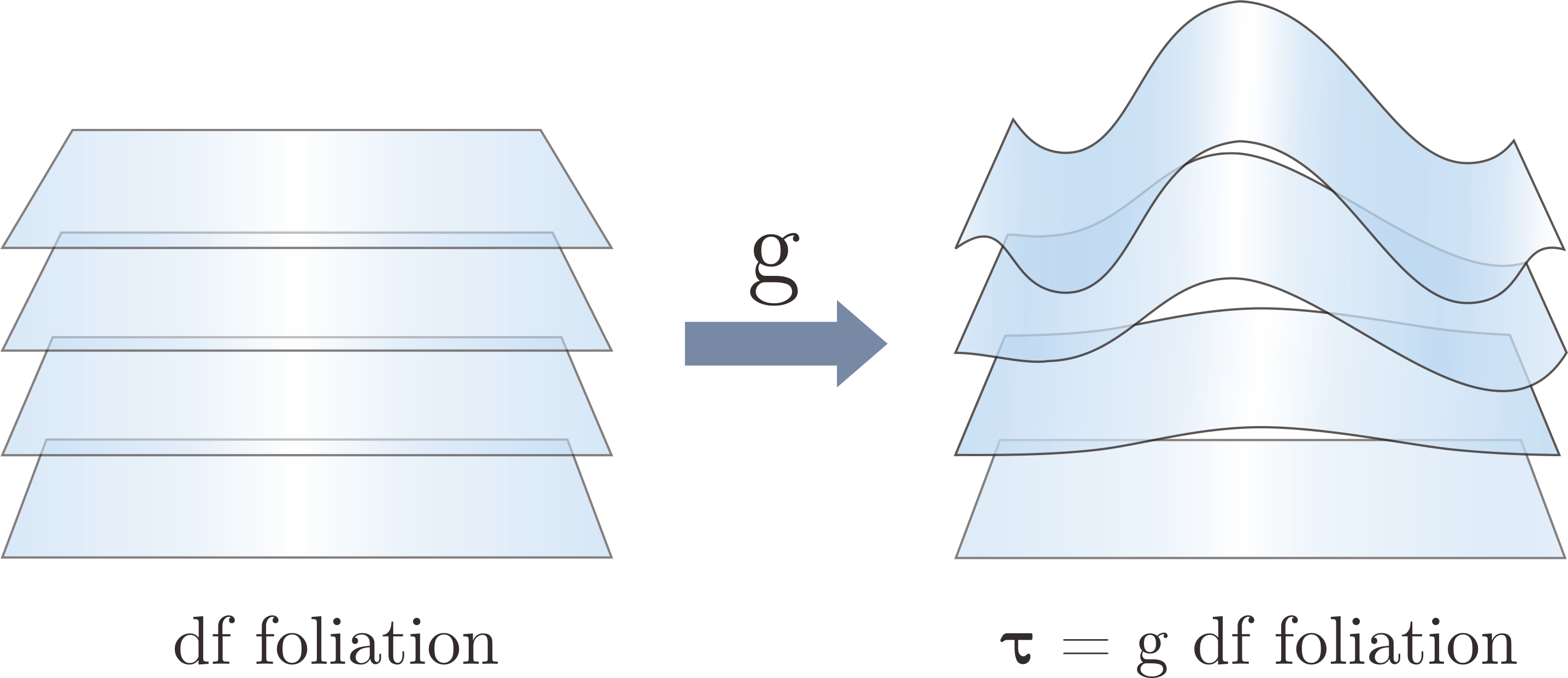}
	\caption{Illustration of $\tau$-hypersurfaces foliation as a smooth deformation from the foliation of $df$ by $g$.  Under such deformation, situations such as Fig.\ref{fig3} will not arise and the hypersurfaces remain stacked with a consistent order.}
	\label{fig4}
\end{figure}

\paragraph{}
The choice of $a$ is not unique, but the obvious choice $a=\frac{dg}{g}=d({\rm ln } \text{ }g)$ is easy to analyze.  In this case $a$ describes how fast and to which direction the magnification factor function $g$ changes.  Note that the torsion ${\rm d}\tau$ is non-zero only when $a$ has components orthogonal to $\tau$, i.e. when the magnification rate of change $dg$ tends to bend the $f$ surfaces to directions tangent to the original $\tau$ direction (see Fig.\ref{fig4}).  On the other hand, if we have for example $g=N$ where $N$ is a constant number everywhere, the time direction is constant\footnote{This corresponds to the situation where the $\tau$ clock is $N$ times faster than the $f$ clock.} and we have the torsionless case ${\rm d}\tau=0$.  The origin of temporal torsion thus goes back to the magnification factor $g$ – now parameterized by $a$. 

\paragraph{}
To reduce the arbitrariness of $a$, we can demand it to be completely orthogonal to $\tau$; this is obtainable by simply taking only the $\tau$-orthogonal component of $d({\rm ln } \text{ }g)$\footnote{$a \wedge \tau$ is unaffected if we subtract the $\tau$-parallel component of $a$.}.  This motivates the following ansatz for $a$:
 
\begin{align}
a_\mu&=(\pounds_{\hat{v}}\tau)_\mu \nonumber\\
&=(\partial_\xi \tau_\mu) \hat{v}^\xi + \tau_\xi (\partial_\mu \hat{v}^\xi ). \label{eq95}
\end{align}

\paragraph{}
Indeed, this definition directly implies its orthogonality with $\hat{v}$, the vector counterpart of $\tau$: 

\begin{align}
\hat{v}^\mu a_\mu&=  \hat{v}^\mu  (\partial_\xi \tau_\mu) \hat{v}^\xi +\hat{v}^\mu \tau_\xi (\partial_\mu \hat{v}^\xi ) \nonumber\\
&=\hat{v}^\mu  \partial_\xi ( \tau_\mu \hat{v}^\mu) \nonumber\\
&=0. \label{eq96} 
\end{align}

\begin{figure}[h!]
	\centering
	\includegraphics[width=6.5cm]{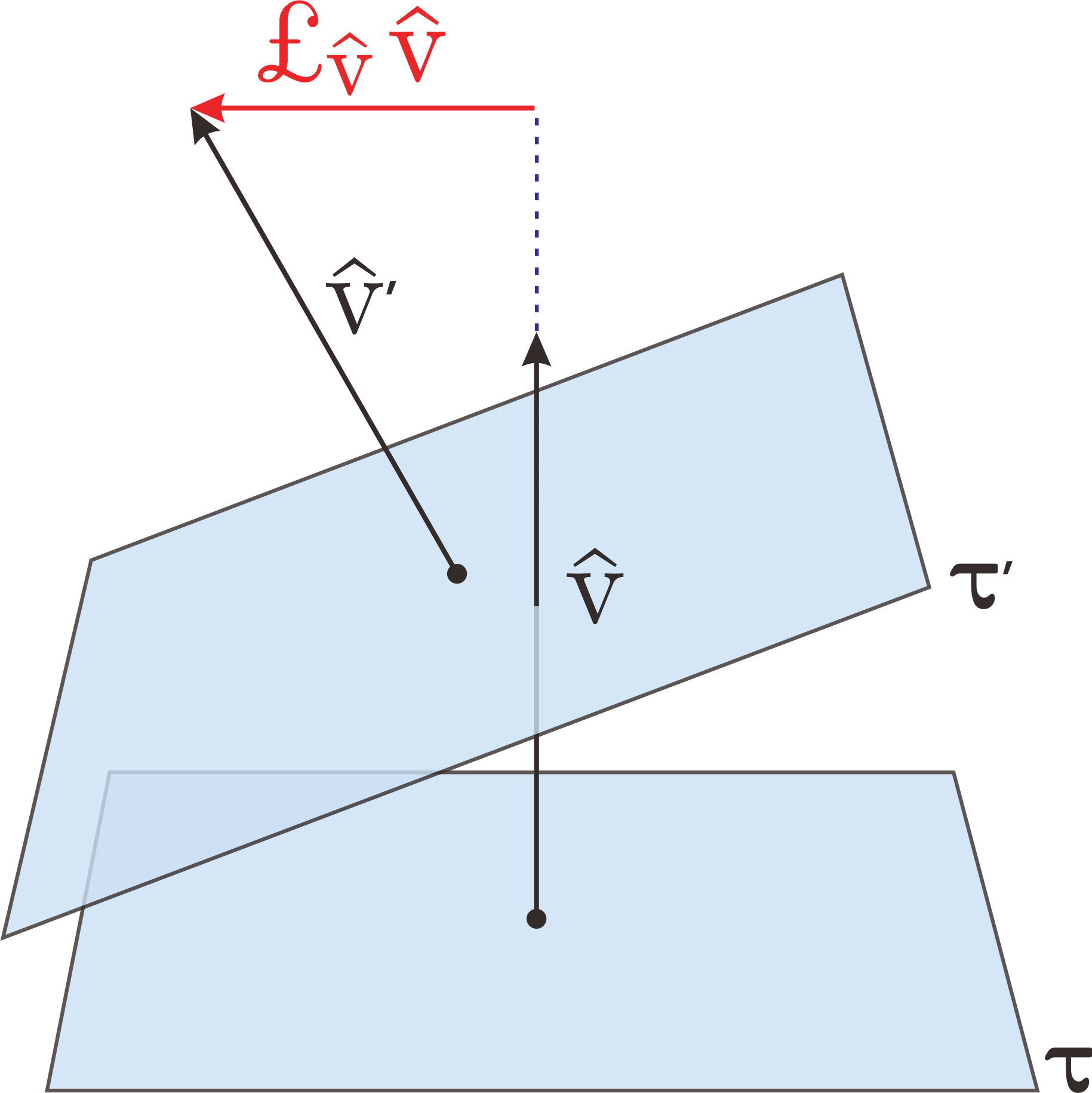}
	\caption{Illustration of the torsion vector $a=\pounds_{\hat{v}}\tau$.  In this figure, $a$'s vector counterpart, $\pounds_{\hat{v}}\hat{v}$ is perpendicular to $\hat{v}$.  Naturally, $a=\pounds_{\hat{v}}\tau$'s inner product with $\hat{v}$ is zero as well .}
	\label{fig5}
\end{figure}

\paragraph{}
Eq.\ref{eq95} is also intuitive; $a=\pounds_{\hat{v}}\tau$ measures the deviation of $\tau$ when one traverses from one time slice to another.  In the presence of temporal torsion, this situation is roughly depicted on Fig.\ref{fig5}.  In the figure, we see that $\pounds_{\hat{v}}\hat{v}$ which is the vector counterpart of $a$ represents a vector proportional to $\tau$'s tendency to change direction i.e. the strength of the torsion.  Most literatures however refers $a$ – a one-form – as the torsion vector instead of $\pounds_{\hat{v}}\hat{v}$.

Finally, it is straightforward to show that $a=\pounds_{\hat{v}}\tau$ does satisfy ${\rm d}\tau= a \wedge \tau$:

\begin{align}
a_{[\mu} \tau_{\nu]}&= (\partial_\xi \tau_{[\mu}) \tau_{\nu]}\hat{v}^\xi + \tau_\xi (\partial_{[\mu} \hat{v}^\xi ) \tau_{\nu]}  \nonumber\\
&=(\partial_{[\mu} \tau_{|\xi|}) \tau_{\nu]}\hat{v}^\xi - (\partial_{[\mu} \tau_{\nu]}) \tau_{\xi}\hat{v}^\xi +  \tau_\xi (\partial_{[\mu} \hat{v}^\xi ) \tau_{\nu]}\nonumber\\
&=(\partial_{[\mu} \tau_{|\xi|} \hat{v}^\xi ) \tau_{\nu]}\hat{v}^\xi +\partial_{[\mu} \tau_{\nu]} \nonumber\\
&=\partial_{[\mu} \tau_{\nu]}. \label{eq97}
\end{align}
 
Where on the second line, we have used the Frobenius condition $\tau_{[\xi} \partial_\mu \tau_{\nu]}=0  \Rightarrow (\partial_\xi \tau_\mu) \tau_\nu - (\partial_\xi \tau_\nu) \tau_\mu= (\partial_\mu \tau_\xi) \tau_\nu - (\partial_\nu \tau_\xi) \tau_\mu + (\partial_\nu \tau_\mu) \tau_\xi - (\partial_\mu \tau_\nu) \tau_\xi$.
 
\paragraph{}
Eq.\ref{eq97} can also be used to prove an interesting corollary: in TTNC, even when ${\rm d}\tau\neq0$, its spatial projection (projection onto the $\tau$ hypersurfaces, which represents the spatial surfaces) is zero.  Hence, in the case of zero spatial contortion, one observes completely torsion free submanifolds embedded in a globally torsional manifold.  This situation is not possible if the Frobenius condition is not satisfied. 

 \begin{align}
h^{\mu \rho} h^{\nu \sigma} \partial_{[\rho} \tau_{\sigma]} &= h^{\mu \rho} h^{\nu \sigma} a_{[\rho} \tau_{\sigma]} \hspace{0.2cm} \text{from Eq.\ref{eq96}}\nonumber\\
&=0 \hspace{0.2cm} \text{due to orthogonality of $h^{\mu \nu} $ and $\tau_\mu$.}
\end{align}

\paragraph{}
Lastly, using $a$, we can define a new quantity called the twist tensor as $\nabla_{[\mu} a_{\nu]}$.  Due to orthogonality of $a$ and $\hat{v}$, and that our torsion is purely temporal, this quantity also equals to $\partial_{[\mu} a_{\nu]}$:

 \begin{align}
\nabla_{[\mu} a_{\nu]}&=\partial_{[\mu} a_{\nu]} - 2 a_\lambda \Gamma^\lambda_{[\rho \sigma]}\nonumber\\
&=\partial_{[\mu} a_{\nu]} - 2 a_\lambda \hat{v}^\lambda  \partial_{[\rho} \tau_{\sigma]} \nonumber\\
&=\partial_{[\mu} a_{\nu]}.
\end{align}

\paragraph{}
Just like ${\rm d}\tau$, the twist tensor also has vanishing spatial projection in case of TTNC:

 \begin{align}
h^{\mu \rho} h^{\nu \sigma}\partial_{[\rho} a_{\sigma]} &= h^{\mu \rho} h^{\nu \sigma} \big( (\partial_{[\rho} \hat{v}^\lambda) (\partial_{|\lambda|} \tau_{\sigma]}) + \hat{v}^\lambda (\partial_{[\rho}  \partial_{|\lambda|} \tau_{\sigma]}) +   (\partial_{[\rho} \tau_{|\lambda|})(\partial_{\sigma]} \hat{v}^\lambda) +      \tau_\lambda \cancel {(\partial_{[\rho}  \partial_{\sigma]} \hat{v}^\lambda) } \big)\nonumber\\
&= h^{\mu \rho} h^{\nu \sigma} \big(  (\partial_\rho \hat{v}^\lambda ) (\partial_\lambda \tau_\sigma - \partial_\sigma \tau_\lambda) +(\partial_\sigma \hat{v}^\lambda ) (\partial_\rho \tau_\lambda - \partial_\lambda \tau_\rho) + \hat{v}^\lambda (\partial_\lambda \partial_\rho \tau_\sigma - \partial_\lambda \partial_\sigma \tau_\rho) \big) \nonumber\\
&= h^{\mu \rho} h^{\nu \sigma} \big(  (\partial_\rho \hat{v}^\lambda ) ( \cancel{a_\lambda \tau_\sigma }- a_\sigma \tau_\lambda) +(\partial_\sigma \hat{v}^\lambda ) (a_\rho \tau_\lambda - \cancel {a_\lambda \tau_\rho }) + \hat{v}^\lambda \partial_\lambda (a_\rho \tau_\sigma - a_\sigma \tau_\rho) \big) \nonumber\\
&=h^{\mu \rho} h^{\nu \sigma} \big( - (\partial_\rho \hat{v}^\lambda ) a_\sigma \tau_\lambda + (\partial_\sigma \hat{v}^\lambda ) a_\rho \tau_\lambda +\cancel{ \hat{v}^\lambda (\partial_\lambda a_\rho) \tau_\sigma } + \hat{v}^\lambda a_\rho (\partial_\lambda \tau_\sigma)  - \cancel{ \hat{v}^\lambda (\partial_\lambda a_\sigma) \tau_\rho }- \hat{v}^\lambda a_\sigma (\partial_\lambda \tau_\rho)   \big) \nonumber\\
&=h^{\mu \rho} h^{\nu \sigma} \big( - (\partial_\rho \hat{v}^\lambda ) a_\sigma \tau_\lambda - \hat{v}^\lambda a_\sigma (\partial_\lambda \tau_\rho) + (\partial_\sigma \hat{v}^\lambda ) a_\rho \tau_\lambda  + \hat{v}^\lambda a_\rho (\partial_\lambda \tau_\sigma) \big) \nonumber\\
&=h^{\mu \rho} h^{\nu \sigma} \big( - a_\sigma (\pounds_{\hat{v}} \tau)_\rho + a_\rho  (\pounds_{\hat{v}}  \tau)_\sigma \big) \nonumber\\
&=0.
\end{align}

Where the Eq.\ref{eq97} has been used in the third line. 

\paragraph{}
Going back to TLNC i.e. $\mathbf{{\rm d}\tau}=0$, all the above results and corollaries of TTNC are also valid since the stronger torsionless condition implies the weaker Frobenius condition.  TLNC however admits a characteristic not found in TTNC, namely the notion of absolute time.      

\begin{figure}[h!]
	\centering
	\includegraphics[width=6.5cm]{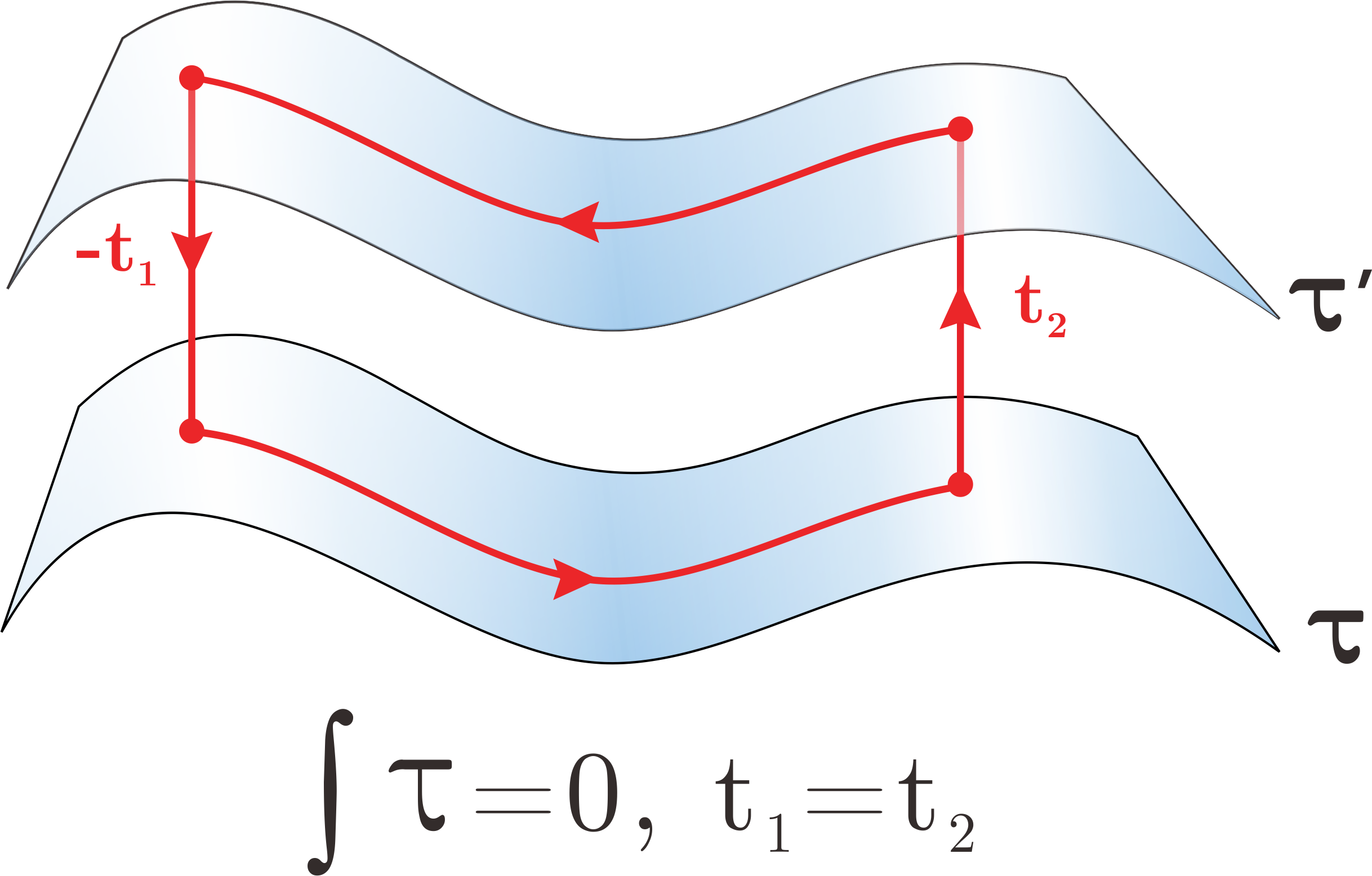} \hspace{1cm}
	\includegraphics[width=6.5cm]{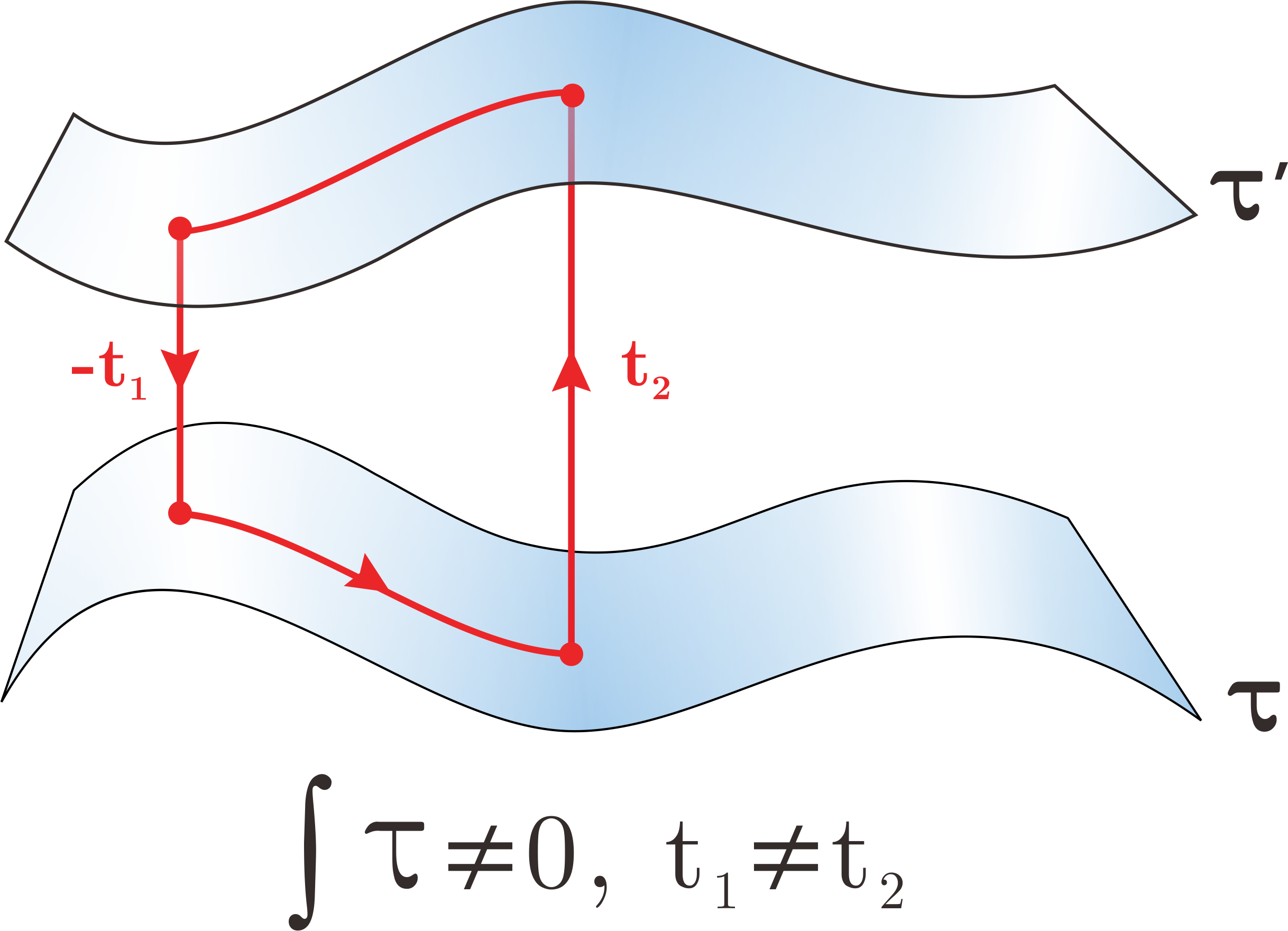}
	\caption{Contrast between TLNC $\tau$-hypersurfaces (left) and TTNC's (right).  Under TLNC, the integral along the closed curve is zero due to $\tau$'s exactness.}
	\label{fig6}
\end{figure}

\paragraph{}
	TLNC's $\mathbf{{\rm d}\tau}=0$ condition implies\footnote{Note that in our manifold, closedness implies exactness.} that $\tau=dt$ for some function $t$ which in turn implies that any time measurement between any two $\tau$ hypersurfaces are absolute.  Any closed curve such as in Fig.\ref{fig6} (left) has a zero loop integral: $\int \tau= \int dt=0$; this means $t_1=t_2$ is always satisfied.  Geometrically, one could imagine TLNC as $\tau$-hypersurfaces being stacked in a consistent order and in parallel.  In TTNC, time measure generally varies between these observers as shown in Fig.\ref{fig6} (right). 

\paragraph{}
On the other hand, both TTNC and TLNC admit the notion of absolute space\footnote{Given zero spatial contortion condition.}.  The "space" in this sense refers to the $\tau$-hypersurface which from our corollary are completely torsionless Riemannian submanifolds.  Thus, following the same argument as above, observers living in these surfaces will have no problem of space measurement ambiguity.\footnote{Once again, please note that by "observers", we mean observers with the corresponding $[e,\tau,m]$ frame.  Boosted observers – in TTNC in particular– might find torsion in their own definition of spatial surfaces and have no notion of absolute space or time.}  

\paragraph{}
TNC – the completely arbitrary condition – might not have many physical applications since it is difficult to discuss spacelike or timelike objects in this case; the $\tau$ hypersurfaces that are the supposedly spatial surfaces might intersect one another and could have arbitrary, meaningless structure.  Such spacetime is unlikely to be physical: on top of the path dependence of both space and time measurements, there is an obvious problem of the lack of causality.

\newpage

%%%%%%%%%%%%%%%%%%%%%%%%%%%%%%%%%%%%%%%%%%%%%%%%%%%%%%%%%%%%%%%%%%%%%%%%%%%%%%%%%%%%%%%%%%%%%%%%%%%%%%%%%%%%%%%%%%%%%%
%%%%%%%%%%%%%%%%%%%%%%%%%%%%%%%%%%%%%%%%%%%%%%%%%%%%%%%%%%%%%%%%%%%%%%%%%%%%%%%%%%%%%%%%%%%%%%%%%%%%%%%%%%%%%%%%%%%%%%
\noindent\fbox{%
    \parbox{\textwidth}{\section*{\Large{Bargmann Vielbein Formalism}}

Another set of rules, complementing the Galilean's vielbein formalism relating the modified but invariant vielbeins $\hat{e}^a_\mu$ and $\hat{v}^\mu$ defined on Eq.\ref{eq71} and Eq.\ref{eq72} can be easily derived 
   
 \begin{align*} 
 &\hat{v}^\mu \tau_\mu=-1        &e^\mu_a \hat{e}^b_\mu=\delta^a_b \\
 &\hat{e}^\mu_a \tau_\mu=0         &\hat{v}^\mu \hat{e}^a_\mu=0,
\end{align*}

with the completeness relation

\begin{equation*}
e^\mu_a \hat{e}^a_\nu=\delta^\mu_\nu+\hat{v}^\mu \tau_\nu.
\end{equation*}    

Albeit being invariants and satisfying the vielbein formalism, the new quantities are not real vielbeins in the sense that they do not satisfy the vielbein postulate.  Indeed, due to the $m_\mu$ and $m^a$, their general covariant derivaties read  

\begin{align*}
D_\mu \hat{v}^\nu &= D_\mu (v^\mu -h^{\nu \sigma} m_\sigma)  = D_\mu (-h^{\nu \sigma} m_\sigma)  = -e^\nu_a  D_\mu(e^\sigma_b \delta^{ab} m_\nu) = -e^\nu_a  D_\mu m^a,  \\
D_\mu \hat{e}^a_\nu &= D_\mu (e^a_\nu -m^a\tau_\nu) =-\tau_\nu  D_\mu m^a.
\end{align*}

Clearly, we do not demand the vielbein postulate for $m^a$ as we want our affine connection $\Gamma^\lambda_{\mu \nu}$ to be defined by the space and time curvatures, not $-m$'s.  Hence, the right hand side of the above equations are non-zero in general.  
\\
\\
Due to the same reason as for the Galilean vielbeins, the new vielbeins should certainly not be used to transform vielbein indexed quantities into coordinate indexed ones.  Likewise, their induced metrics $\hat{g}_{\mu\nu}=\delta_{ab} \hat{e}^a_\mu \hat{e}^b_\nu$ are certainly not index raising and lowering operators. 

    }
}

%%%%%%%%%%%%%%%%%%%%%%%%%%%%%%%%%%%%%%%%%%%%%%%%%%%%%%%%%%%%%%%%%%%%%%%%%%%%%%%%%%%%%%%%%%%%%%%%%%%%%%%%%%%%%%%%%%%%%%
%%%%%%%%%%%%%%%%%%%%%%%%%%%%%%%%%%%%%%%%%%%%%%%%%%%%%%%%%%%%%%%%%%%%%%%%%%%%%%%%%%%%%%%%%%%%%%%%%%%%%%%%%%%%%%%%%%%%%%

%%%%%%%%%%%%%%%%%%%%%%%%%%%%%%%%%%%%%%%%%%%%%%%%%%%%%%%%%%%%
	
\newpage

\section*{Appendix}
\addcontentsline{toc}{section}{\protect\numberline{} Appendix}%

\paragraph{Proof of Eq.\ref{eq21} and Eq.\ref{eq22}}$\text{ }$
\\
\\
We are to proof $R_{\mu \nu \sigma}^\lambda=-e^\lambda_b e_{\sigma a}  (2\partial_{[\mu}\omega^{\text{ } ab}_{\nu]}-2\omega_{[\mu}^{\text{ } \text{ } ca}\omega_{\nu]c} ^{\text{ } b})$ given $R_{\mu \nu \sigma}^\lambda=-2\partial_{[\mu} \Gamma^\lambda_{\nu] \sigma} -2\Gamma^\lambda_{[\mu |\rho|} \Gamma^\rho_{\nu] \sigma}$ and $\Gamma^\lambda_{\mu \nu}=e^\lambda_a (\partial_\mu e^a_\nu -\omega_{\mu \text{ } b}^{\text{ } \text{ } a} e^b_\nu)$.

First, note that 
\begin{align*}
-\partial_\mu \Gamma^\lambda_{\nu \sigma}&= -[(\partial_\mu e^\lambda_a ) (\partial_\nu e^a_\sigma -\omega_{\nu \text{ } b}^{\text{ } \text{ } a}  e^b_\sigma) + e^\lambda_a (\partial_\mu \partial_\nu e^a_\sigma - e^b_\sigma \partial_\mu \omega_{\nu \text{ } b}^{\text{ } \text{ } a}  - \omega_{\nu \text{ } b}^{\text{ } \text{ } a}  \partial_\mu e^b_\sigma)] \\
&= -[e^\lambda_a (\partial_\mu \partial_\nu e^a_\sigma)- (\partial_\mu e^\lambda_a ) (\partial_\nu e^a_\sigma ) - (\partial_\mu e^\lambda_a ) \omega_{\nu \text{ } b}^{\text{ } \text{ } a}  e^b_\sigma - (\partial_\mu \omega_{\nu \text{ } b}^{\text{ } \text{ } a}) e^\lambda_a e^b_\sigma  - (\partial_\mu e^b_\sigma) e^\lambda_a \omega_{\nu \text{ } b}^{\text{ } \text{ } a}]  \\
\text{hence,}\hspace{0.2cm} -\partial_{[\mu} \Gamma^\lambda_{\nu] \sigma} &= -[\partial_{[\mu} e^\lambda_a \partial_{\nu]} e^a_\sigma - \partial_{[\mu} e^\lambda_a \omega_{\nu] \text{ } b}^{\text{ } \text{ } a} e^b_\sigma - e^\lambda_a \partial_{[\mu} \omega_{\nu] \text{ } b}^{\text{ } \text{ } a} e^b_\sigma -e^\lambda_a \partial_{[\mu} e^b_{|\sigma|} \omega_{\nu] \text{ } b}^{\text{ } \text{ } a}].
\end{align*}

On the other hand, 
\begin{align*}
-\Gamma^\lambda_{\mu \rho} \Gamma^\rho_{\nu \sigma}&= -[e^\lambda_a  (\partial_\mu e^a_\rho -\omega_{\mu \text{ } b}^{\text{ } \text{ } a}  e^b_\rho) ][e^\rho_c  (\partial_\nu e^c_\sigma -\omega_{\nu \text{ } d}^{\text{ } \text{ } c}  e^d_\sigma) ] \\
&= -[ e^\lambda_a  e^\rho_c (\partial_\mu e^a_\rho)   (\partial_\nu e^c_\sigma) -  e^\lambda_a  e^\rho_c e^d_\sigma \omega_{\nu \text{ } d}^{\text{ } \text{ } c} (\partial_\mu e^a_\rho)  - e^\lambda_a  e^\rho_c  e^b_\rho \omega_{\mu \text{ } b}^{\text{ } \text{ } a} (\partial_\nu e^c_\sigma)+ e^\lambda_a  e^\rho_c \omega_{\mu \text{ } b}^{\text{ } \text{ } a}\omega_{\nu \text{ } d}^{\text{ } \text{ } c} e^b_\rho e^d_\sigma]\\
&= -[- e^a_\rho  e^\rho_c (\partial_\mu e^\lambda_a)   (\partial_\nu e^c_\sigma) +  e^a_\rho  e^\rho_c e^d_\sigma \omega_{\nu \text{ } d}^{\text{ } \text{ } c} (\partial_\mu e^\lambda_a)  - e^\lambda_a  e^\rho_c  e^b_\rho \omega_{\mu \text{ } b}^{\text{ } \text{ } a} (\partial_\nu e^c_\sigma)+ e^\lambda_a  e^\rho_c \omega_{\mu \text{ } b}^{\text{ } \text{ } a}\omega_{\nu \text{ } d}^{\text{ } \text{ } c} e^b_\rho e^d_\sigma] \\
&= -[(\partial_\mu e^\lambda_a) (\partial_\nu e^a_\sigma)  +  e^b_\sigma  \omega_{\mu \text{ } b}^{\text{ } \text{ } a} (\partial_\mu e^\lambda_a) -e^\lambda_a \omega_{\mu \text{ } b}^{\text{ } \text{ } a} (\partial_\mu e^b_\sigma) + e^\lambda_a e^d_\sigma  \omega_{\mu \text{ } b}^{\text{ } \text{ } a}  \omega_{\nu \text{ } d}^{\text{ } \text{ } b}] \\
\text{hence,}\hspace{0.2cm} -\Gamma^\lambda_{[\mu |\rho|} \Gamma^\rho_{\nu] \sigma}&= -[\partial_{[\mu} e^\lambda_a \partial_{\nu]} e^a_\sigma +   e^b_\sigma \partial_{[\mu} e^\lambda_a \omega_{\nu] \text{ } b}^{\text{ } \text{ } a}  - e^\lambda_a \partial_{[\mu} e^b_{|\sigma|} \omega_{\nu] \text{ } b}^{\text{ } \text{ } a} + e^\lambda_a  \omega_{[\mu \text{ } b}^{\text{ } \text{ } a}\omega_{\nu] \text{ } d}^{\text{ } \text{ } b}]
\end{align*}
 
From the second to the third line above, we have used $e^\lambda_a (\partial_\mu e^a_\rho)= -e^a_\rho (\partial_\mu e^\lambda_a)$ since $\partial_\mu (e^\lambda_a e^a_\rho)=\partial_\mu (\delta^\lambda_\rho)=0$ in this relativistic vielbein system.  Finally, 
\begin{align*}
R_{\mu \nu \sigma}^\lambda=-2\partial_{[\mu} \Gamma^\lambda_{\nu] \sigma} -2\Gamma^\lambda_{[\mu |\rho|} \Gamma^\rho_{\nu] \sigma} &=  2e^\lambda_a \partial_{[\mu} \omega_{\nu] \text{ } b}^{\text{ } \text{ } a} e^b_\sigma+  2e^\lambda_a  \omega_{[\mu \text{ } b}^{\text{ } \text{ } a}\omega_{\nu] \text{ } d}^{\text{ } \text{ } b}\\
 &=  -e^\lambda_b e_{\sigma a}  (2\partial_{[\mu}\omega^{\text{ } ab}_{\nu]}-2\omega_{[\mu}^{\text{ } \text{ } ca}\omega_{\nu]c} ^{\text{ } b})\qed
\end{align*}

\paragraph{Proof of Eq.\ref{eq44} and Eq.\ref{eq45}}$\text{ }$
We need to proof $R_{{\mu}{\nu}{\sigma}}^\lambda=e^\lambda_a \tau_\sigma (2\partial_{[\mu}\omega^a_{\nu]} -2\omega_{[\mu}^{\text{ } \text{ } ab}\omega_{\nu]b}) - e_{\sigma a} e^\lambda_b (2\partial_{[\mu}\omega^{\text{ } ab}_{\nu]}-2\omega_{[\mu}^{\text{ } \text{ } ca}\omega_{\nu]c} ^{\text{ } b})$.  It is as straightforward as the previous proof, yet more terms arise due to the separated temporal quantities.  Furthermore, we have to use the completeness relation $e^\lambda_a e^a_\rho=\delta^\lambda_\rho + v^\lambda \tau_\rho$ instead of contracting directly to identity.  Lastly, $\Gamma^\lambda_{\mu \nu}$ is now updated to $-v^\lambda \partial_\mu \tau_\mu +  e^\lambda_a (\partial_\mu e^a_\nu -\omega^a_\mu \tau_\nu - \omega_{\mu \text{ } b}^{\text{ } \text{ } a} e^b_\nu)$

For simplicity, we denote quantities we have computed for the relativistic vielbeins with subscript "GR" e.g.,

\begin{align*}
(-\partial_\mu \Gamma^\lambda_{\nu \sigma})_{\rm GR}&= -[(\partial_\mu e^\lambda_a ) (\partial_\nu e^a_\sigma -\omega_{\nu \text{ } b}^{\text{ } \text{ } a}  e^b_\sigma) + e^\lambda_a (\partial_\mu \partial_\nu e^a_\sigma - e^b_\sigma \partial_\mu \omega_{\nu \text{ } b}^{\text{ } \text{ } a}  - \omega_{\nu \text{ } b}^{\text{ } \text{ } a}  \partial_\mu e^b_\sigma)] \\
\\
(-\Gamma^\lambda_{\mu \rho} \Gamma^\rho_{\nu \sigma})_{\rm GR} &= -[(\partial_\mu e^\lambda_a) (\partial_\nu e^a_\sigma)  +  e^b_\sigma  \omega_{\mu \text{ } b}^{\text{ } \text{ } a} (\partial_\mu e^\lambda_a) -e^\lambda_a \omega_{\mu \text{ } b}^{\text{ } \text{ } a} (\partial_\mu e^b_\sigma) + e^\lambda_a e^d_\sigma  \omega_{\mu \text{ } b}^{\text{ } \text{ } a}  \omega_{\nu \text{ } d}^{\text{ } \text{ } b}]
\end{align*}

Again, we begin by computing $-\partial_\mu \Gamma^\lambda_{\nu \sigma}$.

\begin{align*}
-\partial_\mu \Gamma^\lambda_{\nu \sigma}&=(-\partial_\mu \Gamma^\lambda_{\nu \sigma})_{\rm GR} - \partial_\mu (-v^\lambda \partial_\nu \tau_\sigma - e^\lambda_a \omega^a_\nu \tau_\sigma) \\
&=(-\partial_\mu \Gamma^\lambda_{\nu \sigma})_{\rm GR} + (\partial_\mu v^\lambda) (\partial_\nu \tau_\sigma) + v^\lambda (\partial_\mu \partial_\nu \tau_\sigma) + (\partial_\mu e^\lambda_a)  \omega^a_\nu  \tau_\sigma + (\partial_\mu  \omega^a_\nu)  e^\lambda_a \tau_\sigma + (\partial_\mu \tau_\sigma )  e^\lambda_a \omega^a_\nu \\
\text{hence,} \hspace{0.2cm} -\partial_{[\mu} \Gamma^\lambda_{\nu] \sigma} &=(-\partial_{[\mu} \Gamma^\lambda_{\nu] \sigma})_{\rm GR} +  \partial_{[\mu} v^\lambda \partial_{\nu]} \tau_\sigma +   \partial_{[\mu} e^\lambda_a  \omega^a_{\nu]}  \tau_\sigma +  \partial_{[\mu}  \omega^a_{\nu]}  e^\lambda_a \tau_\sigma + \partial_{[\mu} \tau_{|\sigma|}  \omega^a_{\nu]}  e^\lambda_a
\end{align*}

The $-\Gamma^\lambda_{\mu \rho} \Gamma^\rho_{\nu \sigma}$ is not so simple.  We shall attempt to simplify the following:  

\begin{align*}
\Gamma^\lambda_{\mu \rho} \Gamma^\rho_{\nu \sigma}&= [(-v^\lambda \partial_\mu \tau_\rho -e^\lambda_a \omega^a_\mu \tau_\rho)+ e^\lambda_a(\partial_\mu e^a_\rho - \omega_{\mu \text{ } b}^{\text{ } \text{ } a}e^b_\rho)] [(-v^\rho \partial_\nu \tau_\sigma -e^\rho_a \omega^a_\nu \tau_\sigma)+ e^\rho_a(\partial_\nu e^a_\sigma - \omega_{\nu \text{ } b}^{\text{ } \text{ } a}e^b_\sigma)]
\end{align*}

To do this, we compute the products term by term:
\begin{align*}
(-v^\lambda \partial_\mu \tau_\rho -e^\lambda_a \omega^a_\mu \tau_\rho)&(-v^\rho \partial_\nu \tau_\sigma -e^\rho_a \omega^a_\nu \tau_\sigma)\\
&=v^\lambda v^\rho (\partial_\mu \tau_\rho) (\partial_\nu \tau_\sigma) + v^\lambda e^\rho_a \omega^a_\nu \tau_\sigma  (\partial_\mu \tau_\rho) - e^\lambda_a \omega^a_\mu (\partial_\nu \tau_\sigma)+ \cancel{e^\lambda_a e^\rho_c \omega^a_\mu \omega^c_\nu \tau_\rho \tau_\sigma}\\
&=v^\lambda v^\rho (\partial_\mu \tau_\rho) (\partial_\nu \tau_\sigma) + v^\lambda e^\rho_a \omega^a_\nu \tau_\sigma  (\partial_\mu \tau_\rho) - e^\lambda_a \omega^a_\mu (\partial_\nu \tau_\sigma)\\
&=\big(-v^\rho \tau^\rho (\partial_\mu v_\lambda) (\partial_\nu \tau_\sigma) + v^\rho (\partial_\mu \tau_\rho v^\lambda) (\partial_\nu \tau_\sigma) \big) + v^\lambda e^\rho_a \omega^a_\nu \tau_\sigma  (\partial_\mu \tau_\rho) - e^\lambda_a \omega^a_\mu (\partial_\nu \tau_\sigma)\\
&= (\partial_\mu v_\lambda) (\partial_\nu \tau_\sigma) + v^\rho (\partial_\mu e^a_\rho e^\lambda_a) (\partial_\nu \tau_\sigma) + v^\lambda e^\rho_a \omega^a_\nu \tau_\sigma  (\partial_\mu \tau_\rho) - e^\lambda_a \omega^a_\mu (\partial_\nu \tau_\sigma)\\
&= (\partial_\mu v_\lambda) (\partial_\nu \tau_\sigma) + v^\rho e^\lambda_a (\partial_\mu e^a_\rho) (\partial_\nu \tau_\sigma) + v^\lambda e^\rho_a \omega^a_\nu \tau_\sigma  (\partial_\mu \tau_\rho) - e^\lambda_a \omega^a_\mu (\partial_\nu \tau_\sigma)\\
\\
e^\rho_a(-v^\lambda \partial_\mu \tau_\rho -e^\lambda_a \omega^a_\mu & \tau_\rho)(\partial_\nu e^a_\sigma - \omega_{\nu \text{ } b}^{\text{ } \text{ } a}e^b_\sigma)\\
&=-e^\rho_a v^\lambda (\partial_\mu \tau_\rho) (\partial_\nu e^a_\sigma - \omega_{\nu \text{ } b}^{\text{ } \text{ } a}e^b_\sigma)\\
&=-v^\lambda e^\rho_a (\partial_\mu \tau_\rho) (\partial_\nu e^a_\sigma)+ v^\lambda e^\rho_a  (\partial_\mu \tau_\rho) \omega_{\nu \text{ } b}^{\text{ } \text{ } a}e^b_\sigma\\
&=-e^\rho_a (\partial_\mu v^\lambda \tau_\rho)  (\partial_\nu e^a_\sigma) + e^\rho_a (\partial_\mu v^\lambda \tau_\rho) \omega_{\nu \text{ } b}^{\text{ } \text{ } a}  e^b_\sigma \\
\\
e^\lambda_a(\partial_\mu e^a_\rho - \omega_{\mu \text{ } b}^{\text{ } \text{ } a}e^b_\rho)(-&v^\rho  \partial_\nu \tau_\sigma -e^\rho_c \omega^c_\nu \tau_\sigma)\\
&=e^\lambda_a ( -v^\rho (\partial_\mu e^a_\rho)(\partial_\nu \tau_\sigma) - (\partial_\mu e^a_\rho) e^\rho_b \omega^b_\nu \tau_\sigma  +\cancel{ \omega_{\mu \text{ } b}^{\text{ } \text{ } a}e^b_\rho v^\rho (\partial_\nu \tau_\sigma)} + \omega_{\nu \text{ } b}^{\text{ } \text{ } a} \omega^b_\nu \tau_\sigma) \\
&=  -v^\rho e^\lambda_a (\partial_\mu e^a_\rho)(\partial_\nu \tau_\sigma) - (\partial_\mu e^a_\rho) e^\lambda_a e^\rho_b \omega^b_\nu \tau_\sigma + e^\lambda_a  \omega_{\nu \text{ } b}^{\text{ } \text{ } a} \omega^b_\nu \tau_\sigma \\
\\
e^\lambda_a e^\rho_c  (\partial_\mu e^a_\rho - \omega_{\mu \text{ } b}^{\text{ } \text{ } a}e^b_\rho)& (\partial_\nu e^c_\sigma - \omega_{\nu \text{ } d}^{\text{ } \text{ } c}e^d_\sigma) \\
&=(-\Gamma^\lambda_{\mu \rho} \Gamma^\rho_{\nu \sigma})_{\rm GR} + e^\lambda_a (\partial_\mu v^\lambda \tau_\rho) (\partial_\nu e^a_\sigma) -e^\lambda_a e^b_\sigma \omega_{\nu \text{ } b}^{\text{ } \text{ } a} (\partial_\mu v^\lambda \tau_\rho)
\end{align*}
Note that the fourth product is exactly the $-\Gamma^\lambda_{\mu \rho} \Gamma^\rho_{\nu \sigma}$ in GR's case, but now we cannot use the previous argument that $e^\lambda_a (\partial_\mu e^a_\rho)= -e^a_\rho (\partial_\mu e^\lambda_a)$.  In the present formalism, we have the "correction" $e^\lambda_a (\partial_\mu e^a_\rho)= -e^a_\rho (\partial_\mu e^\lambda_a) + (\partial_\mu v^\lambda \tau_\rho)$, which produces the two additional terms.\\

Hence, 
\begin{align*}
\Gamma^\lambda_{\mu \rho} \Gamma^\rho_{\nu \sigma}&=(\Gamma^\lambda_{\mu \rho} \Gamma^\rho_{\nu \sigma})_{\rm GR} +  (\partial_\mu v_\lambda) (\partial_\nu \tau_\sigma) +  v^\lambda e^\rho_a \omega^a_\nu \tau_\sigma  (\partial_\mu \tau_\rho) -  e^\lambda_a \omega^a_\mu (\partial_\nu \tau_\sigma) - (\partial_\mu e^a_\rho) e^\lambda_a e^\rho_b \omega^b_\nu \tau_\sigma + e^\lambda_a  \omega_{\mu \text{ } b}^{\text{ } \text{ } a} \omega^b_\nu \tau_\sigma.
\end{align*}

However, 
\begin{align*}
 v^\lambda e^\rho_a \omega^a_\nu \tau_\sigma  (\partial_\mu \tau_\rho) - (\partial_\mu e^a_\rho) e^\lambda_a e^\rho_b \omega^b_\nu \tau_\sigma &=  e^\rho_a \omega^a_\nu \tau_\sigma  (\partial_\mu v^\lambda \tau_\rho) - (\partial_\mu e^b_\rho) e^\lambda_b e^\rho_a \omega^a_\nu \\
&=  e^\rho_a \omega^a_\nu \tau_\sigma  (\partial_\mu v^\lambda \tau_\rho) - e^\rho_a \omega^a_\nu  \tau_\sigma (\partial_\mu e^b_\rho e^\lambda_b) + e^\rho_a  e^b_\rho \omega^a_\nu  \tau_\sigma (\partial_\mu e^\lambda_b)  \\ 
&= \omega^a_\nu  \tau_\sigma (\partial_\mu e^\lambda_a)  
\end{align*}

Substituting back, 
\begin{align*}
-\Gamma^\lambda_{[\mu \rho} \Gamma^\rho_{\nu] \sigma}&=(-\Gamma^\lambda_{[\mu \rho} \Gamma^\rho_{\nu] \sigma})_{\rm GR} -  (\partial_{[\mu} v_{|\lambda|}) (\partial_{\nu]} \tau_\sigma) -  (\partial_{[\mu} \tau_{|\sigma|})  \omega^a_{\nu]} e^\lambda_a   - \omega_{[\mu \text{ } b}^{\text{ } \text{ } a} \omega^b_{\nu]} \tau_\sigma  e^\lambda_a-  \tau_\sigma (\partial_{[\mu} e^\lambda_a) \omega^a_{\nu]}.
\end{align*}

Combining with $-\partial_{[\mu} \Gamma^\lambda_{\nu] \sigma}$ , we finally have
\begin{align*}
R_{\mu \nu \sigma}^\lambda&=-2\partial_{[\mu} \Gamma^\lambda_{\nu] \sigma} -2\Gamma^\lambda_{[\mu |\rho|} \Gamma^\rho_{\nu] \sigma} \\
&=(-2\partial_{[\mu} \Gamma^\lambda_{\nu] \sigma})_{\rm GR} + (-2\Gamma^\lambda_{\mu \rho} \Gamma^\rho_{\nu \sigma})_{\rm GR} + 2 e^\lambda_a \tau_\sigma \partial_{[\mu}  \omega^a_{\nu]}  - 2 e^\lambda_a \tau_\sigma \omega_{[\mu \text{ } b}^{\text{ } \text{ } a} \omega^b_{\nu]}\\
&=(R_{\mu \nu \sigma}^\lambda)_{\rm GR} + e^\lambda_a \tau_\sigma (2 \partial_{[\mu}  \omega^a_{\nu]}  - 2  \omega_{[\mu \text{ } b}^{\text{ } \text{ } a} \omega^b_{\nu]})\\
&= -e^\lambda_b e_{\sigma a}  (2\partial_{[\mu}\omega^{\text{ } ab}_{\nu]}-2\omega_{[\mu}^{\text{ } \text{ } ca}\omega_{\nu]c} ^{\text{ } b}) + e^\lambda_a \tau_\sigma (2 \partial_{[\mu}  \omega^a_{\nu]}  - 2  \omega_{[\mu \text{ } b}^{\text{ } \text{ } a} \omega^b_{\nu]}) \qed
\end{align*}

\paragraph{Proof of Eq.\ref{eq52} and Eq.\ref{eq53}}$\text{ }$ 
\\
We begin with 
\begin{align}
-2\Gamma^\rho_{[\mu \sigma]} h_{\nu \rho} -2 \Gamma^\rho_{[\nu \sigma]} h_{\mu \rho} +2 \Gamma^\rho_{[\mu \nu]} h_{\rho \sigma}   - 2 \tau_\mu e^a_{[\sigma} \omega_{\nu ] a} -   2 \tau_\nu e^a_{[\sigma} \omega_{\mu ] a} = \tau_\mu K_{\sigma \nu} +\tau_\nu K_{\sigma \mu} + L_{\sigma \mu \nu} \tag{A1}\label{eqA1}
\end{align}

First, we contract both sides of Eq.\ref{eqA1} with $v^\nu$ and anti-symmetrize the result in $\{\mu, \sigma\}$ by subtracting 
\begin{align}
-2 v^\nu \Gamma^\rho_{[\nu \sigma]} h_{\mu \rho} +2 v^\nu \Gamma^\rho_{[\mu \nu]} h_{\rho \sigma}   -  2 v^\nu\tau_\mu e^a_{[\sigma} \omega_{\nu ] a} +  2 e^a_{[\sigma} \omega_{\mu ] a} &= v^\nu \tau_\mu K_{\sigma \nu} - K_{\sigma \mu} - v^\nu L_{\nu \mu \sigma} \nonumber \\
\text{with}\hspace{0.2cm}-2 v^\nu \Gamma^\rho_{[\nu \mu]} h_{\sigma \rho} +2 v^\nu \Gamma^\rho_{[\sigma \nu]} h_{\rho \mu}   -  2 v^\nu\tau_\sigma e^a_{[\mu} \omega_{\nu ] a} +  2 e^a_{[\mu} \omega_{\sigma ] a} &= v^\nu \tau_\sigma K_{\mu \nu} - K_{\mu \sigma} - v^\nu L_{\nu \sigma \mu} \nonumber  \\
\nonumber \\
\Rightarrow \hspace{2.1cm}  -  2 v^\nu\tau_\mu e^a_{[\sigma} \omega_{\nu ] a}- 2 v^\nu\tau_\sigma e^a_{[\mu} \omega_{\nu ] a} + 4 e^a_{[\sigma} \omega_{\mu ] a} &=-2 v^\nu L_{\nu [\mu \sigma]} + 2v^\nu \tau_{[\mu} K_{\sigma] \nu} - 2K_{\sigma \mu} \tag{A2}\label{eqA2}
\end{align}

Next, we contract  Eq.\ref{eqA1} with $v^\mu$ and symmetrize the result in $\{\nu, \sigma\}$ by adding
\begin{align}
-2 v^\mu  \Gamma^\rho_{[\mu \sigma]} h_{\nu \rho}  + 2 v^\mu \Gamma^\rho_{[\mu \nu]} h_{\rho \sigma}  + 2 e^a_{[\sigma} \omega_{\nu ] a} -2 v^\mu  \tau_\nu e^a_{[\sigma} \omega_{\mu ] a} &= - K_{\sigma \nu} +v^\mu \tau_\nu K_{\sigma \mu} +  v^\mu L_{ \sigma \mu \nu} \nonumber \\
\text{with}\hspace{0.2cm}-2 v^\mu  \Gamma^\rho_{[\mu \nu]} h_{\sigma \rho}  + 2 v^\mu \Gamma^\rho_{[\mu \sigma]} h_{\rho \nu}  + 2 e^a_{[\nu} \omega_{\sigma ] a} -2 v^\mu  \tau_\sigma e^a_{[\nu} \omega_{\mu ] a} &= - K_{\nu \sigma} +v^\mu \tau_\sigma K_{\nu \mu} +  v^\mu L_{ \nu \mu \sigma} \nonumber  \\
\nonumber \\
\Rightarrow \hspace{4.1cm}  2 v^\mu  \tau_\nu e^a_{[\sigma} \omega_{\mu ] a} - 2v^\mu \tau_\sigma   e^a_{[\nu} \omega_{\mu ] a} &= v^\mu \tau_\nu K_{\sigma \mu} +v^\mu \tau_\sigma K_{\nu \mu}  \nonumber  \\
\text{or,} \hspace{0.2cm}  2 v^\nu  \tau_\mu e^a_{[\sigma} \omega_{\nu ] a} + 2v^\nu \tau_\sigma   e^a_{[\mu} \omega_{\nu ] a} &= -v^\nu \tau_\mu K_{\sigma \nu} -v^\nu \tau_\sigma K_{\mu \nu} \tag{A3} \label{eqA3}
\end{align}

We then contract Eq.\ref{eqA3} once again with $v^\sigma$: 
\begin{align}
	 \cancel{ 2 v^\sigma v^\nu  \tau_\mu e^a_{[\sigma} \omega_{\nu ] a} } - 2v^\nu   e^a_{[\mu} \omega_{\nu ] a} &= \cancel{ -v^\nu  v^\sigma \tau_\mu K_{\sigma \nu} } +v^\nu  K_{\mu \nu} \nonumber  \\
 2v^\nu   e^a_{[\mu} \omega_{\nu ] a} &=-v^\nu  K_{\mu \nu} \nonumber  \\
\text{or,} \hspace{0.2cm}  2v^\nu \tau_\sigma    e^a_{[\mu} \omega_{\nu ] a} &=-v^\nu \tau_\sigma   K_{\mu \nu} \tag{A4} \label{eqA4}
\end{align}

If we substitute Eq.\ref{eqA4} to Eq.\ref{eqA2}, we shall have
\begin{align*}
 4 e^a_{[\sigma} \omega_{\mu ] a} &= -2v^\nu L_{\nu [\mu \sigma]}    - 2K_{\sigma \mu}  \\
\text{or,} \hspace{0.2cm}  2 \omega^a_{[\mu} e_{\nu ] a} &= -v^\lambda L_{\lambda [\mu \nu]}  + K_{\mu \nu},  \qed
\end{align*}

proving Eq.\ref{eq52}.

Finally, we can substitute Eq.\ref{eq52} to Eq.\ref{eqA1} to obtain 
\begin{align*}
-2\Gamma^\rho_{[\mu \sigma]} h_{\nu \rho} -2 \Gamma^\rho_{[\nu \sigma]} h_{\mu \rho} +2 \Gamma^\rho_{[\mu \nu]} h_{\rho \sigma}  &= L_{\sigma \mu \nu } - \tau_\mu v^\lambda L_{\lambda [\nu \sigma]} -  \tau_\nu v^\lambda L_{\lambda [\mu \sigma]} \qed
\end{align*}

Which is exactly Eq.\ref{eq53}.
\\
\\
We can also contract Eq.\ref{eqA1} with $v^\mu$ and anti-symmetrize the result in $\{\mu, \nu\}$ by subtracting
\begin{align*}
-2v^\sigma \Gamma^\rho_{[\mu \sigma]} h_{\nu \rho} - 2 v^\sigma \Gamma^\rho_{[\nu \sigma]} h_{\mu \rho} - 2  v^\sigma \tau_\mu e^a_{[\sigma} \omega_{\nu ] a} -   2 \tau_\nu e^a_{[\sigma} \omega_{\mu ] a} & =  v^\sigma  L_{\sigma \mu \nu} + v^\sigma \tau_\mu K_{\sigma \nu} +  v^\sigma \tau_\nu K_{\sigma \mu} \\
\text{with}\hspace{0.2cm} -2v^\sigma \Gamma^\rho_{[\nu \sigma]} h_{\mu \rho} - 2 v^\sigma \Gamma^\rho_{[\mu \sigma]} h_{\nu \rho} - 2  v^\sigma \tau_\nu e^a_{[\sigma} \omega_{\mu ] a} -   2 \tau_\mu e^a_{[\sigma} \omega_{\nu ] a} & =  v^\sigma  L_{\sigma \nu \mu} + v^\sigma \tau_\nu K_{\sigma \mu} +  v^\sigma \tau_\mu K_{\sigma \nu} \\
\\
\Rightarrow \hspace{4.2cm} -4v^\sigma \tau_\mu  e^a_{[\sigma} \omega_{\nu ] a} - 4 v^\sigma \tau_\nu  e^a_{[\sigma} \omega_{\mu ] a} &= 2v^\sigma  L_{\sigma [\mu \nu]}  \\
\text{or,} \hspace{0.2cm}   4v^\nu \tau_\mu  e^a_{[\sigma} \omega_{\nu ] a} - 4 v^\nu \tau_\sigma  e^a_{[\mu} \omega_{\nu ] a} &= 2v^\nu  L_{\nu [\mu \sigma]}  
\end{align*}

This equation shows us that the temporal component of $L_{\sigma \mu \nu}$ is completely determined by the Newton-Coriolis; this illustrates why $L_{\sigma \mu \nu}$ cannot be identified with the spatial contortion in general. 

\paragraph{Computing $\delta_G \Gamma^\lambda_{\mu \nu}$ for Eq.\ref{eq60} and Eq.\ref{eq61}}$\text{ }$

We are to proof
\begin{align*}
 \delta_G \Gamma^\lambda_{\mu \nu}&=\delta_G\Big[  -  v^\lambda  \partial_\mu \tau_\nu  +  \frac{1}{2}  h^{\lambda \sigma} (\partial_\mu h_{\nu \sigma} + \partial_\nu h_{\mu \sigma}-\partial_\sigma h_{\mu \nu})+  \frac{1}{2}  h^{\lambda \sigma} (\tau_\mu K_{\sigma \nu} +\tau_\nu K_{\sigma \mu} + L_{\sigma \mu \nu})\Big]  \\
=&\frac{1}{2}  h^{\lambda \sigma} \Big[ 2 \tau_\nu \partial_{[\mu} \lambda_{\sigma]} + 2 \tau_\mu \partial_{[\nu} \lambda_{\sigma]}  -2\lambda_\sigma \partial_{[\mu} \tau_{\nu]}+ 2\lambda_\mu \partial_{[\nu} \tau_{\sigma]} + 2 \lambda_\nu \partial_{[\mu} \tau_{\sigma]} \Big]  \\
&+\frac{1}{2}  h^{\lambda \sigma} \big[\tau_\mu \delta_G K_{\sigma \nu}  + \tau_\nu \delta_G K_{\sigma \mu} + \delta_G L_{\sigma \mu \nu}\big]
\end{align*}

$\delta_G$ for the Hartong-Obers parameters term are obvious since both $h^{\lambda \sigma}$ and $\tau_\mu$ are $G$-invariants.  For the "Levi-Civita" term, we note that
 \begin{align*}
\delta_G( -  v^\lambda  \partial_\mu \tau_\nu )&= -\lambda_\sigma h^{\lambda \sigma} \partial_\mu \tau_\nu,
\end{align*}

whereas, 
 \begin{align*}
\delta_G \Big[ \frac{1}{2}  h^{\lambda \sigma} (\partial_\mu h_{\nu \sigma} + \partial_\nu h_{\mu \sigma}-\partial_\sigma h_{\mu \nu})\Big]=\frac{1}{2}  h^{\lambda \sigma} &\Big[\partial_\mu (\lambda_\sigma \tau_\nu + \lambda_\nu \tau_\sigma) + \partial_\nu (\lambda_\sigma \tau_\mu + \lambda_\mu \tau_\sigma) -  \partial_\sigma (\lambda_\nu \tau_\mu + \lambda_\mu \tau_\nu)  \Big]  \\
=\frac{1}{2}  h^{\lambda \sigma} &\Big[ 
(\partial_\mu \lambda_\sigma) \tau_\nu +(\partial_\mu \tau_\nu) \lambda_\sigma + (\partial_\mu \lambda_\nu) \tau_\sigma + (\partial_\mu \tau_\sigma) \lambda_\nu \big)\\
+&(\partial_\nu \lambda_\sigma) \tau_\mu+ (\partial_\nu \tau_\mu) \lambda_\sigma+ (\partial_\nu \lambda_\mu) \tau_\sigma + (\partial_\nu \tau_\sigma) \lambda_\mu \big) \Big]\\
-&(\partial_\sigma \lambda_\nu) \tau_\mu -(\partial_\sigma \tau_\mu) \lambda_\nu - (\partial_\sigma \lambda_\mu) \tau_\nu - (\partial_\sigma \tau_\nu) \lambda_\mu \big) \\
= \frac{1}{2}  h^{\lambda \sigma} &\Big[  \tau_\sigma   (\partial_\mu \lambda_\nu+ \partial_\nu \lambda_\mu) + \tau_\nu (  \partial_\mu \lambda_\sigma  - \partial_\sigma \lambda_\mu )  + \tau_\mu (\partial_\nu \lambda_\sigma- \partial_\sigma \lambda_\nu)\\
 +&\lambda_\sigma (\partial_\mu \tau_\nu +\partial_\nu \tau_\mu)+ \lambda_\nu (\partial_\mu \tau_\sigma -\partial_\sigma \tau_\mu) +\lambda_\mu (\partial_\nu \tau_\sigma -\partial_\sigma \tau_\nu) \Big].
\end{align*}

Hence, 
\begin{align*}
\delta_G & \Big[  -  v^\lambda  \partial_\mu \tau_\nu  +  \frac{1}{2}  h^{\lambda \sigma} (\partial_\mu h_{\nu \sigma} +\partial_\nu h_{\mu \sigma}-\partial_\sigma h_{\mu \nu})  \Big] \\
&= \frac{1}{2}  h^{\lambda \sigma} \Big[  \tau_\sigma   (\partial_\mu \lambda_\nu+ \partial_\nu \lambda_\mu) + \tau_\nu (  \partial_\mu \lambda_\sigma  - \partial_\sigma \lambda_\mu )  + \tau_\mu (\partial_\nu \lambda_\sigma- \partial_\sigma \lambda_\nu) \\
 &\hspace{1.6cm}-\lambda_\sigma (\partial_\mu \tau_\nu -\partial_\nu \tau_\mu)+ \lambda_\nu (\partial_\mu \tau_\sigma -\partial_\sigma \tau_\mu) +\lambda_\mu (\partial_\nu \tau_\sigma -\partial_\sigma \tau_\nu) \Big] \\
&= \frac{1}{2}  h^{\lambda \sigma} \big[ 2\tau_\nu \partial_{[\mu} \lambda_{\sigma]}  +2 \tau_\mu \partial_{[\nu} \lambda_{\sigma]} -2\lambda_\sigma \partial_{[\mu} \tau_{\nu]} + 2 \lambda_\nu \partial_{[\mu} \tau_{\sigma]}+ 2\lambda_\mu \partial_{[\nu} \tau_{\sigma]}\big]. \qed
\end{align*}

This completes the proof.


\newpage


\begin{thebibliography}{1}
\bibitem{Cartan}E. Cartan,  (1923). {\em "Sur les variétés à connexion affine et la théorie de la relativité généralisée (première partie)"}, Ann. Sci. Ecole Norm. Sup. {\bf 40}: 325-412.
\bibitem{Friedrichs}K. O. Friedrichs,  (1927). {\em "Eine Invariante Formulierung des Newtonschen Gravitationsgesetzes und der Grenzüberganges vom Einsteinschen zum Newtonschen Gesetz"}, Mathematische Annalen, {\bf 98}: 566–575.
\bibitem{MTW}      Charles W. Misner, Kip S. Thorne, and John Archibald Wheeler. {\em "Gravitation"}, (W. H. Freeman, 1973)
\bibitem{Havas} P. Havas, (1964). {\em "Four-dimensional formulations of Newtonian mechanics and their relation to the special and general theory of relativity"}, Reviews of Modern Physics, {\bf 36} (4): 938–965.
\bibitem{Kunzle} H. Künzle, (1976). {\em "Covariant Newtonian limts of Lorentz space-times"}, General Relativity and Gravitation, {\bf 7} (5): 445–457.
\bibitem{Dixon}W. G. Dixon, (1975). {\em "On the uniqueness of the Newtonian theory as a geometric theory of gravitation"}, Communications in Mathematical Physics, {\bf 45} (2): 167–182
\bibitem{Horava} P. Horava, (2009). {\em "Quantum gravity at a Lifshitz point"}, Phys. Rev. D. {\bf 79} (8): 084008. 
\bibitem{HartongObers} J. Hartong, N.A. Obers, (2015). {\em "Hořava-Lifshitz gravity from dynamical Newton-Cartan geometry"}, J. High Energ. Phys. 2015, {\bf 155}.
\bibitem{Geracie} M. Geracie, K. Prabhu, Matthew M. Roberts, (2015). {\em "Curved non-relativistic spacetimes, Newtonian gravitation and massive matter"}, Journal of Mathematical Physics {\bf 56}, 103505
\bibitem{Bargmann}V. Bargmann, (1954). {\em "On Unitary Ray Representations of Continuous Groups"}, Annals of Mathematics, Second Series, {\bf 59}, No. 1, pp. 1–46
\bibitem{HartongObers2} J. Hartong, E. Kiritsis, N.A. Obers, (2015). {\em "Lifshitz space–times for Schrödinger holography"}, Physics Letters B. {\bf 746} (C), pp. 318-324.
\bibitem{Schrodinger1}C. R. Hagen, (1972). {\em “Scale and conformal transformations in Galilean-covariant field the-ory”} Phys. Rev. D{\bf5} 377.
\bibitem{Schrodinger2}U. Niederer, (1972). {\em “The maximal kinematical invariance group of the free Schroedinger equation”} Helv. Phys. Acta {\bf 45} 802.
\bibitem{Isham} Chris J. Isham, {\em "World Scientific Lecture Notes in Physics: Volume 61. Modern Differential Geometry for Physicists"}, 2nd ed. (WSPC, 1999)
\bibitem{Nakahara} M. Nakahara, {\em "Geometry, Topology and Physics"}, 2nd ed. (CRC Press, 2003)
\bibitem{Bergshoeff} R. Andringa, E. Bergshoeff, S. Panda, and M. de Roo, (2011). {\em “Newtonian Gravity and the Bargmann Algebra”}, Class. Quant. Grav.{\bf 28} 105011.
\bibitem{DennisHansen} D. Hansen (2016). {\em "On Non-Relativistic Field Theory and Geometry"}. Retrieved from \url{http://www.nbi.dk/~obers/MSc_PhD_files/Dennis_Hansen_MSc.pdf}
\bibitem{InonuWigner} E. Inönü, E. P. Wigner, (1953). {\em "On the Contraction of Groups and Their Representations"}, Proc. Natl. Acad. Sci. {\bf 39} (6): 510–24.
\bibitem{GeracieBargmann} M. Geracie, K. Prabhu, Matthew M. Roberts, (2015). {\em "Fields and fluids on curved non-relativistic spacetimes"}, J. High Energ. Phys. 2015, {\bf 42}.
\bibitem{Dawson}John F. Dawson, {\em "Quantum Mechanics: Fundamental Principles and Applications"}, lecture notes, University of New Hampshirem, October 14, 2009.
\bibitem{Lian2} K. N. Lian (2020). {\em "Path Integral Formulation and Holonomy Groups in Newton-Cartan Schwarzschild Geometry".} arXiv.
\bibitem{Bleeken} D. Van den Bleeken, (2017). {"Torsional Newton-Cartan gravity from the large c expansion of General Relativity"}, Class. Quant. Grav. {\bf 34} no.18, 185004.
\bibitem{Frobenius} G. Frobenius,  (1877). {\em "Uber das Pfaffsche problem"}, J. Reine Angew. Math., {\bf 82}: 230-315.
\bibitem{Frobenius1} C. von Westenholz, {\em "Differential Forms in Mathematical Physics"}, Studies in Mathematics and Its Applications, Volume 3, pp.208-256.  
\bibitem{Frobenius2} P. Renteln, (1978). {\em "Manifolds, Tensors, and Forms An Introduction for Mathematicians and Physicists"}, 1st ed, pp.165-187. (Cambridge University Press, 2013).  
\end{thebibliography}
\end{document}